\documentclass[aps,english,floatfix,tightenlines,twocolumn,nofootinbib]{revtex4-2}
\renewcommand{\thesection}{\arabic{section}}

\usepackage{amssymb}
\usepackage{amsmath,bm}
\usepackage{amssymb}
\usepackage{graphicx}
\usepackage{amsfonts}         
\usepackage{fancybox}

\usepackage[usenames,dvipsnames]{xcolor}%http://en.wikibooks.org/wiki/LaTeX/Colors

\definecolor{Mathematica1}{rgb}{0.368417, 0.506779, 0.709798}
\definecolor{Mathematica2}{rgb}{0.880722, 0.611041, 0.142051}
\definecolor{Mathematica3}{rgb}{0.560181, 0.691569, 0.194885}

\definecolor{TGorange}{RGB}{140,50,0}
\definecolor{JMblue}{RGB}{25,25,125}

\usepackage[usenames,dvipsnames]{xcolor}%http://en.wikibooks.org/wiki/LaTeX/Colors

\usepackage[pagebackref,  % this puts links to the page numbers where refs appear
bookmarks={false}, pdfauthor={John McGreevy, Tarun Grover}, pdftitle={Liquid-solid transitions}]{hyperref}
\hypersetup{colorlinks=true, 
linkcolor=Mathematica1,
citecolor=Mathematica3,
filecolor=OliveGreen, 
urlcolor=Mathematica2,
filebordercolor={.8 .8 1}, 
urlbordercolor={.8 .8 0}}%http://en.wikibooks.org/wiki/LaTeX/Hyperlinks
\usepackage{soul}%strike through text \st{} .  not in math mode.
\setstcolor{Red}
\definecolor{darkgreen}{rgb}{0,0.4,0}
\definecolor{darkred}{rgb}{0.4,0,0}
\definecolor{darkblue}{rgb}{0,0,0.4}
\definecolor{lightblue}{rgb}{.6,.6,0.9}

\newcommand{\cor}{\color{red}}

\definecolor{uglybrown}{rgb}{0.8,  0.7,  0.5}

\definecolor{palatinatepurple}{rgb}{0.41, 0.16, 0.38}
\definecolor{celebrationcolor}{rgb}{0.75,  0.0,  0.9}

\definecolor{Mathematica1}{rgb}{0.368417, 0.506779, 0.709798}
\definecolor{Mathematica2}{rgb}{0.880722, 0.611041, 0.142051}
\definecolor{Mathematica3}{rgb}{0.560181, 0.691569, 0.194885}

\definecolor{shadecolor}{rgb}{0.90,0.90,0.90}
\definecolor{DVcolor}{rgb}{0.95,  0.5,  0.2}
\definecolor{lightbluemuons}{rgb}{0.0,.65,1.0}

\definecolor{chartreuse}{rgb}{0.70, 1.00, 0.00}

\usepackage{tikz}
\usetikzlibrary{shapes}
\usetikzlibrary{trees}
\usetikzlibrary{snakes}
\usetikzlibrary{matrix,arrows} 				% For commutative diagram
\usetikzlibrary{positioning}				% For "above of=" commands
\usetikzlibrary{calc,through}				% For coordinates
\usetikzlibrary{decorations.pathreplacing}  % For curly braces
\usepackage[tikz]{bclogo} 					% For cute logo boxes
\usepackage{pgffor}							% For repeating patterns

\usetikzlibrary{decorations.markings}
\usetikzlibrary{intersections}				% For interesections

\usetikzlibrary{arrows,decorations.pathmorphing,backgrounds,positioning,fit,petri,automata,shadows,calendar,mindmap, graphs}
\usetikzlibrary{arrows.meta,bending}

\tikzset{
    vector/.style={decorate, decoration={snake}, draw},
    fermion/.style={postaction={decorate},
        decoration={markings,mark=at position .55 with {\arrow{>}}}},
    fermionbar/.style={draw, postaction={decorate},
        decoration={markings,mark=at position .55 with {\arrow{<}}}},
    fermionnoarrow/.style={},
    gluon/.style={decorate,
        decoration={coil,amplitude=4pt, segment length=5pt}},
    scalar/.style={dashed, postaction={decorate},
        decoration={markings,mark=at position .55 with {\arrow{>}}}},
    scalarbar/.style={dashed, postaction={decorate},
        decoration={markings,mark=at position .55 with {\arrow{<}}}},
    scalarnoarrow/.style={dashed,draw},
	vectorscalar/.style={loosely dotted,draw=black, postaction={decorate}},
}

\def\centerarc[#1](#2)(#3:#4:#5)% Syntax: [draw options] (center) (initial angle:final angle:radius)
    { \draw[#1] ($(#2)+({#5*cos(#3)},{#5*sin(#3)})$) arc (#3:#4:#5); }

\usepackage{enumitem}

\usepackage{slashed}

\usepackage[font=small,labelfont=bf]{caption}
\DeclareCaptionFont{tiny}{\tiny}
\usepackage{epstopdf}
\usepackage{wasysym}

\usepackage[yyyymmdd,hhmmss]{datetime}   
\usepackage{framed}  % needed for \begin{shaded}
 \usepackage{mdframed}
 \usepackage{wrapfig}
 \usepackage{yfonts}  

\def\sign{\text{sign}}

\newmdenv[%
        backgroundcolor=lightgray,
    linecolor=black,
    outerlinewidth=2pt,
]{boxedandshaded}

\def\parfig#1#2{
\parbox{#1\textwidth}
{\includegraphics[width=#1\textwidth]{#2}}
}

\numberwithin{equation}{section}

\renewcommand{\theequation}{\arabic{section}.\arabic{equation}}

\def\nd{{ \vphantom{\dagger}}}

\newcommand{\vev}[1]{\langle #1 \rangle}

\newlength{\extraspace}
\newlength{\extraspaces}
\def\be{\begin{equation}}
\def\ee{\end{equation}}

\newcommand{\bea}{\begin{eqnarray}}
\newcommand{\eea}{\end{eqnarray}}

\def\dbar{{\mathchar'26\mkern-12mu \dd}}

\def\eps{\epsilon}
\def\half{{1\over 2}}

\def\tr{{\rm tr}}

\def\Im{{\rm Im\hskip0.1em}}

\def\vev#1{\left\langle{#1}\right\rangle}

\def\CD{{\cal D}}

\def\CN{{\cal N}}
\def\CO{{\cal O}}%AEL

\def\II{\relax{I\kern-.10em I}}

\def\IZ{\mathbb{Z}}
\def\IB{\relax{\rm I\kern-.18em B}}

\def\ID{\relax{\rm I\kern-.18em D}}
\def\IE{\relax{\rm I\kern-.18em E}}
\def\IF{\relax{\rm I\kern-.18em F}}
\def\IG{\relax\hbox{$\inbar\kern-.3em{\rm G}$}}
\def\IGa{\relax\hbox{${\rm I}\kern-.18em\Gamma$}}
\def\IH{\relax{\rm I\kern-.18em H}}
\def\II{\relax{\rm I\kern-.18em I}}
\def\IK{\relax{\rm I\kern-.18em K}}
\def\inbar{\,\vrule height1.5ex width.4pt depth0pt}

\def\simgt{\hskip0.05in\relax{ 
\raise3.0pt\hbox{ $>$
{\lower5.0pt\hbox{\kern-1.05em $\sim$}} }} \hskip0.05in}

\def\lp10{\ell_p^{10}}
\def\lp11{\ell_p^{11}}
\def\R11{R_{11}}

\def\frac#1#2{{#1 \over #2}}

\def\Ione{\hbox{$1\hskip -1.2pt\vrule depth 0pt height 1.53ex width 0.7pt
                  \vrule depth 0pt height 0.3pt width 0.12em$}}

\newdimen\tableauside\tableauside=1.0ex
\newdimen\tableaurule\tableaurule=0.4pt
\newdimen\tableaustep
\def\phantomhrule#1{\hbox{\vbox to0pt{\hrule height\tableaurule width#1\vss}}}
\def\phantomvrule#1{\vbox{\hbox to0pt{\vrule width\tableaurule height#1\hss}}}
\def\sqr{\vbox{%
  \phantomhrule\tableaustep
  \hbox{\phantomvrule\tableaustep\kern\tableaustep\phantomvrule\tableaustep}%
  \hbox{\vbox{\phantomhrule\tableauside}\kern-\tableaurule}}}
\def\squares#1{\hbox{\count0=#1\noindent\loop\sqr
  \advance\count0 by-1 \ifnum\count0>0\repeat}}
\def\tableau#1{\vcenter{\offinterlineskip
  \tableaustep=\tableauside\advance\tableaustep by-\tableaurule
  \kern\normallineskip\hbox
    {\kern\normallineskip\vbox
      {\gettableau#1 0 }%
     \kern\normallineskip\kern\tableaurule}%
  \kern\normallineskip\kern\tableaurule}}
\def\gettableau#1 {\ifnum#1=0\let\next=\null\else
  \squares{#1}\let\next=\gettableau\fi\next}

\def\({\left(}
\def\){\right)}

\def\ii{{\bf i}}

\def\dd{\text{d}}

\def\lsim{\mathrel{\mathstrut\smash{\ooalign{\raise2.5pt\hbox{$<$}\cr\lower2.5pt\hbox{$\sim$}}}}}
\def\gsim{\mathrel{\mathstrut\smash{\ooalign{\raise2.5pt\hbox{$>$}\cr\lower2.5pt\hbox{$\sim$}}}}}

\def\overleftrightarrow#1{\vbox{\ialign{##\crcr
     $\leftrightarrow$\crcr\noalign{\kern-0pt\nointerlineskip}
     $\hfil\displaystyle{#1}\hfil$\crcr}}}
     
     \def\overleftarrow#1{\vbox{\ialign{##\crcr
     $\leftarrow$\crcr\noalign{\kern-0pt\nointerlineskip}
     $\hfil\displaystyle{#1}\hfil$\crcr}}}

\def\eg{{\it e.g.}}
\def\ie{{\it i.e.}}

\def\gSO{\textsf{SO}}

\def\gU{\textsf{U}}

\newif{\ifeq}           % defines a new condition @eq tested by the conditional \ifeq
\eqtrue                 % if uncommented, declares @eq to be true
\newcounter{lecturecounter}
 \usepackage[indentafter]{titlesec}
 \titleformat{\section}{\normalfont\small\bfseries\centering}{\thesection.}{0.5em}{\MakeUppercase}\titlespacing*{\section}{0pt}{20 pt}{10 pt}

\usepackage[normalem]{ulem}

\def\inplane{\sqcap}
\def\outplane{\slashed{\sqcap}}
\def\dkinplane{\Delta k_\inplane}
\def\dpinplane{\Delta p_\inplane}
\def\dqinplane{\Delta q_\inplane}
\renewcommand{\theequation}{\arabic{equation}}

\numberwithin{equation}{section}
\renewcommand{\theequation}{\arabic{section}.\arabic{equation}}

\usepackage{etoolbox}
\makeatletter
\patchcmd\NAT@citexnum{\let\NAT@last@num\NAT@num}{\MakeLinkTarget[cite]{}\Hy@backout{\@citeb\@extra@b@citeb}\let\NAT@last@num\NAT@num}{}{\fail}
\makeatother

\begin{document}
%\maketitle
%\section{}
%\subsection{}

%\title{Brief Article}
%\author{The Author}
%\date{}							% Activate to display a given date or no date

%\title{Squaring the circle: continuous liquid metal to solid insulator transitions}
%\title{Death of a Fermi surface by freezing}
\title{A critical theory for solidification of a liquid Fermi liquid}

\author{Tarun Grover}
\author{John McGreevy}

\affiliation{Department of Physics\\University of California San Diego}

\begin{abstract}

We give a simple description of a zero-temperature phase transition between a liquid metal and a solid.
The critical point has a Fermi surface as well as a Bose surface, a sphere in momentum space of gapless bosonic excitations.  We find a fixed point of the renormalization group governing such a non-Fermi liquid, using an expansion in the codimension of both the Fermi and Bose surfaces.  
We comment on 
the nature of the solid phase and possible physical realizations.

\end{abstract}

%\today
\date{\today}

\maketitle

\section{Introduction}
%{\bf Introduction.} 

A Fermi surface is a dramatic low-energy manifestation of quantum mechanics. 
Despite its plethora of low-energy modes, a Fermi surface is a robust feature of a phase of matter. 
The question of how a Fermi surface can be destroyed \cite{Senthil_2008,Senthil_2008_criticalFS} 
or damaged 
\cite{abanov2000spin,Metlitski:2010pd, 
Metlitski:2010vm,Dalidovich:2013qta,Lee:2017njh}, as parameters are varied, is then an extremely interesting one.  
The robustness of a Fermi surface is tied to the combined effects of particle number conservation and translation symmetry via Luttinger's theorem~\cite{luttinger1960,Oshikawa00a}. It is natural to ask about the nature of the phase and the associated phase transition as one breaks either of these two symmetries spontaneously. The perturbations associated with particle number symmetry breaking are marginally relevant, and lead to the parametrically-low-temperature phenomenon of superconductivity. In contrast, phase transitions driven by translational symmetry breaking remain less well explored, particularly for short-range interacting fermions without a pre-existing lattice, such as in ultracold Fermi gases. Here, we propose a candidate theory for such a transition. A key feature of our theory is that it consists of Fermi surface coupled to a continuum of gapless bosonic modes (a `Bose surface'). Using a controlled renormalization group analysis, we identify a fixed point that governs this theory and find that there are no well-defined electron-like quasiparticles at the critical point.

\begin{figure}
%$$ \includegraphics[width=.4\textwidth]{fig-2d-geometry.pdf}
$$
\parfig{.24}{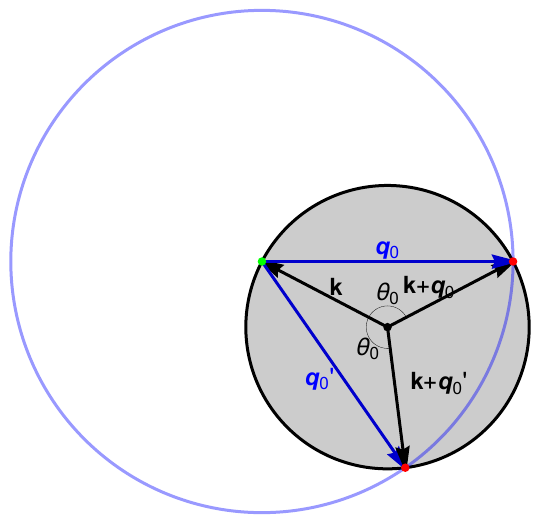}
%\parfig{.24}{fig-squaring-the-circle.pdf}
\parfig{.24}{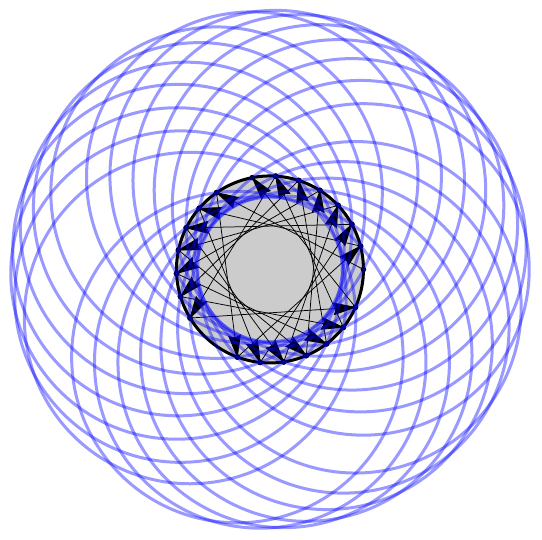}
$$
\caption{\label{fig:2d-geometry}
Left: At the critical point in 2d, each point $\vec k$ (green dot) on the Fermi surface (black circle) is coupled to two other momenta $\vec k + \vec q_0, \vec k + \vec q_0'$ ($|\vec q_0| = |\vec q_0'|=q_0$), 
the intersections (red dots) of the FS with the Bose surface centered at $\vec k$ (blue circle).   
%The figure is drawn to scale for the square lattice with $k_F/q_0$ chosen so that there is one particle per unit cell. 
Right: 
For generic values of $q_0/k_F = 2 \sin {\theta_0\over 2}$, 
each point on the FS is coupled to every other by a series of couplings to the gapless boson modes.
% That is, the angle between points connected by a vector of length $q_0$ is %($2 \sin^{-1}(\sqrt{\pi}/2)$) 
% an irrational multiple of $2\pi$.    
Shown here are a series of points on the FS, with the (blue) circle of points a distance $q_0$ away.  Each one is connected to the next by a vector of length $q_0$. 
% At the energetically-preferred value of $q_0/k_F$, the Fermi surface in the liquid phase (grey) and the Brillouin zone of the solid phase (blue) have the same area $\pi k_F^2 = q_0^2$.
}
\end{figure}

For simplicity, we focus on neutral fermions in the continuum with short-range interactions, so that they can form a Fermi liquid.  
Now we imagine tuning some parameter (say, pressure) so that the system has a tendency to spontaneously break translation invariance. Here we can appeal to the logic of 
%Refs.~
\cite{brazovskii1975phase, alexander1978should} for the form of the Landau-Ginzburg effective action for the order parameter $\rho(x)$ conjugate to the density.  The key result is that at the critical point, \textit{a whole sphere's worth of boson modes becomes gapless}, as happens also in \eg~\cite{dyugaev1976contribution,
swift1976fluctuations,
swift1977hydrodynamic,
PhysRevB.16.4137-swift-leitner, Mukamel_1980,
ling-PhysRevB.24.2718, 
dyugaev1982effects,
dyugaev1982crystalline,
hohenberg-brazovskii,
shiwa-brazovskii,
schmalian2004quantum,
Zhang_2023,sedrakyan2013composite,sedrakyan2014absence,sedrakyan2015statistical,maiti2019fermionization,sur2019metallic,Lake:2021omz}. 
 We will refer to this as a `Bose surface'. Explicitly, the Brazovskii-Alexander-McTague action is\footnote{Here and below, by analogy with $\hbar \equiv { h \over 2\pi}$, we employ the useful notation $ \int \dbar^dk \equiv \int { d^dk \over (2 \pi)^d}$.} $S_\text{BAM}[\rho]=$  
\be \int \dbar \omega~ \dbar^dq  \rho_q \rho_{-q} \( r + (\vec q^2 - q_0^2)^2 + \omega^2 \) + \int d\tau d^dx V(\rho) . 
\label{eq:SAT}\ee
where we will discuss the form of $V(\rho)$ later.
The fermions, with kinetic energy $S_\psi = \int d\tau \int \dbar^d k \psi^\dagger_{k\sigma}(\tau)\psi^\nd_{k\sigma}(\tau) \( \ii \omega - \left(k^2/2m - \mu\right) \)$, will couple to the density field via an interaction of the form $S_{\psi-\rho} = \int d^dx\,d\tau\,\, \psi^{\dagger}_\sigma(x,\tau) \psi^\nd_\sigma(x,\tau) \rho(x,\tau)$. 
$\sigma=1..s$ is a spin index. 
The full action is  $S = S_\psi + S_{\psi-\rho} + S_{\text{BAM}}$. 
 (See App.\ref{app:field_theory} for more detail.)  Because of the $\rho \psi^\dagger \psi$ interaction, when $r \approx 0$, the soft modes $\rho(x)$ couple all points on the Fermi surface to other points  (see Fig.\ref{fig:2d-geometry}), unlike 
at a spin density wave transition in a pre-existing lattice, where only `hot spots' on the Fermi surface are coupled to each other, and also unlike nematic or ferromagnetic instability of a Fermi surface where different points of the Fermi surface do not couple strongly with each other.  
Beyond this value of the tuning parameter $r$, a set of modes of $\rho$ with nonzero momentum condense and produce a lattice. 
Generically, if this transition is continuous or very weakly first order, then the conservative conclusion (see App.\ref{Sec:peierls-argument}) is that the solid phase will be a metal with both electron and hole Fermi surfaces \cite{ashcroft1978solid} (in $d=1$, the solid phase is an insulator; for general $d$, we discuss the possibility of a solid insulator in \S\ref{sec:RG} below). Our zeroth-order picture of the phase diagram is then:
$$ \parfig{.45}{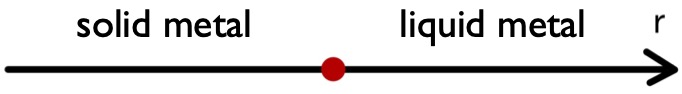}$$

{\bf General considerations.} 
%\section{General considerations}
A few points we must emphasize: (1) The liquid-solid transition, when dictated by the $S_{\text{BAM}}$ term alone, is commonly believed to be first order \cite{brazovskii1975phase,dyugaev1976contribution,swift1976fluctuations,swift1977hydrodynamic,PhysRevB.16.4137-swift-leitner, Mukamel_1980,
dyugaev1982effects,dyugaev1982crystalline,
ling-PhysRevB.24.2718, hohenberg-brazovskii,
Chaikin, kivelson-spivak,
shiwa-brazovskii}.  
The origin of this belief is twofold. 
First, in the absence of any symmetries beyond particle number conservation, rotations and translations, a cubic term is allowed in the effective action, which will lead to a first-order transition~\cite{Chaikin}. We focus on scenarios where either (i) the cubic term is fine-tuned to zero, making the theory multicritical, or (ii) an additional unitary $\mathbb{Z}_2$ symmetry forbids it. 
One possible physical realization of case (ii) is if the order parameter were the spin density
$ \psi^{\dagger} \sigma^z \psi$ at a spin-density wave transition. 
In case (i), this multicritical point contains all the universal information about the weakly first-order transition in its neighborhood.
We emphasize that in case (i), rotational symmetry enforces that the coupling for the cubic term is a number, and not a function ---unlike the quartic term in our theory or forward scattering in Fermi liquid theory. Consequently, the multicritical point can be reached by tuning an $\CO(1)$ number of relevant couplings. Later, we will briefly also explore the possibility of a weakly first-order transition between a liquid metal and solid \textit{insulator} in the vicinity of this multicritical point.
%The effects of fluctuations on such a transition at finite temperature have been studied extensively, beginning with Brazovskii 
%\cite{brazovskii1975phase,dyugaev1976contribution,swift1976fluctuations,swift1977hydrodynamic,PhysRevB.16.4137-swift-leitner, Mukamel_1980,
%dyugaev1982effects,dyugaev1982crystalline,
%PhysRevB.24.2718, hohenberg-brazovskii,shiwa-brazovskii}.  
The second reason the transition is believed to be first order, even if one prohibits a cubic term, is the following.  At $T >0$ (in low enough dimensions ($d<5$) but with any finite number of components of the order parameter) and near the would-be transition, the fluctuations renormalize the effective $r$ by an infrared (IR)-divergent negative amount -- the would-be critical theory is not self-consistent.   
Interestingly, at $T = 0$, this calculation has a different conclusion when we take into account 
the coupling between the order parameter and the Fermi surface. As we show in App.~\ref{appendix:continuous-transition}, at least within a self-consistent mean-field approximation, the fluctuations are IR finite at $T = 0$, and thus we may hope that the resulting transition might be continuous in the presence of the aforementioned $\mathbb{Z}_2$ symmetry. (2) The above description makes manifest that $q_0 = |\vec G|$ is the peak of the static structure factor in the liquid ($r>0$) phase. Near the transition, the location of the peak in the static structure factor $S(q)$ of the liquid determines the magnitude of the ordering wave vector.  In turn, according to the theory reviewed above, this determines the lattice spacing of the resulting solid phase.  {\it However}, the location of this peak of $S(q)$ in the liquid phase is {\it not} determined just by the density of the liquid.  If it were, for example if the peak were at $2\pi$ divided by the average interparticle spacing, we would arrive at a contradiction, since the density must be continuous across a second-order transition, but the density of any lattice with reciprocal lattice vectors of magnitude $|G|$ is different from that of a liquid with average interparticle spacing $2\pi/|G|$.  
(3) The lattice type is determined by  the quartic and higher-order terms in the potential $V(\rho)$ \cite{alexander1978should, Chaikin}. (4) The critical theory comes with two intrinsic length scales, $q_0$ and $k_F$, allowing for hyperscaling violation. 
Though one can be eliminated by a choice of units, universal properties of the critical theory will depend on their dimensionless ratio $q_0/k_F$.
(5) The role of $\rho$ in the theory may be played either by the density of the fermions themselves (as above), or by some other degree of freedom.

Below we will identify a certain fixed-point description of the field theory at $r=0$ and perform a scaling analysis of its perturbations.  The conclusion will be that a certain $\rho^4$ term is relevant 
%[cite Brazovski\cite{brazovskii1975phase} ? no. this is a distinct problem.] 
and that therefore this theory is naively infinitely multicritical (this is because, in analogy with BCS instability of a Fermi liquid, the coupling in the aforementioned $\rho^4$ term will be a function of certain angles parametrizing it, and therefore, potentially one may need to tune an infinite number of parameters). There are then two possibilities for the fate of the theory.  
One is that there is a nearby weakly-coupled fixed point with similar phenomenology, where all $\rho^4$ couplings flow to a finite value. This is analogous to the Wilson-Fisher fixed point arising from the gaussian fixed point for small $\epsilon = 4 - D$.  
As described below, and detailed in App.\ref{appendix:scaling}, we identify a codimension expansion 
\cite{2009PhRvL.102d6406S, Dalidovich:2013qta,Lee:2017njh, PhysRevB.16.4137-swift-leitner} in which we can indeed find such a fixed point. In the absence of a cubic term, this theory is a single-parameter tuned transition between a liquid metal and solid metal.
The second possibility is that the system runs to strong coupling; we discuss this possibility at the end of the paper.
%In Appendix \ref{app:nematic-mft} we describe the effect of this strong interaction on the phase diagram, and explain why the critical field theory described above may nevertheless be useful even in this case.

%Based on the above discussion, we arrive at the following point of view.
%Despite the fact that $q_0/k_F$ appears to be a marginal coupling in our effective field theory, we believe that it is in fact fixed by microscopic energetic considerations. 
%We can compare this situation to the Peccei-Quinn solution of the strong CP problem \cite{Peccei:1977hh}.  Naively the QCD theta angle is a dimensionless coupling in the Standard Model.  The Peccei-Quinn idea is that it is actually the zeromode of a dynamical variable, whose value is therefore determined energetically (to be zero).  In this high-energy physics context, it is somewhat of a leap to suggest that the theta angle is dynamical and requires the addition of previously-undiscovered degrees of freedom; in our system where we know the microscopic degrees of freedom, it is much less controversial.   

\section{Renormalization group and universal properties}
%{\bf Renormalization Group and universal properties.}
\label{sec:RG}
Above we motivated the action $S = S_{\text{BAM}} + S_\psi + S_{\psi-\rho}$
 for a liquid metal to solid transition. This action omits the boson self-interaction, which we will soon see is important. However, let us first follow a large-$N$ RG for this action which is similar in spirit to \cite{shamit-matrix-large-N}. We replace  the fermion field  by an $N$-component vector $\psi_\alpha$,  the boson field by an $N\times N$ matrix $\rho_{\alpha\beta}$, and the interaction takes the form $ \psi_\alpha^\dagger \rho_{\alpha\beta} \psi_\beta^\nd$. One way to package the analysis is in terms of self-consistent Schwinger-Dyson equations for the fermion ($G$) and boson ($D$) Green's functions (see App.\ref{appendix:self-consistency} for details). Approaching the transition from the liquid side, the self-energies $\Pi, \Sigma$ are as follows \cite{Abanov_2001, Abanov_2001a, Abanov_2003,  PhysRevLett.93.255702, Metlitski:2010vm, Hartnoll:2011ic, Sachdev:2012if}:
the Landau damping correction to the boson self-energy is of the form $\Pi(\epsilon, q) = {|\epsilon| \over q} \simeq  {|\epsilon| \over |G|}$. This singular term in $\epsilon$ dominates over the bare $\eps^2$ kinetic term in  \eqref{eq:SAT} and we can drop the latter.
 The leading correction to the fermion self-energy right at the transition scales as $ \Sigma(\omega,k) \sim  
\omega^{z-1 \over z}=\sqrt{\omega}$, as previously seen in various avatars in \cite{schmalian2004quantum, Zhang_2023,dyugaev1976contribution,dyugaev1982effects}, and therefore,
 the whole critical theory has dynamical critical exponent $z=2$, in this approximation, like the SDW critical theory (in the same approximation).   The singular part of the self energy is independent of the fermion momentum, like the case of the Ising nematic transition or Fermi surface coupled to gauge field \cite{Metlitski:2010pd} but unlike the case of the spin-density wave transition \cite{Metlitski:2010vm}. This correction to the self energy means that near the transition the Green's function has the form $G(\omega,k) \simeq { Z(r) \over \ii \omega - v_F k_\perp}$
 with  $ Z = 2 \sqrt{r} \to 0, v_F = v_F^0 2 \sqrt{r} \to 0$, 
 and therefore the effective mass diverges as ${m^\star \over m} \propto {1\over v_F} \sim {1\over \sqrt{r}}$. 

Next, we return to the question of boson self-interaction, which we and others 
\cite{schmalian2004quantum, Zhang_2023,dyugaev1976contribution,dyugaev1982effects} have ignored so far. As in the analogous analysis of the BCS theory \cite{Polchinski:1992ed, shankar-RG}, the most relevant part of the quartic boson self-interaction is 
 the forward-scattering part\footnote{Because the boson is real, forward scattering, back-to-back scattering and the BCS channel are all the same interaction.}\footnote{As in the Fermi liquid theory, the purely forward scattering interaction is not local.  It is an approximation to a fully local fixed-point theory that also involves nearly-forward scattering, and, in the case of the Fermi liquid, has been studied in \cite{2024PhRvB.109d5143M,Ye:2021usa}.  We defer the construction of an analogous fully local version to the future.  Similar comments apply to the Yukawa interaction.}: $S_\text{forward} = \int \dbar^d q_1 \dbar^d q_2 d\tau \,\rho_{q_1,\tau} \rho_{q_2,\tau} \rho_{-q_1,\tau} \rho_{-q_2,\tau} u(q_1,q_2)$, so that the full action is $S = S_{\text{BAM}} + S_\psi + S_{\psi-\rho} + S_\text{forward}$. The boson self-interaction is in fact relevant both at the Gaussian fixed-point $S_0 = S_{\textrm{BAM}} + S_\psi$ as well as at the aforementioned large-$N$ fixed point, and therefore, cannot be neglected. We make progress on this full action by perturbing around the Gaussian fixed-point via a `co-dimension expansion' \cite{2009PhRvL.102d6406S,Dalidovich:2013qta, Lee:2017njh}: we assume that the Fermi surface and Bose surface have codimension $c$ in momentum space, and they lie in the same $(d-c)$-dimensional subspace of the $d$-dimensional momentum space. We restrict ourselves to $d-c = 1$, so that the Fermi surface is one-dimensional (`nodal-line'). The physical value of $c$ for a Fermi surface is of course one. We find when $c = 3$, $S_\text{forward}$ is marginal, while $S_{\psi-\rho}$ is irrelevant for generic kinematics (all four-fermion interactions, including forward scattering, are also irrelevant). By perturbing around the Gaussian fixed-point at $c = 3$, we find (App.\ref{appendix:g-is-dangerously-irrelevant}) that the RG flow of $u_\ell = \frac{1}{2\pi}\int u_{\text{forward}}(\theta) e^{i \ell \theta}$ is $\beta_{u_\ell} 
= \eps u_\ell - { 4N_d \gamma  } u_\ell^2$, 
with $N_d = {1\over 16 \pi^3}$,
so that there is a stable
fixed point at $ u^{*}_\ell = { \eps  \over 4 N_d \gamma } $
which we identify as the liquid metal-solid metal transition within the present scheme. The $\ell$-independence of $u^{*}_\ell$ implies that at the critical point, only $u_{\text{forward}}(\theta=0)$ is non-zero. Notably, $S_{\psi-\rho}$ played no role at the critical point described above, and therefore, it is \textit{dangerously irrelevant} - in the solid phase fermions are gapped out precisely due to $S_{\psi-\rho}$. Therefore, in this theory, the fermions do not acquire an anomalous dimension, although the fermion density operator does: this is not a Landau Fermi liquid.

One issue with the preceding calculation is that in fact there are special kinematics for which the Yukawa interaction scales faster, analogous to forward scattering.
To improve on this theory, we next isolate the most singular part of $S_{\psi-\rho}$ by considering an interaction of a form that constrains the fermions to scatter by an angle $\pm \theta_0$ (Fig.~\ref{fig:2d-geometry}) when they emit a bosonic mode,  keeping them on the Fermi surface when the boson is critical.  Schematically, this modified interaction takes the form $S'_{\psi-\rho} = g \int \dbar^Dk_1 ~\dbar^D k_2 ~\dbar^D q 
\delta^D ( -k_1 + k_2 + q) \cdot \bar \Psi_{k_1} M \Psi_{k_2} \rho_{q}
~\sqrt{\delta(\theta_1- \theta_2, \theta_0)}$ where $D$ denotes the total space-time dimension, and $\sqrt{\delta(\theta, \theta_0)} \equiv 
\sqrt{\delta(\theta - \theta_0)} + \sqrt{ \delta(\theta + \theta_0)}$ constrains the scattering to an angle $\pm \theta_0$. 
The matrix $M$ is chosen so that the interaction has a symmetry that sends $\rho_q \to -\rho_q, \bar \Psi_{k_1} M \Psi_{k_2} \to -\bar \Psi_{k_1} M \Psi_{k_2}$, thus ruling out a  term cubic in $\rho_q$.

Remarkably, the modified version of the boson-fermion interaction, $S'_{\psi-\rho}$, is also marginal exactly at $c = 3$, thereby allowing a perturbative RG calculation where both boson self-interaction and boson-fermion interaction play a crucial role. The salient features of the RG for the full action $S = S_{\text{BAM}} + S_\psi + S'_{\psi-\rho} + S_\text{forward}$ are as follows (App.\ref{appendix:sqrt-delta-scheme}). (1) Most importantly, we find a stable fixed point for all the couplings in $S$ above (besides the tuning parameter $r$), including the ratio of fermion to boson velocity, $v_F/v_B$. This fixed-point describes the phase transition from liquid Fermi-liquid to solid Fermi liquid we are after. 
We note that our varying-codimension theory is completely local; we find that without this property, a fixed point does not exist.  
(2) Unlike a Fermi liquid or the aforementioned fixed-point where  only $u^\star(0)$ survives, now $u^{*}(\theta)$ is non-zero for \textit{all} $\theta$, with two prominent peaks at $\theta = 0, \theta_0$ (see Fig.~\ref{fig:brazovskii-wins}). 

 \begin{wrapfigure}[15]{R}{0.1\textwidth}
 \vskip-.3in
\centering
 \parbox{.1\textwidth}{\begin{tikzpicture}[line width=1.0 pt, scale=.5, rotate=90]
%   \draw[fermion] (0,0) circle (1);
      \draw [fermion]      (2,2) -- node[midway,above ]{$k$} (1,1);
      \draw[fermion] (1,1)  -- node[midway,below left]{$k_1$}  (0,0) ;%node[midway,above]{$\omega-\epsilon, q-k$};
      \draw [fermion]    (0,0) -- node[midway, below right]{$k_1+q$} (1,-1);
      \draw[fermion] (1,-1) --node[midway,above]{$k' = k+q$} (2,-2) ;%node[midway,above]{$\omega-\epsilon, q-k$};

    \draw[dashed] (-2,0)--node[midway,right]{$q$}(0,0) ; %node[midway, below]{$(\epsilon, k)$};
     \draw [line width = .5pt, ->] (-1.5, -.1) --  (-.5, -.1);
      %-- node[below] {$\epsilon, k$} 
%     ++ (0.4, 0); 
    
    \draw[dashed] (1,-1)--node[midway,above]{$k_1-k$}(1,1) ; %node[midway, below]{$(\epsilon, k)$};
         \draw [line width = .5pt, ->] (1.1, -.5) --  (1.1, .5);

%         \draw [line width = .5pt, ->] (0.3, -.3) 
          %-- node[below] {$\epsilon, k$} 
%         ++ (0.4, 0); 
%   \draw[fermionbar] (30:1)--(0,0);
%   \draw[vector] (140:1)--(0,0);
%   \draw[vector] (-140:1)--(0,0);
    \end{tikzpicture}
     }
\caption{\label{fig:vertex-correction}Given that $k, k', q$ all lie on the critical surfaces, it is not possible to put $k_1, k_1+q, k_1-k$ on the critical surfaces.}
\end{wrapfigure}
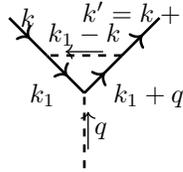
(3) Due to kinematical constraints (see Fig.~\ref{fig:vertex-correction}), the vertex corrections for $S'_{\psi-\rho}$ are \textit{exactly zero}, a statement similar to but stronger than Migdal's theorem. (4) The critical velocity ratio $|v_F|/|v_B|$ is fully determined by $q_0/k_F$,  $|v_F|/|v_B| = \sin(\theta_0/2) \equiv {q_0 \over 2 k_F }$. 

(5) The anomalous dimensions for both the fermion and the boson are non-zero, underlining that there are no well-defined quasiparticles at the critical point, and the critical matter is a non-Fermi liquid. 
The anomalous dimensions for the fermion and boson also depend on the dimensionless parameter: 
\be \eta_\psi = { 6 \eps \over 12 + s \cos^2 { \theta_0 \over 2} },
~~~ \eta_\rho = { s \eps \cos^2 {\theta_0 \over 2} \over 12 + s \cos^2 { \theta_0 \over 2} } 
\ee
where $s$ is the number of spin components.
We also calculated the universal exponent associated with the correlation length divergence, $\eta_r$.
We refer to App.\ref{appendix:sqrt-delta-scheme}
 for details of the RG calculation.

(6) The non-interacting part of the action for fermions within our co-dimension expansion involves a `projected Dirac algebra', despite being non-relativistic. Schematically, it looks like $\int \dbar ^D k \bar \Psi  \ii \( \omega \Gamma_0^\theta + k_z \Gamma_z^\theta + v_F k_\inplane \Gamma_\inplane^\theta \) \Psi$, where  $\{\Gamma_\mu^\theta, \Gamma_\nu^\theta \}  = 4 \delta_{\mu\nu} P_-(\theta)$ 
where $\mu,\nu \in \{ \omega, z, \inplane\} $ and 
$P_-(\theta) $ is a projector onto the low-energy bands. We anticipate that this observation might be useful for studying the physics of nodal-line semimetals.

\textbf{Liquid metal to solid insulator transition:} 
The theory we discuss  describes a transition out of a liquid metal to a state that is, within mean field theory, a solid metal. This is because in $d > 1$, when a metal with non-nested Fermi surface\footnote{One possibility for a direct liquid metal to solid insulator transition (still ignoring the cubic coupling) is if the Fermi surface immediately on the solid side of the transition is nested.  In this case, the four-fermion interactions are dangerously irrelevant at the transition and at any $r<0$ take the system to an insulating phase.  Since the lattice structure as well as the lattice spacing of the solid is dynamically selected, this mechanism for realizing an insulator may be favorable. } is subjected to a weak periodic potential, the resulting band structure consists of electron and hole Fermi surfaces \cite{ashcroft1978solid} (this is true even at an integer filling, where one obtains a ``compensated metal''). In our description, we assumed that the $\rho^3$ term is absent either due to fine-tuning or symmetry constraints. Now, let us instead consider the case where symmetry does not prohibit a $\rho^3$ term, nor is it fine-tuned to zero. In this scenario, an interesting possibility is that the discontinuous jump in the order parameter leads to a band structure where the maximum energy in the lowest band is lower than the minimum energy in the second lowest band. If this happens, then we expect that for energetic reasons analogous to $d = 1$ (App.\ref{Sec:peierls-argument}), $q_0$ will be dynamically selected so that the resulting state is a solid \textit{insulator}, i.e., the lowest band is fully filled. 
This is consistent with the observation that solids are typically commensurate.
Further, if the first-order transition is sufficiently weak, then we expect that our critical theory will remain applicable at length scales $(k_F)^{-1} \ll \ell \ll \xi$ where $\xi$ is determined by the coefficient of the $\rho^3$ term. One will then observe a transition from a liquid metal to a solid insulator with vestiges of our theory at intermediate length scales. 
In this case, the value of our parameter $q_0/k_F$ will be fixed by equating the volume of the Brillouin zone with the volume of the Fermi surface; for the square lattice, this is the condition for squaring the circle, $q_0 = \sqrt{\pi} k_F$.

To explore this possibility, let us consider the Landau theory of a first order transition with Landau free energy, schematically, $f = r \rho^2/2 - w \rho^3 + u \rho^4$. At the first order transition, one finds, $\rho_c \sim w/u, r_c \sim w^2/u$. To obtain a solid, one requires, $ g \rho_c \gtrsim \textrm{bandwidth of the metal} \sim (k_F)^2/2m \Rightarrow g w/u \gtrsim k_F^2/2m$. Using the action $S_{\rm{BAM}}[\rho]$, one may estimate the correlation length $\xi$ at the first-order transition $\xi \sim q_0/\sqrt{r_c} \sim q_0 \sqrt{u}/w$. Therefore, so as to be able to observe the critical behavior at intermediate length scales $\ell$, $(k_F)^{-1} \ll \ell \ll \xi$, one requires $k_F/q_0 \ll  g m/\sqrt{u}$. Since $k_F/q_0 > 1/2$, this implies that one may be able to observe signatures of our critical theory at such a first order transition from liquid metal to solid insulator, if the bare couplings $g, u$ satisfy $gm/\sqrt{u} \gg 1$.

\section{Discussion}
%\textbf{Discussion:} 
The field theory we have discussed in this paper is 
very gapless, 
in that it has both a Fermi surface worth of gapless modes, as well as a Bose surface of gapless modes. This will result in a large heat capacity. We recall that in a scale-invariant theory in $d$ spatial dimensions with dynamical exponent $z$, without hyperscaling violation, the heat capacity is fixed by dimensional analysis to scale like $ c_V \sim T^{d/z}$.  
With hyperscaling violation, $d$ is replaced by 
$d_\text{eff}$, the effective number of dimensions in which the modes propagate, and the dimensions are made up by powers of the hyperscaling violation parameter, here $k_F$ and $q_0$.  
In the case of both Fermi and Bose surfaces, $d_\text{eff} = 1$.  Within RPA, our fixed point has $z=2$ and we would conclude that $c_V \propto T^{1/2}$ at low temperatures, larger even than the behavior in the metallic phase. 
% Within the codimension expansion, we find $z=1$, and it is only the coefficient of the linear-$T$ behavior that would be modified at the transition.
The coefficient of the logarithmic violation of the area law for the entanglement entropy will also jump at the critical point, and at least within a mean-field treatment of our critical theory, entanglement for a subregion of size $\ell$ will scale as $S \sim c_F k_F \ell \log(k_F \ell) + c_B q_0 \ell \log(q_0 \ell)$ where $c_{F,B}$ are positive constants.

{\bf Symmetries of the fixed point field theory.}
In other examples of Fermi surfaces coupled to gapless bosons (at $q=0$), 
the different points on the Fermi surface decouple from each other in the infrared, and one can use a description that focusses on a patch of the Fermi surface (and its antipode) \cite{Metlitski:2010pd,Mross:2010rd}.  We emphasize that at this critical point, this approximation fails dramatically -- each point on the Fermi surface is coupled by gapless boson modes to other, distant points on the Fermi surface.
Because of this failure of different points on the Fermi surface to decouple from each other, 
we can ask whether our critical theory falls in the category of  ersatz Fermi liquid \cite{2021PhRvX..11b1005E, 2021PhRvL.127h6601E},
%. Recall that an ersatz Fermi liquid is 
defined to be a system (in two spatial dimensions) with a Fermi surface that, like a Fermi liquid, has a $\mathsf{LU}(1)$ symmetry, associated with independent fermion number conservation at each point on the Fermi surface.
In App.~\ref{appendix:LU1}, we describe an unsuccessful attempt to implement such a symmetry in our system.
If there is no such symmetry, the system would have to be called a {\it non-ersatz non-Fermi liquid}. %\textcolor{red}{examples of which are not yet known. According to Sung-Sik's email, there are known examples of such liquids.}.  

{\bf Possibility of intermediate phases:} 
Our field-theory is most reliable when $\epsilon = 3-c$ is small, and therefore, similar to any other $\epsilon$-expansion, to really know whether our conclusions continue to hold true when $c = 1$ requires either an exact solution to the problem when $c= 1$ or numerical simulation of lattice models. In the absence of such results, it is worthwhile to discuss other possibilities. A noteworthy possibility, akin to the hexatic phase in the classical theory of melting \cite{halperin1978theory}, and also analogous to the possible intermediate phases in the context of quantum Wigner crystals with long-range interactions \cite{kivelson-spivak}, is a nematic phase that breaks rotational invariance but does not break translational symmetry. The order parameter for such a phase may be written as a diagonal, traceless matrix with components $Q_{\vec{q}} \propto \langle \rho_{\vec{q}} \rho_{-\vec{q}}\rangle  - 
{1\over N} \sum_k \langle \rho_{\vec k} \rho_{\vec -k} \rangle$, where $|\vec{q}| \approx q_0$, and $N$ is the number of points on the Bose surface. The mean-field theory of such a nematic phase is essentially identical to that for liquid-crystals \cite{Chaikin}. The Landau free energy is given by $f = r \tr \, Q^2 - w \tr \, Q^3 + u_1 \tr\, Q^4
+ u_2 (\tr\, Q^2)^2$, and will generically exhibit a first-order transition due to the presence of a cubic term. In the symmetry broken phase, $Q_{\vec{q_0}}  \neq 0$, where $\vec{q_0}$ is a unique vector (or perhaps a small set of vectors) with magnitude $q_0$. We note that our field theory will still be applicable in the neighborhood of the nematic phase. In this light it is interesting to note that an intermediate nematic phase is indeed found in 
\cite{AlAs-wigner-solid}, between a metallic phase and a Wigner solid phase in ultraclean AlAs quantum wells.

Another possibility \cite{brazovskii1975phase} is an intermediate stripe phase, where only one $\rho_{\vec G}$ condenses. A transition to such a stripe phase can also be continuous and is governed by the same theory we studied.

For bosons with particle number conservation and dispersion similar to the BAM action, the possibility of gapped topological phases or exotic metallic phases has been pointed out in \cite{sedrakyan2013composite,sedrakyan2014absence,sedrakyan2015statistical,maiti2019fermionization,sur2019metallic,Lake:2021omz}. Two key differences distinguish our setup: (i) the boson field $\rho$ in our problem is real, i.e., the total boson particle number is not conserved; and (ii) the fermions play a crucial role in our problem, whereas the models in \cite{sedrakyan2013composite,sedrakyan2014absence,sedrakyan2015statistical,maiti2019fermionization,sur2019metallic} involve only bosons. Nonetheless, exploring implications of these results for possible intermediate phases might be interesting.

{\bf Potential experimental systems:} An ideal system for our theory would be short-range-interacting neutral fermions in the continuum at finite density that undergo a solidification transition at $T=0$ with an order parameter that is odd under an internal $\mathbb{Z}_2$ symmetry (so that the cubic term is not allowed). Such a symmetry could be associated with discrete rotation in the spin-space, and it could also arise as a layer-exchange symmetry in a bilayer system. As discussed above, in the absence of a $\mathbb{Z}_2$ symmetry, one may still be able to observe signatures of our critical theory at intermediate scales if the coefficient of the cubic term is small. Since a cubic term tends to favor a triangular lattice, one might look for such a multicritical point in the neighborhood of a transition to a square lattice. This can be explored numerically using specially designed systems that favor square lattice over triangular ~\cite{schweigert1999enhanced,jagla1999minimum,marcotte2011unusual,jain2014dimensionality}. In terms of the tunability of interactions, cold atomic fermionic gases such ${}^{40}$K and ${}^{6}$Li  provide an ideal playground, and may show a solidification transition \cite{matveeva2012liquid} \footnote{Although these systems also possess dipolar interactions, which are likely a relevant perturbation at the critical fixed point we studied, the strength of the short-range interaction can be made larger compared to the dipolar interaction via a Feshbach resonance \cite{regal2003tuning,schunck2005feshbach,zhang2004p,gunter2005p,top2021spin}.}.
Another possibly-relevant system is hydrogen \cite{mcmahon2012properties}: when the pressure $P \gtrsim 300$ GPa, and temperature $T$ is in the range $10^3 \rm{K} \lesssim  T \lesssim 10^4 \rm{K} \ll T_F \approx 10^5 \rm{K}$, it is  estimated that one can drive a transition from a liquid hydrogen where electrons and protons are both delocalized, to a metallic phase where protons form a lattice and electrons move freely in the resulting periodic potential (here $T_F$ denotes the Fermi temperature of the electrons, the protons can essentially be treated classically in this temperature range). Notably, this is a two-component system, and the density fluctuations $\rho$ that enter the action $S_{\rm{BAM}}$ correspond to those of the protons, and not the electrons. It is reasonable to expect that the interactions will be screened in this system since either side of the transition is a metal.

Finally, it may be worthwhile to consider systems where the crystallization occurs in the presence of a pre-existing lattice. A pre-existing periodic potential will be a relevant perturbation for our theory, but if the corresponding lattice spacing 
%$a$ 
is much less than the period of the incipient crystallization ($2\pi/q_0$ in our notation), then the corresponding cross-over length scale will be large. For example, Wigner crystallization in a 2d electron gas %(2DEG)
or in transition metal dichalogenides 
%(TMDs) 
in the absence of magnetic field (see, e.g., \cite{yoon1999wigner,shayegan2022wigner}) takes place in the background of an existing lattice. The inter-electronic distance in the regime where Wigner crystallization occurs is typically much larger than the lattice spacing of the underlying lattice. 
Bilayer 2DEG systems also potentially offer a natural $\mathbb{Z}_2$ symmetry associated with the exchange of the two layers. If the order parameter is odd under layer-exchange symmetry and carries non-zero momentum, then the cubic term in the BAM action will be prohibited, and the field theory will be identical to the one we studied. Even without a layer-exchange symmetry, and in the absence of inter-layer tunneling, bilayer 2DEG systems have a natural $\gU(1)$ symmetry that acts as $c \to e^{\ii \theta Z}c$, where $Z$ acts on the bilayer pseudospin \cite{narasimhan1995wigner, joglekar2006wigner,chen2006pinned,zhou2021bilayer}. At a phase transition between a Fermi liquid and a pseudospin-ferro Wigner crystal (also called exciton supersolid) with order parameter $ \langle c^{\dagger} \left( X + \ii  Y \right) c\rangle$ \cite{narasimhan1995wigner, joglekar2006wigner,chen2006pinned}, cubic terms would again be disallowed.  However, the order-parameter is now a complex, and not a real scalar, and one would need to revisit our analysis. We leave this interesting problem to the future. Another important issue worth exploring is the role played by the spin-degree of freedom. In a single layer 2DEG, the Wigner crystal is likely a Mott insulator (and not a metal or a band insulator), a feature shared by other systems such as bulk He-3 \cite{roger1983magnetism,cross1985magnetism} or thin films of He-3 (see e.g.~\cite{casey2003evidence}). Therefore, the spin degree of freedom could play an important role at the transition (see e.g.~\cite{Musser:2021ldk,Musser_2022} for a theory of Wigner-Mott transitions).
A class of systems where the spin degree of freedom seems to be frozen out is 
the subject of recent experiments on multilayer graphene (\eg~\cite{han2024signatureschiralsuperconductivityrhombohedral, long-ju-pentalayer-graphene}), where  as a function of the displacement field, at low fillings and in small magnetic fields, a transition from a metallic spin- and valley-polarized phase to a highly resistive phase has been observed, consistent with a Wigner crystal whose lattice spacing adjusts to the filling.
% Particularly interesting is the fact that in this system, the active electrons are in a layer very far from the source of the Moir\'e potential, and indeed some theoretic attempts to describe these experiments find 
% \cite{AHCI,AHCII} that the Moir\'e potential is not even required to produce the observed phases.
% Thus, it is conceivable that this transition can be regarded as breaking continuous translation and rotation symmetries.  
% We note that the conductivity data of \cite{long-ju-pentalayer-graphene} on the liquid side has features on approaching the transition that don't look like smearing of the transition by disorder, possibly militating against a strongly first-order transition.  
% Perhaps one can also look for signs of nematic phases in the resistivity data near the transition.

\begin{acknowledgments}
{\bf Acknowledgements.} 
We are grateful to T.~Senthil for long-ago inspiration to think about the destruction of Fermi surfaces, and for useful albeit brief discussions.
We thank R.~Shankar and Ganpathy Murthy for helpful correspondence,
Sasha Chernyshev, Sung-Sik Lee, Daniel Parker, and Darius Shi for helpful comments, and 
Leon Balents, Matthew Fisher, Leo Radzihovsky, Subir Sachdev, Dam Son and Ashvin Vishwanath for useful feedback about the cubic term. 
This work was supported in part by
funds provided by the U.S.~Department of Energy
(D.O.E.) under cooperative research agreement 
DE-SC0009919, 
and by the Simons Collaboration on Ultra-Quantum Matter, which is a grant from the Simons Foundation (652264, JM).
JM received travel reimbursement from the Simons Foundation;
the terms of this arrangement have been reviewed and approved by the University of California, San Diego in accordance with its conflict of interest policies.

\end{acknowledgments}

\appendix
\renewcommand{\theequation}
{\Alph{section}.\arabic{equation}}

\section{Field Theory for the transition} \label{app:field_theory}

In this appendix we describe the degrees of freedom
 in the neighborhood of a 
 solidification transition of a liquid Fermi liquid,
as well as
their coupling.

{\bf Solidification without fermions.} Let us first review the Landau-Ginzburg theory for the solidification transition in the absence of any fermions. The microscopic density $\rho(x)$ at point $x$ may be written as
\be 
\rho(\vec{x}) =  \rho_0 + \frac{1}{\sqrt{V}}\sum_k e^{i \vec{k}\cdot \vec{x}}\,\, \rho_{\vec{k}}\label{eq:densitymicro}
\ee 
where $\rho_0$ is the average density and the 
sum over $k$ runs over 
%set $\{\vec{k}\}$ corresponds to 
all allowed momenta in the continuum. Near the transition, the system will have the tendency to order at a certain \textit{magnitude} of the wavevector. For example, in the absence of any fermions, the term quadratic in the Landau free energy will take the form \cite{alexander1978should}
\be 
\sum_k \, \, A_k \, \rho_{\vec{k}} \, \, \rho_{-\vec{k}} \label{eq:landau}
\ee 
where $A_k$ depends only on $k = |\vec{k}|$, and will  have a minima at some $k = |\vec{G}|\equiv q_0$ close to the transition.  
The value of $|\vec{G}|=q_0$ is visible in the liquid phase as the maximum of the spin structure factor.
 Therefore, at the leading order within the Landau theory, one may approximate the Eq.~\ref{eq:landau} as
\be 
\sum_{\vec{G}}\int dR\,\, A_{|R \vec{G}|} \,\,\rho_{R \vec{G}}\, \rho_{-R \vec{G}} \label{eq:landauapprox}
\ee 
where $R \in O(d)$ is an orthogonal matrix, and $R \vec{G}$ denotes the vector resulting from the action of $R$ on $\vec{G}$. The sum $\sum_{\vec{G}}$ runs over the lattice in the reciprocal space defined by $\{\vec{G} = \sum_i n_i \vec{G}_i\}$ where $n_i \in \mathbb{Z}$, and $\vec{G}_i$ are the wavevectors corresponding to the preferred lattice (e.g. for a square lattice $\vec{G}_1 = \frac{2\pi}{a}\hat{x}, \vec{G}_2 = \frac{2\pi}{a}\hat{y}$). Correspondingly, the expression for the density (Eq.\ref{eq:densitymicro}) at the leading order may be approximated as
\be 
\rho(x) \approx  \rho_0 + \frac{1}{\sqrt{V}}	\sum_{\vec{G}}\int dR \,\,e^{i ({R \vec{G}})\cdot \vec{x}} \,\, \rho_{{R \vec{G}}}(x) \label{eq:densitycoarse}
\ee 
where $\rho_{{R \vec{G}}}(x)$ are allowed to depend on the position $x$ and are slowly varying functions of $x$. That is, their dominant Fourier modes live near zero momentum.

{\bf Symmetries.} Under translation by $\vec{a}$, $\rho(\vec{x}) \rightarrow \rho(\vec{x}+\vec{a})$. Therefore, $\rho_{{\vec{K}}}(x) \rightarrow e^{i \vec{K}\cdot \vec{x}} \rho_{{\vec{K}}}(x) $, where $\vec{K} = R \vec{G}$ (we use capital letters for the momenta belonging to the set $\{R \vec{G}\}$). Under rotation by a matrix $S \in O(d)$, $\rho(\vec{x}) \rightarrow \rho(S \vec{x})$, and therefore
$\rho_{{\vec{K}}}(x) \rightarrow \rho_{{S\vec{K}}}(x)$.

The Landau-Ginzburg action up to leading, quadratic order may then we written as:
\bea
S_\rho & =& \int d^dx d\tau \left((\rho(x,\tau)-\rho_0)^2 + \left(\vec{\nabla} \rho(x,\tau)\right)^2\right)  \nonumber
\\ 
 &\approx  & \int d^dx\,d\tau\,
 \sum_{\vec{G}}\int dR\,\, 
 \left(\rho_{R \vec{G}}(x,\tau)\, \rho_{-R \vec{G}}(x,\tau) 
 \right.  \nonumber
 \\   
 && 
 \left. 
 ~~~~~~~~~~~~~~~~~+ |\vec{\nabla} \rho_{R \vec{G}}(x,\tau)|^2\right)~~.
 \nonumber
\eea
One may similarly write down higher-order terms, see, e.g., Ref.\cite{alexander1978should}.

\begin{figure}
$$ \includegraphics[width=.4\textwidth]{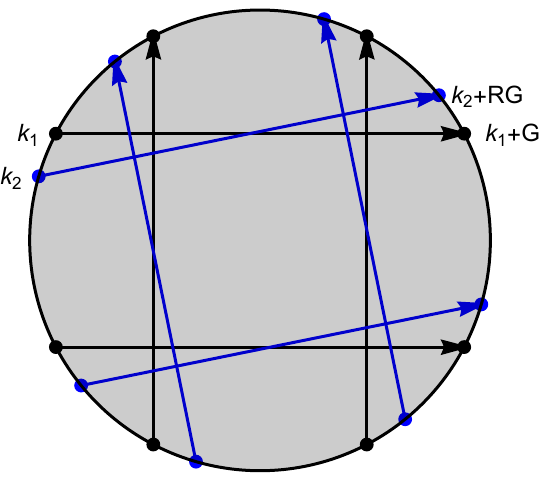}$$
\caption{\label{fig:destroy-FS}At the transition, every point $\vec k$ on the Fermi surface is connected by a vector of the form $R\vec G$ to another point on the Fermi surface,  $ \vec k + R\vec G$, where $\vec G$ is a lattice generator and $R$ is a rotation matrix.  The lengths of the lattice vectors $\vec G$ are determined by demanding that the fermions exactly fill the square lattice.   
}
\end{figure}

{\bf Adding back fermions.} At the leading order, the coupling between the order parameter and the fermions will be
\be 
S_{\psi-\rho} = \int d^dx\,d\tau\,\, \psi^{\dagger}_\sigma(x,\tau) \psi_\sigma^\nd(x,\tau) \rho(x,\tau)~.
\ee 
Using the low-energy expression for the density, Eq.\ref{eq:densitycoarse}, this becomes,
\bea 
S_{\psi-\rho} &\approx& \sum_{\vec{G}}\int dR\, d^dk_1\, d^dk_2 \,\, d\tau\,\, 
\\ && \rho_{R\vec{G}}(\vec{k}_1,\tau) \psi_\sigma^{\dagger}(\vec{k}_2,\tau) \psi_\sigma^\nd(\vec{k}_2-\vec{k}_1 +R\vec{G},\tau)
\nonumber
\eea
In this expression, $\vec{k}_i$ are all close to zero momentum. Due to integration over $R$, all points on the Fermi surface will be coupled to the soft modes $\{\rho_{R\vec{G}}\}$, as illustrated in Fig.~\ref{fig:destroy-FS} for the case of the transition in 2d to a square lattice.

\section{Contrast between $d = 1$ and $d>1$}
\label{Sec:peierls-argument}

 \begin{figure}
 $$ \includegraphics[width=.4\textwidth]{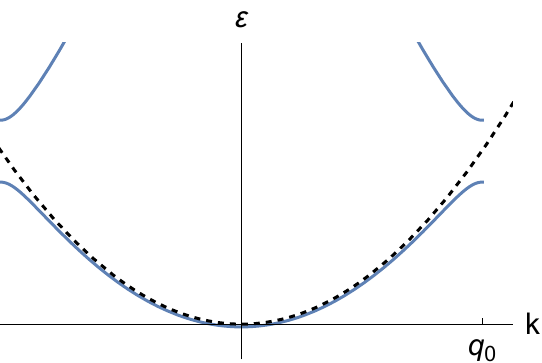}$$
 \caption{\label{fig:peierls-argument}The dashed curve is the unperturbed single-particle dispersion, $\epsilon(k) = { k^2 \over 2 m}$, shown over the region $k \in \{-\pi/a, \pi/a\}$, the Brillouin zone of the incipient lattice.  
 In blue is the folded bandstructure resulting from $\rho>0$ in the coupling 
 $ \rho \cos \left( 2 \pi x / a \right) c^\dagger_x c_x $. Note that the minimum of the second band lies above the maximum of the first band. }
 \end{figure}
%For non-interacting fermions, one must fill states up to $|k| = { \pi \over a_\star}$. Clearly when $a \simeq a_\star$, the state with $\rho>0$ has lower energy than the state with $\rho=0$.  Furthermore, as a function of $a$, the energy is smallest when $a=a_\star$.

%{\bf Nature of the solid phase.}
%Having given a broad description of the degrees of freedom and the fate of the theory close to the phase transition, now we must ask the apparently-simpler question of the nature of the solid phase, assuming that the transition between the liquid metal and the solid phase is indeed continuous (this corresponds to the first scenario described above). In particular, is the solid phase a metal or an insulator at $T = 0$? For $r>0$ and any value of the other parameters, we have an isotropic Fermi liquid in the continuum. As $r$ becomes negative, a collection of modes of $\rho$ with wavenumber of magnitude $q_0$ will condense and create a periodic potential for the fermions.  A mean-field picture of the solid phase can be obtained by treating the electrons as free fermions moving in this potential. A natural question is: what determines the lattice spacing $a = {2\pi \over q_0}$ in the solid phase? 

In one dimension, 
within a mean-field picture, energetics would pin $q_0$ to a value so that the resulting state is a band insulator. The argument is essentially same as that for the Peierls instability, but for completeness, we will briefly review it. Let us denote the order parameter as $\rho_{q_0}$, so that $\rho_{q_0} = 0$ in the liquid phase, and $\rho_{q_0}$ increases continuously from zero as one enters the solid phase. The mean-field Hamiltonian for fermions is then $\sum_{k} \frac{k^2}{2m} \psi^{\dagger}_k \psi^\nd_k + \left(\rho_{q_0} \psi^{\dagger}_k \psi^\nd_{k+q_0} + h.c.\right)$. Our aim is to find the value of $q_0$ that minimizes the ground state energy at a fixed density of electrons, i.e., at a fixed Fermi wavevector $k_F$. When $\rho_{q_0} \ll k^2_F/2m$, one can use linear response theory to estimate the lowering of the energy due to the crystallization. It is simply given by $\rm{Re}(\chi(q_0,\omega =0)) |\rho_{q_0}|^2$, where $\chi(q,\omega)$ is the  Lindhard susceptibility. The static Lindhard susceptibility $\rm{Re}(\chi(q_0,\omega =0))$ for a 1d Fermi has  a logarithmic divergence at $|q| = 2k_F$: $\rm{Re}(\chi(q_0,\omega =0)) \sim \log(\frac{q + 2k_F}{q-2k_F})$. Therefore, the energy will be minimized when $|q| = 2k_F$, resulting in a band insulator\footnote{Of course, this is a mean-field calculation that neglects fluctuations -- stand-alone solids with long-range order cannot exist in $d = 1$ due to the Mermin-Wagner theorem.}. A crucial input in this conclusion is that the bands generated due to the periodic potential do not overlap, e.g., the minimum of the second band lies above the maximum of the first band.

In dimensions larger than one, however, it is not possible to produce an insulating bandstructure with infinitesimal $\rho_{q_0}$
\cite{ashcroft1978solid}.  Rather, 
the bands necessarily overlap, and at any value of $q_0/k_F$ there are both particle and hole Fermi surfaces for small $\rho_{q_0}$ as shown in the figure below:
$$ \parfig{.4}{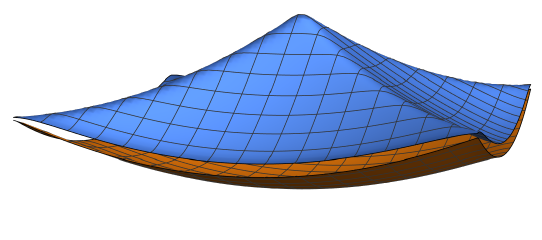}$$

\begin{figure}
$$ \parfig{.23}{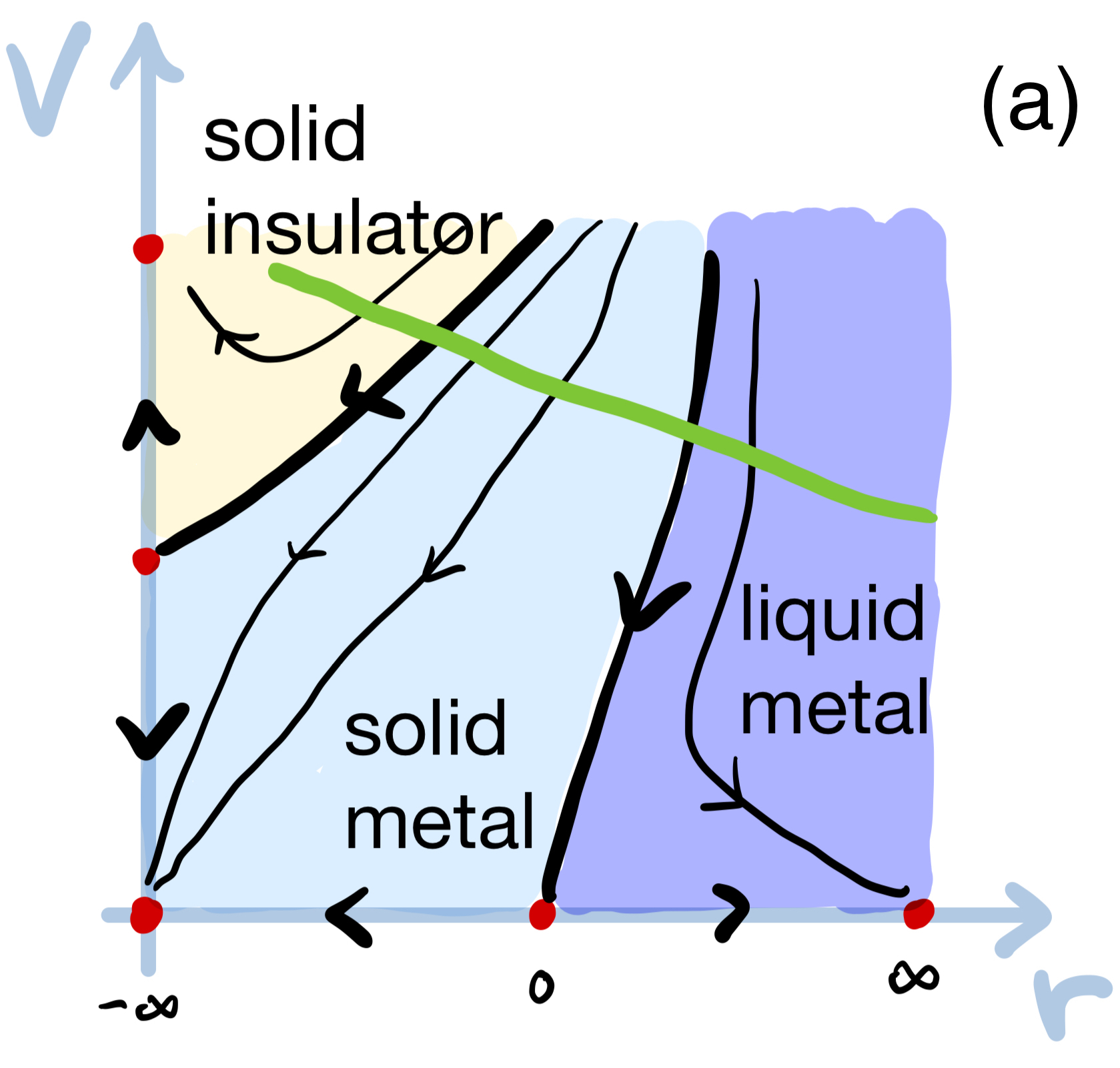}
~~~ \parfig{.23}{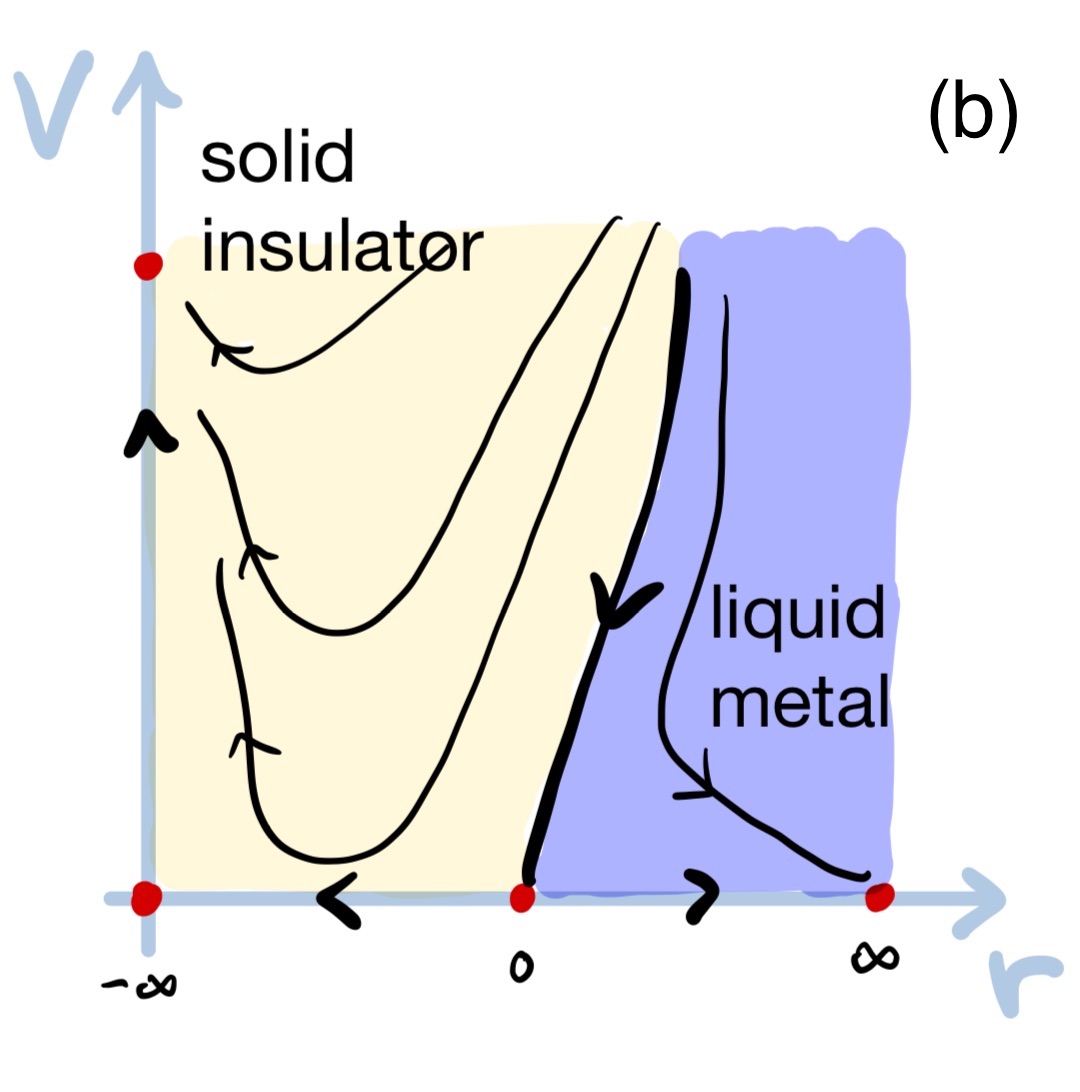}$$
\caption{\label{fig:phase-diagram-with-V} (a) Expected phase diagram of the model perturbed by short-ranged four-fermion interactions $V$, such as will be generated by the fluctuations of $\rho$ neglected in mean field theory.  The green line represents a possible trajectory of a microscopic Hamiltonian across the transition, which by continuity, inevitably sees an intermediate solid metal phase.  (b) If the FS of the solid metal were nested, the direction of the arrows in the $V$ direction in that region would be reversed.  In this case we find a direct transition from liquid metal to sold insulator.  In both figures, we set the cubic coupling to zero.
}
\end{figure}

%% ADD HERE DISCUSSION OF PHASE DIAGRAM WITH V AND r
In the mean-field theory, we completely neglected the repulsive four-fermion interactions $\sim \int V \left(\psi^{\dagger} \psi\right)^2$. Such interactions are irrelevant at the critical theory for solidification we described in the main text (see Eq.\ref{eq:four-fermion} in Appendix~\ref{appendix:scaling}). They are also marginally irrelevant in either a liquid or a solid Fermi liquid, assuming absence of nesting. We now sketch the RG flow of our theory in the $(r, V)$ plane where $r$ is the tuning parameter in the BAM action (Eq.\ref{eq:SAT}); as in the main text, we set the $\rho^3$ coupling to zero. Our assumptions are: (i) There exists a microscopic realization where the tuning parameter effectively moves one along a continuous path in the $(r, V)$ plane (ii) The Mott transition between a solid metal and a solid insulator can be obtained by tuning $V$ (one theory for such a transition was described in Ref.\cite{Senthil_2008,Senthil_2008_criticalFS} - the solid, (electrical) insulator in this theory is unconventional and has a Fermi surface of neutral spinons). Fig.~\ref{fig:phase-diagram-with-V}(a) shows the RG flow under these assumptions. The RG flow we draw describes a two-step process: liquid metal $\to$ solid metal $\to$ solid insulator (indicated by the green line in Fig.~\ref{fig:phase-diagram-with-V}(a)). As briefly mentioned in the main text, a possibility for a direct transition between a liquid metal and a solid insulator is the following. Suppose that, when $r$ is infinitesimal and negative, the system dynamically selects a nested Fermi surface, which is unstable towards an insulating phase for infinitesimal $V$. In this scenario, the four-fermion interaction is dangerously irrelevant at the transition. We sketch this scenario in Fig.~\ref{fig:phase-diagram-with-V}(b).

\section{Can the transition be continuous?}
\label{appendix:continuous-transition}

It is commonly believed that melting transitions are always first order \cite{Chaikin, brazovskii1975phase, brazovskii1987theory}.  
As we explained in the main text, there are two layers to this statement.  One is that there can be a cubic term in the BAM Landau-Ginzburg theory. Even if the cubic term is somehow removed, there remains the fact that fluctuations of the density order parameter can drive the transition first order.  One simple way to understand this is by the following mean-field calculation.
To get oriented, first consider an Ising transition, with Landau free energy
\be F[\phi] = \half \int \dbar^d q \( r + q^2 \) \phi_q \phi_{-q} + \int d^d x g \phi^4 (x). 
\label{eq:phi-four}\ee
The mean-field approximation gives a self-consistency condition: 
\be \vev{n^2} = G(x,x) = \int \dbar^d q G(q) \simeq \int \dbar^dq  { 2 T \over r + q^2 + 12 g \vev{n^2} } . \ee
The transition occurs when 
\be 0 = \chi^{-1} = T G^{-1}(q=0) = 
r + 24 T g \int \dbar^d q {1\over \chi^{-1} + q^2 } \propto {\Lambda^{d-2} \over d-2} .\ee
For $d \leq 2$ this correction drives the critical temperature to zero.  
Thus we detect the lower critical dimension.

Now consider the case where the structure factor has a minimum at $q^2=q_0^2 > 0$: 
\be F[\rho] = \half \int \dbar^d q \( r + (q^2-q_0^2) \) \rho_q \rho_{-q} + \int d^d x g \rho^4 (x). \ee
The same logic (for $T>0$) now gives 
\bea 0 &= \chi^{-1} = T G^{-1}(q=0)  
\\ &=  r + 24 T g \int \dbar^d q {1\over \chi^{-1} + (q^2-q_0^2) } \propto g q_0^{d-3} \int_0 {dq_\parallel \over q_\parallel^2} . \nonumber \eea
Here $\vec q = q_0 \hat \Omega + \vec q_\parallel, \vec q_\parallel \cdot \hat \Omega = q_\parallel$, and we used the expansion $ (\vec q^2 - q_0^2)^2 = (q-q_0)^2 (q+q_0)^2 \simeq 4 q_0^2 q_\parallel^2 $, since  $q_\parallel \ll q_0$.  
This is an IR divergence in any dimension, which indicates that the transition cannot happen continuously at finite temperature, but rather becomes weakly first-order.  We note that in such a situation, properties in the neighborhood of the transition can still usefully be studied using field theory.  

Two effects change the result in our case.  
First, we consider a transition at zero temperature, so we include coherence in the time direction.  
For the Ising case, adding a $\dot \phi^2 $ kinetic term to \eqref{eq:phi-four} changes the shift in the critical temperature to $ {\Lambda^{D-2}\over D-2}$, where $D = d+1$ is the number of spacetime dimensions.
Second, we include the effects of the fluctuations of all the gapless Fermi surface degrees of freedom on the dynamics of the order parameter.  The most important such effect is the Landau damping term 
$\delta \Pi = \gamma {| \omega |\over q} = \gamma { |\omega| \over |G|} $, 
which dominates over any local-in-time kinetic terms for $\rho$.  
Thus we consider the Euclidean $T=0$ action
\be S[\rho] = \int \dbar^d q~ \dbar\omega  \( r + (q^2 - q_0^2)^2 + \gamma { |\omega| \over |G|} \) 
+ \int d^d x \int dt g \rho^4 . \ee
The analogous calculation now gives 
\begin{align} 0& =\chi^{-1} = G(\omega=0, q=q_0) 
\\ & = r + 12 g \int \dbar^d q~ \dbar \omega { 1\over \gamma {|\omega| \over |G|} + (q^2 - q_0^2 )^2 } 
\\ & = r + 12 g k_0^{d-1} K_d  \int {dq_\perp d\omega \over {|\omega| \over |G| } + q_\perp^2 q_0^2 } \label{eq:zeroTfluct}
\end{align}
which is IR finite:
%\be \int_{-\infty}^{\infty} d\epsilon  \int_{- \Lambda^2}^{\Lambda^2} dq_\perp {1\over |\epsilon| + q^2 } 
\be \int_{-\Lambda}^{-\Lambda} d\epsilon  \int_{-\infty}^{\infty} dq_\perp {1\over |\epsilon| + q^2 } 
=  \pi \sqrt{ \Lambda} . \ee

One may repeat the above analysis at a non-zero temperature. Firstly, we recall that due to fluctuations of the Goldstone modes, a solid can exist at a non-zero temperature only in $d \geq 3$ (Mermin-Wagner theorem). So the question is whether the finite temperature liquid-solid transition can be continuous at a non-zero temperature, again assuming that there are no cubic terms in the Landau theory. Now the integral over $\omega$ in Eq.~\ref{eq:zeroTfluct} will get replaced by a sum over discrete Matsubara frequencies $\omega_n = 2 \pi n T$. That is, one needs to solve 
\be 
 r + \left( 12 g q_0^{d-1} K_d\right) \left( 2 \pi T\right)  \sum_{n=-\infty}^{\infty} \int {dq_\perp \over {2 \pi T |n| \over |G| } + q_\perp^2 q_0^2 } = 0
\ee 
The contribution from $n = 0$ to the  integral on the RHS diverges in the IR as $T L$, where $L$ is linear system size. This indicates that in $ d \geq 3$, when a solid is allowed to exist at a non-zero temperature, the finite-temperature transition can only be first-order.

\section{Self-consistent RPA solution}
\label{appendix:self-consistency}

The goal of this appendix is show that the forms given in the main text for the boson and fermion green's functions 
in RPA 
satisfy self-consistently the Schwinger-Dyson equations:
 \bea
 \label{eq:SD}
 G^{-1}(\omega, k) &=&  \ii \omega \eta + \Sigma(\omega,k)  + v_F k_\perp ,
 \\ 
 D^{-1}(\epsilon, q) &=& { r +  \Pi(\epsilon, q)  +   (q^2 - q_0^2)^2 }, 
 \nonumber
 \\ \Sigma(\omega,k) & =& \int \dbar\epsilon ~\dbar^d q D(\epsilon,q) G(\epsilon-\omega, q-k) ~~~\nonumber
 \\ \Pi(\epsilon,q) & =& \int \dbar\omega ~\dbar^d k G(\omega,k)G(\omega-\epsilon, k-q) ~. \nonumber
 \eea
The diagrams included in these Schwinger-Dyson equations are the ones selected at leading order by the large-$N$ generalization of the theory \cite{shamit-matrix-large-N} mentioned above.
Although the boson self-energy is naively suppressed by a power of $N$ (since it does not contain a free index loop), it is included because of its singular frequency dependence.  
Part of what we need to show is that the fermion self-energy is momentum-independent.  
We will also explore deformations of the problem (analogous to dimensional regularization, or 
the expansion of \cite{Nayak:1993uh, Mross:2010rd}, or the codimension expansion \cite{2009PhRvL.102d6406S,Dalidovich:2013qta, Lee:2017njh}), around which one could try to develop a controlled approximation.

\begin{figure}
$$ \includegraphics[width=.4\textwidth]{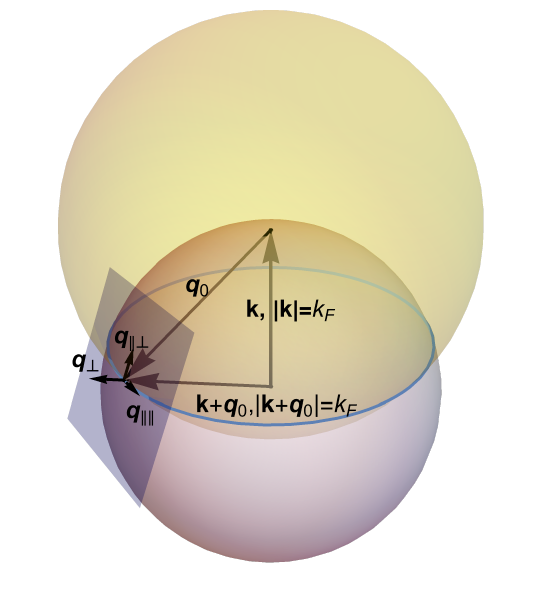}$$
\caption{\label{fig:FS-in-3d}
At the critical point in 3d, each point $\vec k$ on the Fermi surface (pink sphere) is coupled to a ring of other momenta $\vec k + \vec q_0$, 
the intersection (blue ring) of the FS with the bose surface centered at $\vec k$ (yellow sphere).   
The figure is drawn to scale for the cubic lattice.  Also indicated are our coordinates for the momenta of the boson: $\vec q_\perp$ is perpendicular to the FS, 
while the two vectors $ \vec q_{\parallel\perp}$ and $\vec q_{\parallel\parallel}$ are tangent to the FS.  
$\vec q_{\parallel\parallel}$ is also tangent to the intersection circle, while $\vec q_{\parallel\perp}$ is the normal to the intersection circle.  
}
\end{figure}

To do the integrals, we can employ a useful trick from \cite{Metlitski:2010vm}.  
Anticipating the outcome that the fermion self-energy will be a singular power of $\omega$ less than one, so that at low energies the bare fermion kinetic term $\ii \omega$ may be neglected, we may use the propagator 
\be\label{eq:bare-fermion-propagator} 
G_0(\omega, q)^{-1} = \ii \omega \eta - v_F q_\perp ,\ee
where $\eta = 0^+$ is an infinitesimal.  Then we can use the identity 
\be \label{eq:on-shell} { 1\over \ii \omega \eta + x } = {P\over x } + \sign(\omega) \ii \pi \delta(x) \ee
(for real $x$).  
In \eqref{eq:bare-fermion-propagator}, $q_\perp$ denotes the distance between $\vec q$ and (the nearest point on) the Fermi surface. 

We will also want to show that our Green's functions satisfy the self-consistency conditions \eqref{eq:SD}.  
For that purpose, we must include the singular fermion self-energy in $G^{-1}$, 
\be G(\omega, q)^{-1} = \ii \omega \eta + \Sigma(\omega) - v_F q_\perp ,\ee
which contributes to the imaginary part, 
and this trick will not work.   
So we will also describe below a second way to do the integrals.

{\bf Boson self-energy.}
The contribution to the boson self energy for $q \to q_0$, which only involves the fermion propagators, is essentially the same as in other cases where a Fermi surface has a cubic coupling to a gapless boson.  
\bea \Pi(\epsilon, k) &= 
\parbox{.3\textwidth}{\begin{tikzpicture}[line width=1.0 pt, scale=.7]
%	\draw[fermion] (0,0) circle (1);
	  \draw [fermion]      (0,0)  arc [radius=1, start angle=0, end angle= 180] ;%node[midway,above]{$\omega-\epsilon, q-k$};
    \draw[ line width = .5pt, ->]  (-.8,1.3) -- (-1.2,1.3) node[midway,above]{$\omega-\epsilon, q-k$};
	  \draw [fermion]      (-2,0)  arc [radius=1, start angle=180, end angle= 360] ; %node[midway,below]{$\omega, q$};
    \draw[ line width = .5pt, ->]  (-1.2,-1.3) -- (-.8,-1.3) node[midway,below]{$\omega, q$};
  
	\draw[vector] (-2,0)--(-2.5,0) ; %node[midway, below]{$(\epsilon, k)$};
	  \draw [line width = .5pt, ->] (-2.6, -.3) -- node[below] {$\epsilon, k$} ++ (0.4, 0); 
	
	\draw[vector] (0,0)--(0.5,0) ; %node[midway, below]{$(\epsilon, k)$};
		  \draw [line width = .5pt, ->] (0.3, -.3) -- node[below] {$\epsilon, k$} ++ (0.4, 0); 

%	\draw[fermionbar] (30:1)--(0,0);
%	\draw[vector] (140:1)--(0,0);
%	\draw[vector] (-140:1)--(0,0);
	\end{tikzpicture}
 }
 \\ & 
= \int \dbar \omega \dbar^d q G(\omega, q) G(\omega-\epsilon, q - q-k) . 
\eea
Let us first do the integral in $d=2$ dimensions.  
Using the bare fermion propagator, $G_0$, 
and the trick described above, 
this is 
\be \Pi(\epsilon, k) = - {1\over 8 \pi v_F^2} \int d\omega \int dq_\perp dq_\parallel  \delta(q_\perp) \delta((k-q)_\perp) . \ee
The contributions arising from the principal part term in \eqref{eq:on-shell} do not contribute singular terms, and vanish exactly if particle-hole symmetry holds. 
For $|k| \simeq q_0$, $q_1 \equiv q_\perp$ and $q_2 \equiv (q-k)_\perp$ are linearly independent momenta.  Changing variables to  $ q_1 \equiv q_\perp, q_2 \equiv (k-q)_\perp$, 
$ dq_\perp dq_\parallel = \alpha dq_1 dq_2$ (for some constant $\alpha$)
the two delta functions
saturate the two momentum integrals, and we find
\bea \Pi(\epsilon, k \simeq q_0 ) - \Pi(0) & = - { \alpha \over 2 \pi v_F^2 } \int d\omega  \( \sign(\omega)\sign(\omega -\epsilon) - 1 \) \nonumber
\\ & = {\alpha\over \pi v_F^2 }  |\epsilon| . \eea
the familiar form of Landau damping for a boson at nonzero wavenumber.  
We note that as in \cite{Metlitski:2010vm}, the prefactor of $|\epsilon|$ is not the same as the one we would have found had we included the curvature of the Fermi surface in the propagators, which is proportional to the volume of the Fermi surface and the same as $\Pi(0)$ by Kramers-Kronig.   However, the value of this coefficient does not change the physics and the method is reliable for universal quantities.  

In $d>2$ spatial dimensions the only difference is that coordinates along the intersection of the Fermi and Bose surfaces ($q_{\parallel\parallel}$ in Fig.~\ref{fig:FS-in-3d}) do not appear in the integrand, and they produce an innocuous factor $V_\text{BS}$ of the volume of the intersection locus.  

Let us redo the integral with the full fermion propagator, including the singular self-energy.
Then
\be \Pi(\epsilon, k) = 
\int \dbar \omega \dbar^d q 
{1\over F(\omega) + v_F q_\perp } 
{1\over F(\omega- \epsilon) + v_F q_2} 
%{ 1 \over \ii \eta \omega + \Sigma(\omega) + v_F q_\perp } 
%{ 1\over \ii \eta (\omega - \epsilon)  + \Sigma(\omega - \epsilon) + v_F q_2 } 
\ee
where $q_2 \equiv (q-k)_\perp$, 
and $F(\omega) \equiv \ii \eta \omega + \Sigma(\omega)$.  
For $|k| \simeq q_0$, we can again change integration variables $  dq_\perp dq_{\parallel} = \alpha dq_1 dq_2 $ with $q_1 \equiv q_\perp$.
The key input is that $\Im \Sigma(\omega) \propto \ii \sign(\omega) $, 
so that $F(\omega) = \ii \sign(\omega) f(\omega)$ with $f(\omega)>0$ (plus an irrelevant real part),
and each of the integrals over $q_{1,2}$ is of the form
\be \int {\dbar q_i \over \ii \sign(\nu) f(\nu) + v_F q_i } ={ \ii \sign\( \nu\) \over v_F}   . \ee
The rest of the calculation is as above.
%\bea \Pi(\omega, k \simeq q_0 ) - \Pi(0) & = - { \alpha \over 2 \pi v_F^2 } \int d\omega  \( \sign(\omega)\sign(\omega -\epsilon) - 1 \) \nonumber
%\\ & = {\alpha\over \pi v_F^2 }  |\epsilon| . \eea

{\bf Fermion self-energy.}  Next we consider the contribution to the self-energy of the fermion from the Landau-damped bosonic mode with a sphere of minima in its dispersion relation.
\bea
\Sigma(\omega,k) &= 
\parbox{.3\textwidth}{
 \begin{tikzpicture}[line width=1.0 pt, scale=.7]
%	\draw[fermion] (0,0) circle (1);
	  \draw [fermion]  (-1, 0) -- (1,0);
	  \draw [fermion]  (-1.5, 0) -- (-1,0);
	  \draw [line width = .5pt, ->] (-2, -.3) -- node[below] {$\omega,k$} ++ (0.4, 0); 
	  \draw [line width = .5pt, ->] (-.5, -.3) -- node[below]{$ \omega-\epsilon,k-q$} ++(1,0); 
	  \draw [line width = .5pt, ->] (1.8, -.3) -- node[below] {$\omega,k$} ++ (0.4, 0); 
	  \draw [fermion]  (1, 0) -- (1.5,0);
	  \draw [vector]  %{<[scale=1.5,          length=5,          width=3]}-,line width=1.5pt, opacity=.4] 
    (1,0)  arc [radius=1, start angle=0, end angle= 180];
    \draw[ line width = .5pt, ->]  (-.2,1.3) -- (.2,1.3) node[midway,above]{$\epsilon, q$};
	\end{tikzpicture}
}
\\ & = g^2 \int \dbar \epsilon ~\dbar^d q D(\epsilon, q) G(\omega+\epsilon, q+k)  
.\eea
Here the $\rho$ propagator can be approximated as  
\be 
D(\epsilon, q)  = {1\over r + 4 ( \vec q_0 \cdot \delta \vec q)^2 + { |\epsilon| \over \Gamma } } ~,
%D(\epsilon, q)  = {1\over r + 4 ( \vec q_0 \cdot \delta \vec q)^2 + { |\epsilon|^{2/z} \over \Gamma } } ~.
\ee
where $\Gamma \sim q_0$ is a constant.
In this expression, we have decomposed the boson wavevector as $ \vec q  = \vec q_0 + \delta \vec q$,
where $|\vec q_0 |= q_0 = |\vec G|$ is a point on the bose surface, and the vector $\delta \vec q$ is arbitrary but small.
The key point will be that only one linear combination of momenta $\vec q_0 \cdot \delta \vec q \equiv q_b$ appears in the boson propagator.

%First we focus on the behavior for $|k|=k_F$, and 
We wish to understand the singular behavior in $\omega$ and in $r$, the deviation from the critical point, and we wish to understand the momentum dependence.  
We will see that, as in other examples of Fermi surface coupled to critical boson, but unlike the SDW case, 
the self-energy is regular as a function of the deviation of the fermion momentum from the Fermi surface (meaning independent of the momentum in our approximation to the integrals). 

Again we will do the integral in two ways, first using the trick with the bare fermion kinetic term $\propto \eta \to 0$.  
Using \eqref{eq:on-shell} for the fermion propagator, we can use the delta function to do the $q_\perp$ integral, which will set 
$0 = (q-k)_\perp = q_\perp - k_\perp - \cos\theta_0 q_\parallel$.   
Here $\theta_0$ is the angle between points on the FS connected by a vector of length $q_0$ (Fig.~\ref{fig:2d-geometry}).  
In $d=2$ we have 
%Doing the $q_\perp$ and (for $d>2$) $\vec q_\perp$ integrals, this is 
\bea
& \Sigma(\omega,k)  = 
\\ &   \nonumber
{g^2\over (2\pi)^2} \int d\epsilon dq_\perp dq_\parallel { 1\over r + {|\epsilon| \over \Gamma} + ( \alpha q_\parallel + \beta q_\perp ) ^2 } 
{ 1 \over \ii \eta (\omega- \epsilon)  - v_F( k-q)_\perp } 
\\ & = \nonumber
\ii {g^2\over (2\pi)^2 v_F}
\int d\epsilon  dq_\parallel {  \sign(\omega- \epsilon) \over r + 
{|\epsilon| \over \Gamma} + ( \alpha q_\parallel + \beta \( k_\perp + \cos \theta_0 q_\parallel \) ^2 } ~.
%{g^2 \Lambda^{d-2}\over v_F} \int \dbar \epsilon \int \dbar q_\parallel
%{\text{sign}(\omega + \epsilon) \over ( q_\parallel ^2 + { |\epsilon| \over \Gamma } }
%{\text{sign}(\omega + \epsilon) \over q_\parallel^2 + { |\epsilon|^{2/z} \over \Gamma } }
\eea
The term from the principal part integral again does not contribute any singular terms and vanishes in the approximation of particle-hole symmetry.  
Now we can change variables from $q_\parallel$ to $q_1 \equiv  \alpha q_\parallel + \beta \( k_\perp + \cos \theta_\star q_\parallel \)$,
and we see that all dependence on the momentum disappears.  
The crucial difference from the case of the spin density wave (SDW) is the form of the boson propagator; in the SDW case, $D^{-1}$ is a sum of squares of the deviations of the momentum of the soft mode in each direction, whereas here there is only one momentum direction transverse to the Bose surface.  
In $d>2$, the integrand is also independent of the $q_{\parallel\parallel}$ integrals and they again produce a factor of the volume of the intersection of the Bose and Fermi surfaces (see Fig.~\ref{fig:FS-in-3d}).
The result of the frequency integral is 
\bea
\Sigma(\omega,k=k_F) &=   \nonumber
\ii {2 g^2 V_\text{BS}\over v_F (2\pi)^{d+1}}\Gamma \int \dbar q_1 \text{sign}(\omega) \log {  |\omega| + r + q_1^2 \over q_1^2 } 
\\ & = \ii {2 g^2 V_\text{BS} \over v_F (2\pi)^{d}} \text{sign}(\omega) \sqrt{ r + |\omega| } . 
\label{eq:fermion-self-energy-integral}
\eea
Thus, as a function of the tuning parameter $r$, the self-energy at $\omega=0$ goes like $\sqrt{r}$.  

Let us reconsider the fermion self energy in the case where we include the singular self-energy in the fermion propagator, to check for self-consistency.  In this case we cannot use the trick \eqref{eq:on-shell}.  Instead, we will do the $q_\perp$ integral by contours.  
The crucial fact is again that $ \Sigma(\omega) = \ii \sign(\omega) f(\omega)$.   The result is 
\bea & \Sigma(\omega,k)  = 
\\ &   \nonumber
\ii {g^2\over (2\pi)^2} \int d\epsilon dq_\perp dq_\parallel { 1\over r + {|\epsilon| \over \Gamma} + ( \alpha q_\parallel + \beta q_\perp ) ^2 } 
{ 1 \over \Sigma(\omega- \epsilon)  - v_F( k-q)_\perp } 
\\ & = \nonumber
\ii {g^2\over (2\pi)^2 v_F}
\int d\epsilon  dq_\parallel {  \sign(\omega- \epsilon) \over r + 
{|\epsilon| \over \Gamma} + 
\( \alpha q_\parallel + \beta \( k_\perp + \cos \theta_0 q_\parallel  + \Sigma(\omega-\epsilon)/v_F\)  \)^2 } ~.
%{g^2 \Lambda^{d-2}\over v_F} \int \dbar \epsilon \int \dbar q_\parallel
%{\text{sign}(\omega + \epsilon) \over ( q_\parallel ^2 + { |\epsilon| \over \Gamma } }
%{\text{sign}(\omega + \epsilon) \over q_\parallel^2 + { |\epsilon|^{2/z} \over \Gamma } }
\eea
Again, we can change the integration variable to 
$q_1 \equiv \alpha q_\parallel + \beta \( k_\perp + \cos \theta_\star q_\parallel  + \Sigma(\omega-\epsilon)/v_F\) $, and the result is again 
given by \eqref{eq:fermion-self-energy-integral}.
%\be \Sigma(\omega,k) 
%%= 
%\ii {2 g^2 V_\text{BS}\over v_F (2\pi)^{d+1}}\Gamma \int \dbar q_1 %\text{sign}(\omega) \log {  |\omega| + r + q_1^2 \over q_1^2 } 
%\ee as before.  
Thus the form of the fermion self-energy is self-consistent.  
The apparent constant shift of the location of the Fermi surface, which seems to violate Luttinger's theorem, is an artifact of the regularization of the integral.  For a discussion of such issues, the result of which does not change the universal conclusions, see eg App.~A of \cite{Mross:2010rd}.

We also want to study deformations of this integral, in search of a point about which we can develop a controlled expansion.  
First consider modifying the Landau damping term by $ |\epsilon| \to |\epsilon|^{2/z}$ \cite{schmalian2004quantum}.  
Here we will set $d=2$, $r=0$ and $k=0$ for simplicity.  
The $q_1$ integral can be done by scaling $ q_1  \equiv y |\epsilon|^{1/z}/\sqrt{\Gamma}$: 
\bea
& \Sigma(\omega,k)   =
\\ &   \nonumber
{g^2\over (2\pi)^2 v_F}
\int d\epsilon  dq_\parallel { \sign(\omega- \epsilon) \over r + |\epsilon| /\Gamma + ( \alpha q_\parallel + \beta \( k_\perp + \cos \theta_0 q_\parallel \) ^2 } ~.
\\ & =  \nonumber
{g^2 \over v_F (2\pi)^{d+1}} 
\sqrt{\Gamma} \int_{-\infty}^\infty { dy \over 1 + y^2 } 
\int d\epsilon {\text{sign}(\omega + \epsilon)}  |\epsilon|^{-1/z} 
\\ & = \nonumber
2 \pi \sqrt{ \Gamma} {g^2 \Lambda^{d-2}\over v_F} 
{ z \over z - 1 }  \text{sign}(\omega) |\omega|^{z-1 \over z } . 
\eea
When $ z \to 1$, this becomes $ \Sigma \propto  \log |\omega| + ... $.

More useful may be the generalization where we modify the spatial kinetic term of the $\rho$ field by
\be 
D(\epsilon, q)^{-1}  =  r + q_\perp^x + { |\epsilon|^{2} \over \Gamma } 
\ee 
for some variable $x$.  
By the same methods, this gives 
\bea
&&\Sigma(\omega,k=k_F) 
\\ 
& = &  \nonumber
{g^2 \Lambda^{d-2}\over v_F (2\pi)^{d+1}} 
\sqrt{\Gamma} \int_{-\infty}^\infty { dy \over 1 + y^x } 
\int d\epsilon {\text{sign}(\omega + \epsilon)}  |\epsilon|^{1/x-1} 
\\ & =& {g^2 \Lambda^{d-2}\over v_F (2\pi)^{d+1}}  \text{sign}(\omega) |\omega|^{1/x} f(x) 
\eea
where 
\be f(x) \equiv \pi \csc\( { \pi \over x }\) \( 1 - e^{ 2 \pi \ii  { \left\lfloor \half - {x\over 2 } \right\rfloor \over x }  }\)  .\ee
This function has the property that as $ x \to1$, we find 
\be \Sigma(\omega, k=k_F) \propto \omega \log \omega + \cdots ,\ee
a marginal Fermi liquid correction to the self energy.  
The expansion about $x=1$ may therefore provide a controlled approximation analogous to that found by 
\cite{Nayak:1993uh}, and used in \cite{Mross:2010rd}, to repair problems in the large-$N$ expansion of other non-Fermi liquids found in \cite{Lee:2009epi,Metlitski:2010vm}.  

%%%%%%%%%%%%%%%%%%%%%%%%%%%%%%%%%%%%%

\section{A free fixed point}
\label{appendix:scaling}

We will consider a situation where both the Fermi surface and Bose surface have codimension $c$ in momentum space.  We assume that they lie in the same $d-c+1$-dimensional subspace of the $d$-dimensional momentum space. 
As in 
\cite{2009PhRvL.102d6406S,Dalidovich:2013qta, Lee:2017njh},
our motivation and immediate goal is to identify a value of $c$ where our interactions become marginal, analogous to the upper critical dimension.
Bose surfaces with codimension $c>1$ have been studied in \cite{PhysRevB.16.4137-swift-leitner}.
We decompose the momentum of the fermion field as 
\be \vec k = k_F \hat \Omega + \vec k_{\perp F}\ee
where $k_F \hat \Omega$ is the 
point on the Fermi surface closest to $\vec k$ (this is unambiguous for our round Fermi surfaces).
$\vec k_{\perp F} \perp \hat \Omega$ is perpendicular to all the vectors tangent to the Fermi surface, and has $c$ independent components.
In the important special case when $c =1$, this can be written as $ \vec k =\hat \Omega( k_F + k_\perp)$. We also make the analogous decomposition for the boson momentum about the Bose surface, 
\be \vec q = q_0 \hat \Omega + \vec q_{\perp B}~.\ee
We seek a scaling symmetry of the form
\bea 
\label{eq:scaling-rule}
\psi_{\omega, \vec k = k_F \hat \Omega + \vec k_{\perp F}} &  \mapsto \lambda^{\Delta_c} \psi_{\lambda^{z_c} \omega,  k_F \hat \Omega + \lambda \vec k_{\perp F}} ,
\\  \rho_{\omega, \vec q =  q_0 \hat \Omega + \vec q_{\perp B}} & \mapsto\lambda^{\Delta_\rho}  \rho_{\lambda^{z_\rho}\omega, k_F \hat \Omega +  \lambda \vec q_{\perp B}} ~~,
\eea
analogous to the scaling symmetry of the Fermi liquid \cite{Polchinski:1992ed, shankar-RG}.
%We have defined the scaling rule $\lambda$ by the transformation of the component of the fermion momentum transverse to the FS, $k_\perp \equiv \sum_{i=1}^{c_c} \sqrt{(k_\perp^i)^2}$, so that we can make contact with Polchinski's scaling analysis of the Fermi liquid \cite{Polchinski:1992ed}.  
Note that scaling rule involves $k_{\perp F} \equiv  |k_{\perp F}|  = \sqrt{\sum_{i \in \perp} (k_{\perp F})_i^2 } $, the distance from the vector $\vec k$ to the Fermi surface,
and the analogous property of the momentum of the Bose field.
While it is tempting to speak loosely and say that we scale the momenta, of course it is the fields that transform under the scale transformation.

To constrain the possible form of the effective action, we would like to begin our RG flow with a theory that is local in space and time.  
In order to have a local action with codimension $>1$ gapless fermion modes, we must introduce spin indices
\cite{2009PhRvL.102d6406S,Dalidovich:2013qta, Lee:2017njh}.  
We will denote the spinful fermion field as $\Psi \equiv (\psi, ...) $; it has $s$ spin components.  We will take $s\to 1$ or $2$ at the end of the calculations, which rely only on the Clifford algebra 
$ \{ \Gamma^\mu, \Gamma^\nu \} = 2 \delta^{\mu\nu}$, where $\Gamma^\mu$ are a collection of $D$ $s \times s$ matrices.  
No spin indices are necessary for the boson, since $\int_q q_{\perp B}^2 \rho_q \rho_{-q}$ is already the Fourier transform of a local functional.

The (local) critical action we wish to study is
$ S[\Psi, \bar \Psi, \rho] = S_\Psi + S_\rho + S_{g} + S_u + S_r$ 
where we define the individual terms next. 
The kinetic term for the fermions will have the schematic form
\be S_\Psi \equiv \int \dbar \omega ~\dbar^d k ~\bar \Psi \ii \Gamma^\mu \( K_{\perp F}\)_\mu \Psi,\label{eq:Q-space-action}\ee
 generalizing a nodal line
 (we postpone a detailed discussion to \S\ref{sec:kinematics-of-nodal-lines}).
This problem has an approximate relativistic symmetry that rotates the frequency and the momenta perpendicular to the Fermi surface, $\gSO(c+1)$.
Following \cite{Dalidovich:2013qta}, we use capital letters $K^\mu$ to denote vectors of this $\gSO(c+1)$, 
$ \(K_{\perp F} \)_\mu \equiv (\omega, \vec k_{\perp F} , 0)_{\mu} $.
(Note that $ \vec k_{\perp F}$ can be further decomposed into a component in the linear subspace containing the Fermi surface, and a component perpendicular to this subspace; this distinction will be important below.)

The boson kinetic term is: 
\be S_\rho \equiv \half \int\dbar \omega~\dbar^d q \rho_q \rho_{-q} Q_{\perp B} ^2  \ee
where $Q_{\perp B}^\mu \equiv (\omega, \vec q_{\perp B}, 0 )^\mu $. 
This is an approximation to the momentum space representation of the local action
\bea \int d^D x \rho_x 
& \( ( \ii \partial_\tau)^2 + \sum_{\alpha = d-c+2}^d  
( \ii \partial_\alpha)^2 
\right. \nonumber  
\\ & \left. 
+ \( \sum_{i = 1}^{d-c+1} ( \ii \partial_i)^2 - q_0^2 \) ^2 \) \rho_x ~.\eea
The interaction terms are 
\be S_g \equiv \int d^D x g \bar  \Psi(x) M \Psi(x) \rho(x) , ~~~S_u \equiv \int d^D x u  \rho^4(x) ~.\ee
$M$ is a matrix of dimensionless numbers.
We will see that the four-fermion interaction is irrelevant at the critical value of $c$, so we omit it from the outset.

Now consider the scaling under \eqref{eq:scaling-rule} of each of these terms in turn. 
In order for $S_\Psi$ to be scale invariant we must take $z_\Psi = 1$.
Defining $\tilde K_\perp \equiv \lambda K_\perp$,
\bea 
S_\Psi[\Psi] &  =& 
%\int \dbar\omega ~\dbar^d k \bar \Psi_{\omega,k} \Psi^\nd_{\omega,k} \( \Sigma(\omega) + v_F k_\perp \) 
%\\ & = & 
  k_F^{d-c}\int \dbar ^{d-c} \hat \Omega ~\dbar^{c+1} K_\perp 
\bar \Psi_{\omega,k} \Psi_{\omega,k}  K_{\perp F} \cdot \Gamma \nonumber
\\  \nonumber & \mapsto& \lambda^{- 1 - c + 2 \Delta_\Psi } 
k_F^{d-c_c} \int \dbar^{d-c} \hat \Omega ~\dbar^{c+1}\tilde K_\perp 
\bar \Psi_{\tilde \omega,\tilde k} \Psi_{\tilde \omega,\tilde k} 
\\ & \cdot& ~~~\( \lambda^{-1 } \tilde K_{\perp F} \cdot \Gamma \)  
\\ & = & \lambda^{ - 2 - c + 2 \Delta_\Psi } S_\Psi[\Psi] ~.
\eea
Thus, in order for the kinetic term to be marginal, we must scale $\Psi$ with exponent 
$ \Delta_\Psi = \Delta_\psi = { c+ 2  \over 2} $.  

\bea 
S_\rho[\rho] 
& = &   \half q_0^{d-c}\int \dbar ^{d-c} \hat \Omega ~\dbar^{c+1} Q_\perp 
\rho_q \rho_{-q} Q_\perp^2
\\ &  \mapsto & 
\lambda^{- c - 1 + 2 \Delta_\rho} \half q_0^{d-c}\int \dbar ^{d-c} \hat \Omega ~\dbar^{c+1} \tilde Q_\perp 
\lambda^{-2} \tilde Q_\perp^2 \rho_{\tilde q} \rho_{\tilde q}
\nonumber
\\ & = & \lambda^{- c - 3 + 2 \Delta_\rho} S_\rho[\rho]  
\eea
so
$ \Delta_\rho = { c + 3 \over 2 } $.  

Now let's consider the $\rho^4$ interaction.
The scaling of $\rho^4$ term is similar to that of the four-fermion term in Fermi liquid theory
\cite{Polchinski:1992ed, shankar-RG}.
\bea 
\nonumber
S_{\rho^4} [\rho] & = &\prod_{i=1}^4  \( \int \dbar^dq_i~\dbar\omega_i 
 \rho_{q_i} \) \delta \( \sum_i \omega_i\) \delta^d\(\sum_i q_i\)
\\ & \mapsto &
\lambda^{ - 4 (1+c ) + 4 \Delta_\rho + 1 + \delta } S_{\rho^4}  \eea
where $\delta$ is from  the scaling of the momentum delta function.
So $S_{\rho^4}$ scales as $\lambda $ to the power
\be
\Delta_{\rho^4} =  - 3 - 2 c + 4 + \delta .
\label{eq:quartic-scaling}
\ee
For generic kinematics, $\delta= c-1$ and 
\be
\Delta_{\rho^4}^\text{generic} = 2-c ,
\ee
relevant for $c < 2$.  
When the momenta are back-to-back or forward (these produce the same interaction by Bose statistics), 
then the tangent spaces to the Bose surface are parallel and one extra scaling variable is constrained by the delta function.
Then the scaling of the delta function is enhanced to $\delta = c$ and 
\be
\Delta_{\rho^4}^\text{forward} =  3 -c
\ee
relevant for $c < 3$.

Now let's look at the Yukawa term: 
\bea 
S_{g} [\Psi, \rho] & = & \int \dbar^Dk_1~ \dbar^D k_2 ~\dbar^Dq 
\bar \Psi_{k_1} M \Psi_{k_2}^\nd \rho_q  \delta^D\(-k_1 + k_2 + q \)  \nonumber
\\ & \mapsto  &
\lambda^{\Delta_{g}} \int \dbar^D \tilde k_1~ \dbar^D \tilde k_2 ~\dbar^D \tilde q  \nonumber
\bar \Psi_{\tilde k_1} M  \Psi_{\tilde k_2} \rho_{\tilde q}^\nd   \\ & & ~~~\cdot  \delta^D\(-\tilde k_1 + \tilde k_2 + \tilde q \)~
\eea
where
\be \Delta_{g} \equiv  - (3 c + 3 ) + 2 \Delta_\Psi + \Delta_\rho + \delta_3 .\ee
%$M$ is a matrix of dimensionless numbers.
Here $\lambda^{\delta_3}$ is the transformation of the momentum delta function;   
the scaling of the delta function counts the number of $\perp$ components that it constrains.

Now what is $\delta_3$?  It seems that $\delta_3 = c+1$ for the interactions of the critical modes.
When the two fermion momenta $k_{1,2}$ are on the Fermi surface, and are separated by a vector $q$ on the Bose surface (as in Fig.~\ref{fig:2d-geometry}), 
the delta function constrains $d-c$ non-scaling variables and $c+1$ scaling variables.
In contrast, for generic momenta, only $c$ scaling variables are constrained, and the generic interaction is therefore less relevant.
However, if we impose the condition that the interaction only involves modes precisely on the respective critical surfaces (analogous to forward scattering), there is no undetermined momentum in the in-plane directions.
Loop diagrams involving the Yukawa vertex would therefore not produce logarithms.  

We will find below (in App.~\ref{appendix:g-is-dangerously-irrelevant}) that if we do not restrict the momenta appearing in the vertex at all, loop diagrams involving the Yukawa vertex also do not produce logarithms for $c=3$.  We tentatively conclude that the Yukawa coupling $g$ is (dangerously) irrelevant at the upper critical dimension for $u$.  
In what sense is the irrelevant coupling $g$ dangerously irrelevant?   If we set $g$ to  $0$, the boson completely decouples from the Fermi surface, while for any finite $g$ it has a large effect on the infrared behavior.

However, in App.~\ref{appendix:sqrt-delta-scheme} we describe a scheme to partially constrain the vertex, which does produce logarithmic corrections at the inferred upper critical dimension.

We should consider interactions involving other powers of the boson.  
The term $ S_r[\rho] \equiv \int d^D x \rho_x^2 r $ is the relevant term that we tune through the transition.
The scaling of $\rho^3$ term is
\bea 
\nonumber
S_{\rho^3} [\rho] & = &\prod_{i=1}^3  \( \int \dbar^dq_i~\dbar\omega_i 
 \rho_{q_i} \) \delta \( \sum_i \omega_i\) \delta^d\(\sum_i q_i\)
\\ & \mapsto &
\lambda^{ - 3 (1+c ) + 3 \Delta_\rho + 1 + \delta } S_{\rho^4}  \eea
where $\delta$ is from  the scaling of the momentum delta function.
So $S_{\rho^3}$ scales as $\lambda $ to the power
\be
\Delta_{\rho^3} =  - {3 \over2} c + {5 \over 2} + \delta .
\label{eq:cubic-scaling}
\ee
For generic kinematics, $\delta = c-1$, this gives 
$ \Delta_{\rho^3}|_\text{generic} = 0$.  
We note that for generic kinematics, introducing any dependence on the momentum would increase the scaling dimension and make it less relevant, so the generic-kinematics interaction is not a function of angles like the forward-scattering interaction.  
For special kinematics, where all three momenta are exactly on the Bose surface (and therefore must form an equilateral triangle), $\delta = c$ and we find 
$ \Delta_{\rho^3}|_\text{special} = 1$, 
this interaction is relevant.  
These couplings we must either forbid by symmetry or tune to zero; it is not clear to us whether in a physical realization of the problem the generic and special kinematics represent different couplings that must be tuned independently.  
We believe there are no special kinematics where the boson self-couplings of degree larger than four are marginal or relevant.

We can also consider the four-fermion interactions.  
We find that even the forward-scattering interaction is irrelevant at $c=3$:
\bea 
\nonumber
S_{\psi^4} [\psi] & = &\prod_{i=1}^4  \( \int \dbar^{d+1} k_i
%~\dbar\omega_i 
  \) \psi_1^\dagger \psi_2^\dagger \psi_3 \psi_4 
  %\delta \( \sum_i \omega_i\) 
  \delta^{d+1}\(\sum_i k_i\)
\\ & \mapsto &
\lambda^{\Delta_{\psi^4}} S_{\psi^4}. \label{eq:four-fermion}\eea
For forward scattering the delta function scales as $\delta = c$, and we find 
\be\Delta_{\psi^4}  = - 4 (1+c ) + 4 \Delta_\psi + 1 + \delta 
= 1 - c \buildrel{c=3}\over {=} - 2. 
\nonumber
\ee
Thus, we ignore all four-fermion interactions.

In conclusion, in the main case of interest, $c = 1$, this IR theory is multicritical.  
However, we can control it by an expansion about $c=3$, where $u(\theta)$ are a marginal perturbations and $g$ is (dangerously) irrelevant.

For the boson self-interaction, we restrict to the forward-scattering interaction for now: 
\bea 
S_u &=&  \int \dbar^D q_1 \dbar^D q_2  \rho_{q_1} \rho_{q_2} \rho_{-q_1} \rho_{-q_2} u(q_1,q_2)~.
\label{eq:forward-scattering} 
\eea
% \delta^D \(\sum_i q_i \) \cdot 
% \\  && 
% u(\hat m_i)\Theta\( \left| \sum_i \hat m_i \right| < \delta p \)~.
% \eea
In the case $d-c=1$ where the Fermi and Bose surfaces are one-dimensional, rotation invariance, which we assume, then reduces the interaction $u(q_1,q_2) = u(\theta)$ to a function of the angle $\theta$ between these two vectors.  Bose statistics then implies $u(\theta) = u(\theta+\pi) = u(- \theta)$.  

We note that the interaction \eqref{eq:forward-scattering}, which is picked out by its relevance under scaling, is non-local in real space.  
It is intended as an approximation to a local interaction that is not delta-function localized on a subspace of momentum space, but rather involves some `wiggle room'.

\section{Kinematics of nodal lines}
\label{sec:kinematics-of-nodal-lines}

We focus on the case of $(d-c=1)$-dimensional nodal surfaces, \ie~nodal lines.  We begin with the case of a nodal line of codimension two $c=2$ and later describe the case of higher codimension.
Consider a collection of five $4\times 4$ matrices
$ \alpha_{x,y,z}, \beta, \gamma^0 $ 
satisfying
\be\label{eq:nodal-line-algebra1}
\{\gamma^0, \alpha_i\} =0 ,~\{\gamma^0, \beta \}=0 ,~\{ \alpha_i , \alpha_j \} = 0,~\{ \alpha_z, \beta\} =0
\ee
but
\be
\label{eq:nodal-line-algebra2}
[ \alpha_x , \beta ]= 0 , ~[ \alpha_y , \beta ]= 0 .
\ee
A set of matrices that accomplishes this is
(see equation (17) of \cite{2016ChPhB..25k7106F})
\be \label{eq:fang-basis} \alpha_x = \sigma_x ,~\alpha_y = \sigma_y \tau_y ,~\alpha_z = \sigma_z ,~\beta = \sigma_x \tau_x,~\gamma^0 = \sigma_y \tau_x .\ee 

\begin{figure}
$$ \parfig{.5}{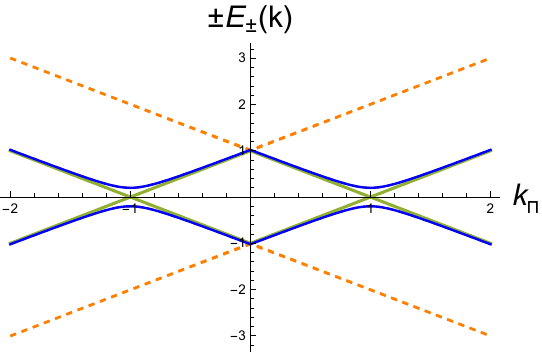} $$
\caption{\label{fig:nodal-line-spectrum}The spectrum of \eqref{eq:nodal-line-H} at $k_z=0$ (blue and orange) and $k_z = 0.2$ (green and orange).  The effective theory we develop below includes only the middle bands (in blue and green) with energy $\pm E_-(k)$.
}
\end{figure}

Then
\be\label{eq:nodal-line-H} H \equiv \Psi^{\dagger} \left ( k_i \alpha_i+ r \beta\right) \Psi \ee
has spectrum
$$ E_\pm(k)^2 = k_z^2+ (k_\perp \pm r)^2$$where $k_\perp \equiv \sqrt{k_x^2+ k_y^2} $.  
The middle branch $E_-$ 
(see Fig.~\ref{fig:nodal-line-spectrum})
has a nodal ring at $k_z=0, k_\perp = r$ where the dispersion can be approximated as the relativistic form
$$E_-(k)^2 \simeq k_\outplane^2+ k_\inplane^2
$$where $k_\outplane \equiv k_z$ and $k_\inplane \equiv k_\perp - r $.

The Hamiltonian $H$ admits the following to unitary particle-hole symmetry (denoted PH): 
\be 
\textrm{PH}: \Psi \to \tau_y \Psi^{\dagger}, \ii \to \ii \label{Eq:PHsym}
\ee 
This symmetry forbids both the chemical potential term $\Psi^{\dagger} \Psi$, as well as the term $ \Psi^{\dagger} \gamma^0 \Psi \equiv \bar \Psi \Psi$ (note that if $\Psi \to U (\Psi^{\dagger})^T, \ii \to \ii$, then a fermion bilinear $\Psi^{\dagger} N \Psi \to - \Psi^{\dagger} (U^{\dagger} N U)^T \Psi$). Further, the transpose of a derivative, i.e., $ \partial^T$, equals  $-\partial $. Here we have assumed that $N$ is traceless (if it were not, one can subtract a constant from it to make it traceless). Later, we will forbid the terms cubic in the density fluctuation field $\rho$ by choosing $\rho$ to transform as a fermion bilinear $\bar \Psi M\Psi$ that is odd under the PH symmetry.
%This implies that if we require that the density fluctuation field $\rho$ transforms in the same manner as  $\bar \Psi M\Psi$ (or alternatively, as $\Psi^{\dagger} M \Psi$), then the lowest order boson-fermion vertex will be $\bar \Psi M\Psi \rho$ (or alternatively, $\Psi^{\dagger} M \Psi \rho$), and furthermore, terms cubic in $\rho$ will be disallowed. From now on, we will enforce the PH symmetry, which implies that the chemical potential (= the coefficient of $\Psi^{\dagger} \Psi$ in $H$) is zero. 

An action associated with this Hamiltonian is
$$S = \int \bar \Psi \ii ( \gamma^0 \omega+ \gamma^i k^i+ \Upsilon r ) \Psi 
$$
where $\Upsilon \equiv - \ii \gamma^0 \beta$ and $\gamma^i \equiv - \ii \gamma^0 \alpha_i$
With this definition, all the gammas (as well as $\Upsilon$) square to one and are hermitian.
The explicit gamma matrices in the basis \eqref{eq:fang-basis} are 
\be 
\gamma_0 = \sigma^y \tau^x, 
\gamma_x = - \sigma^z \tau^x,
\gamma_y = - \sigma^0 \tau^z,
\gamma_z = - \sigma^x \tau^x,
\Upsilon = - \sigma^z \tau^0. \ee

The propagator for $\Psi$ is
\bea G &=& - \ii (  \gamma^0 \omega+ \gamma^i k^i+ \Upsilon r )^{-1}= \sum_{\alpha = \pm} { N_\alpha(k)\over \omega^2+ E_\alpha (k)^2} \nonumber
\\ & \approx&  { N(k)\over \omega^2+ E_- (k)^2} \equiv G_2(k). \label{eq:def-of-G2}\eea
We expand in the neighborhood of the nodal ring at $k_x^2 + k_y^2 = r^2, k_z = 0$ by writing
\be  (k_x, k_y) = \hat n ( r + k_\inplane), \hat n = (\cos\theta, \sin\theta)  ~,\ee
in terms of which the denominator is $ \omega^2 + E_- (k)^2 \approx \omega^2 + k_z ^2 + k_\inplane^2 $.  
In this regime, the numerator matrix $N(k) \equiv N_-(k)$ is 
\bea
2 \ii N(k) &=& 
% - \omega \gamma_0 + \ii k_z \gamma_z + \ii k_\inplane \hat n \cdot \vec \gamma 
% - \ii \Upsilon k_\inplane + \ii \omega \hat n \cdot \vec D_0 + k_z \hat n \cdot \vec D_z  
% \\ & = & 
% \nonumber
% \omega \( - \gamma_0 + \ii \hat n\cdot \vec D_0 \) 
% + k_\outplane^i \( \ii \gamma^i + \hat n \cdot \vec D_i \) 
% + k_\inplane \( \ii \hat n \cdot \vec \gamma - \ii \Upsilon \) 
% \\ & \equiv & 
% \nonumber
k_\outplane^i \Gamma_i^\theta + k_\inplane \Gamma_\inplane^\theta 
\eea
where 
the $\outplane$ directions include the time direction, and 
\be \Gamma_\mu^\theta \equiv \gamma_\mu 
- \vec D_i \cdot \hat n, ~~~~
\Gamma_\inplane^\theta \equiv \vec \gamma \cdot \hat n - \Upsilon ~.
\ee
In the basis \eqref{eq:fang-basis} above, 
% \be \vec D_0 \equiv \ii ( - \sigma^y \otimes \sigma^0 , \sigma^x \otimes \sigma^y ), 
% \vec D_z \equiv ( \sigma^x \otimes \sigma^0 , \sigma^y \otimes \sigma^y ) . \ee
% OR
\be \vec D_0 \equiv  (  \sigma^y \tau^0 , - \sigma^x \tau^y ), 
\vec D_z \equiv ( \sigma^x \tau^0 , \sigma^y\tau^y ) . \ee
This collection of matrices satisfies several nice properties, which we study next.

\subsection{Projected Dirac algebra}

In order to do the renormalization procedure, we will also need to know the form of the action that produces 
the approximate propagator 
\eqref{eq:def-of-G2}
% \be G_2 \equiv {N(k) \over \omega^2 + E_-(k)^2} \ee
which propagates only the branch with the nodal ring.
To understand this, first observe that the objects 
\be 
\Gamma_0^\theta \equiv  \gamma_0 - \hat n \cdot \vec D_0 , ~~
\Gamma_z^\theta\equiv \gamma^z - \hat n \cdot \vec D_z, ~~
\Gamma_\inplane^\theta  \equiv  \hat n \cdot \vec \gamma -  \Upsilon
\ee
satisfy an effective Clifford algebra 
\be \{ \Gamma_\mu^\theta, \Gamma_\nu^\theta \}  = 4 \delta_{\mu\nu} P_-(\theta) \ee
where $\mu,\nu \in \{ \omega, z, \inplane\} $ and 
\be P_-(\theta) \equiv \half \( \Ione - \hat n \cdot \vec \gamma  \Upsilon \)  \ee
is the hermitian rank-2 projector ($P_-(\theta)^2 = P_-(\theta)$) into the eigenspace associated with $E_-(k)$ (\ie~the range of $N(k)$).  
We call this algebra the projected nodal Dirac algebra.
We observe that 
\be P_-(\theta) N(k) P_-(\theta) = N(k) ,\ee
that is, the image of the propagator is entirely in the $E_-(k)$ subspace.

Therefore an effective action for just the $E_-$ branch is 
\be S_2[\Psi] = \int \dbar ^D k \bar \Psi  \ii \( \omega \Gamma_0^\theta + k_z \Gamma_z^\theta + v_F k_\inplane \Gamma_\inplane^\theta \) \Psi . \label{eq:effective-kinetic-term} \ee
The exact propagator determined by $S_2$ is 
$G_2$ in \eqref{eq:def-of-G2} (with $k_\outplane \to v_F k_\outplane$).  
\eqref{eq:effective-kinetic-term} bears a strong resemblance to the naive guess that we initially used, but which was not local.  
The difference is just in the algebra satisfied by these $\Gamma$s.   

The relative coefficient in the $\Gamma_\mu^\theta$ between the two terms is crucial for the property $\left(\Gamma_\mu^\theta\right)^2 = 2 P_- $.  
The self-energy $\Sigma$ that we generate at one loop will appear to violate this property. 
On the other hand, the high-energy bands clearly decouple and cannot be reintroduced by loop corrections. 
In order to understand the loop corrections to the effective action, it will be important to treat these projectors carefully.  Recall that given the bare propagator $G_2$, the self-energy $\Sigma$ (the sum of 1PI diagrams with one incoming and one outgoing fermion) corrects the full propagator $G_F$ as follows: 
\bea G_F(k) &=& G_2(k) + G_2(k) \Sigma(k) G_2(k) 
\\ & & \nonumber ~~~~+ G_2(k) \Sigma(k) G_2(k) \Sigma(k) G_2(k) + \cdots  
\nonumber \\ &=& G_2(k) { 1\over 1 + \Sigma(k) G_2(k) } ~.\eea
Thus, the self-energy only appears sandwiched between the bare propagator $G_2$, whose image is that of $P_-(\theta)$.  Thus $\Sigma$ may be replaced everywhere by 
\be \Sigma(k) \to P_-(\theta) \Sigma(k) P_-(\theta) ~\ee
in all calculations, which preserves the low-energy nodal subspace.

For future use, we define 
the deformed combinations 
\bea 
\Gamma_0^\theta(w) \equiv  &\gamma_0 &- w \hat n \cdot \vec D_0 , ~~
\Gamma_z^\theta(w)\equiv  \gamma^z - w \hat n \cdot \vec D_z, ~~ \nonumber
\\ \label{eq:deformed}
\Gamma_\inplane^\theta(w)  &\equiv& w \hat n \cdot \vec \gamma - \Upsilon~
\eea
which will appear in the loop calculations below.  
Their projections into the $\pm E_-$ subspace are 
\bea 
\tilde \Gamma_0^\theta(w) &\equiv& P_-(\theta) \Gamma_0(w)P_-(\theta), ~~
\tilde \Gamma_z^\theta(w) \equiv P_-(\theta)\Gamma_z(w)P_-(\theta), ~~ \nonumber
\\
\tilde \Gamma_\inplane^\theta(w) & \equiv& P_-(\theta) \Gamma_\inplane^\theta(w) P_-(\theta) .
\eea
These satisfy 
\be \{ \tilde \Gamma_\mu^\theta(w) , \tilde \Gamma^\theta_\nu(w) \} 
= \delta_{\mu\nu} 4 (1+w)^2 P_-(\theta) . \ee
Indeed, this relation follows from the fact that the projections of the deformed matrices are related to the original ones by the simple relation
\be\label{eq:project-gamma-w} 
\tilde \Gamma^\theta_\mu(w) = { 1 + w \over 2 } \Gamma^\theta_\mu. \ee

%\JM{I THINK IT IS GOING TO WORK!}

\subsection{Trace identities}
In this subsection we write $k_\inplane$ in place of $\Delta k_\inplane$ to avoid clutter. 
The numerator matrix $N(k)$ satisfies the following nice relation, similar to that for the numerator of the ordinary Dirac propagator: 
\bea\nonumber
 \tr N(k) N(p) &=& -2 (k^0 p^0 + k_z p_z + k_\inplane p _\inplane) \cos^2 \( { \theta_k - \theta_p \over 2 } \)
\\  \label{eq:trace-identity2}
&=&  -2 (\vec k \cdot \vec p) \cos^2 \( { \theta_k - \theta_p \over 2 } \)
. \eea
In the last expression, we treat $ \vec k = (k^0, k_z, k_\inplane)$ as a 3-component vector.
Unlike the ordinary Dirac numerator, 
\bea
\label{eq:trace-identity3} 0 &\neq &\tr N(k) N(p) N(q)  
\\\nonumber &=& 2  \cos \( { \theta_k - \theta_p \over 2 } \) \cos \( { \theta_p - \theta_q \over 2 } \) \cos \( { \theta_q - \theta_k \over 2 } \) \cdot 
\\\nonumber & \cdot & 
\( k_z p_\inplane q_0 - k_\inplane p_z q_0 - k_z p_0 q_\inplane + k_0 p_z q_\inplane + k_\inplane p_0 q_z - k_0 p_\inplane q_z \)
\\ & = & 
2 \cos \( { \theta_k - \theta_p \over 2 } \) \cos \( { \theta_p - \theta_q \over 2 } \) \cos \( { \theta_q - \theta_k \over 2 } \)
\vec k \cdot \vec p \times \vec q  \nonumber
 \eea
 where in the last expression we (again) regard each vector as a three component object 
 $\vec k = (k^0, k_z, k_\inplane)$.  
 However, we will forbid a cubic interaction in our problem.  
The formula 
%\bea \tr N(k) N(p) N(q) N(w) 
%&=& \cos \( { \theta_k - \theta_p \over 2 } \) \cos \( { \theta_p - \theta_q \over 2 } \) \cos \( { \theta_q - \theta_w \over 2 } \) 
% \cos \( { \theta_w - \theta_k \over 2 } \)  \cdot  \nonumber
%\\ & \cdot & \( ( k \cdot p)( q \cdot w )+ (w \cdot k )( p \cdot q )- (k \cdot q )( p \cdot w) \)  
%\eea
\bea\label{eq:trace-identity4} && 
\tr N(k_1) N(k_2) N(k_3) N(k_4) 
\\\nonumber &=& 2 \cos \( { \theta_{12} \over 2 } \) \cos \( { \theta_{23} \over 2 } \) \cos \( { \theta_{34} \over 2 } \) 
 \cos \( { \theta_{41} \over 2 } \)  \cdot  
\\ & \cdot & \( ( k_1 \cdot k_2)( k_3 \cdot k_4 )+ (k_4 \cdot k_1 )( k_2 \cdot k_3 )- (k_1 \cdot k_3 )( k_2 \cdot k_4) \)  
\nonumber
\eea
is also similar to that for the ordinary Dirac numerator.
Notice that the cosines depend on the angle differences in cyclic order around the trace.
With our constraint on the angles in the vertex, each of these will turn into a factor of $\cos {\theta_0\over 2}$.

% For future reference when considering the coupling $\rho \psi^\dagger \psi$, we also note that 
% \bea \tr\gamma^0  N(k) \gamma^0 N(p) &=& -2 (k^0 p^0 - k_z p_z - k_\inplane p _\inplane) \cos^2 \( { \theta_k - \theta_p \over 2 } \).
% \eea
% \be
% \gamma^0 N(k^0, k_z, k_\inplane, \theta) \gamma^0 
% = N(k^0, -k_z, -k_\inplane, \theta)~.
% \ee

Define the chirality operator for $c=2$
\be\label{eq:chirality2} \Gamma \equiv \gamma_0 \gamma_x \gamma_y \gamma_z .\ee
For future reference (since we will use the coupling $\rho \bar \psi \Gamma \psi$), we also note that 
\bea \tr\Gamma  N(k) \Gamma N(p) &=& 2 (k^0 p^0 + k_z p_z - k_\inplane p _\inplane) \cdot 
\\ \nonumber && ~~~~\sin^2 \( { \theta_k - \theta_p \over 2 } \).
\eea
\bea\nonumber 
&&\Gamma N(k^0, k_z, k_\inplane, \theta) \Gamma 
= - N(k^0,k_z, -k_\inplane, \theta+ \pi)\\ 
&&=~~~~ N(-k^0,-k_z, k_\inplane, \theta+ \pi)~.
\eea

\subsection{The case $c=3$}

To make a nodal surface with codimension three by the method above requires $ 8 $-component spinors. 
We use $\mu^{a=0,x,y,z}$ to denote the Pauli operators acting on the new index.
In terms of the $4 \times 4$ matrices \eqref{eq:fang-basis} used above for $c=2$, we take 
% \be 
% \alpha_i^8 = \alpha_i \otimes \sigma^x, i = x,y,z,~~~
% \beta^8 = \beta \otimes \sigma^x, 
%  \gamma_0^8 = \gamma_0 \otimes \sigma^x, 
% \alpha_w = \sigma^0 \otimes \sigma^0 \otimes \sigma^z \ee
% OR
\be 
\alpha_i^8 = \alpha_i \mu^x, i = x,y,z,~~~
\beta^8 = \beta \mu^x, 
 \gamma_0^8 = \gamma_0 \mu^x, 
\alpha_w = \sigma^0\tau^0 \mu^z \ee
These satisfy the same set of relations
\eqref{eq:nodal-line-algebra1}, \eqref{eq:nodal-line-algebra2}
as for $c=2$.
The explicit gamma matrices $ \gamma_i = - \ii \gamma_0 \alpha_i$ are 
\bea 
\gamma_0 &=& \sigma^y \tau^x \mu^x, 
\gamma_x = - \sigma^z \tau^x \mu^0,
\gamma_y = - \sigma^0 \tau^z \mu^0, 
\\ \nonumber
\gamma_z &=& - \sigma^x \tau^x \mu^0,
\gamma_w = \sigma^y \tau^x \mu^y,
\Upsilon = - \sigma^z \tau^0 \mu^0. \eea
In the full 8-dimensional Clifford algebra, there are 7 independent generators satisfying $ \{ \gamma_\mu, \gamma_\nu \} = 2 \delta_{\mu\nu}$.  In addition to 
$\gamma_0,\gamma_x, \gamma_y,\gamma_z,\gamma_w$ above, 
the objects
$ \gamma_v \equiv \sigma_y \tau_x \mu_z$
and 
$\gamma_u \equiv \sigma_0 \tau_y \mu_0 $
also satisfy 
$ \{ \gamma_\mu, \gamma_\nu \} = 2 \delta^{\mu\nu}$.
We can regard
\be\label{eq:chirality3} \gamma_u = - \ii \gamma_0 \gamma_x \gamma_y \gamma_z \gamma_w \gamma_v \equiv \Gamma  \ee
as the chirality operator for $c=3$.  
Note that 
\be \Upsilon = - \ii \gamma_0 \gamma_z \gamma_w \gamma_v. \ee

The energy spectrum of 
$ H = \alpha_i k_i + r \beta $ 
is again $ E_\pm^2 = \vec k_\outplane^2 + \( | \vec k_\inplane| \pm  r \)^2 $ 
(where $\{ k_\outplane\} = \{ k_z, k_w \}, \{ k_\inplane\} = \{k_x, k_y \} $)
but now with a two-fold degeneracy of each level.
So again, we will focus on the neighborhood of the nodal ring by expanding 
\be k_\inplane = \hat n \( r + k_\inplane\),  \text{ with }k_\inplane, |\vec k_\outplane|, \omega \text{~small} \ee
in terms of which the energy satisfies 
$ E_-(k)^2 = k_\inplane^2 + k_\outplane^2$.  

The projected propagator is 
\be G_-(k)  = { N(k) \over \omega^2 + k_\inplane^2 + k_\outplane^2 } \ee
with 
\be 2 \ii N(k) = \vec k_\outplane \cdot \vec \Gamma_\theta 
+ k_\inplane \Gamma_\theta^\inplane ~. \ee
The explicit matrices are 
\be \Gamma^\theta_\mu =  \gamma_\mu - \hat n \cdot \vec D_\mu , ~~
 % \Gamma_\theta^z = \gamma^z - \hat n \cdot \vec D_z ,~~
 %  \Gamma_\theta^w = \ii \gamma^w + \hat n \cdot \vec D_w ,~~
   \Gamma^\theta_\inplane =  \hat n \cdot \vec \gamma -\Upsilon 
  \ee
%\alpha_x = \sigma^x \otimes \sigma^0 \otimes \sigma^x, 
%\alpha_y = \sigma^y \otimes \sigma^y \otimes \sigma^x 
where 
\bea\nonumber
\vec D_0 &\equiv& (  \sigma^y \tau^0 \mu^x, -\sigma^x \tau^y \mu^x ) , 
 \vec D_z \equiv  ( \sigma^x \tau^0\mu^0 , \sigma^y \tau^y\mu^0 ),  \\ 
\vec D_w &\equiv& ( -  \sigma^y \tau^0 \mu^y, \sigma^x \tau^y \mu^y ) ~.
\eea
As for the $c=2$ case, these matrices, for fixed $\theta$, satisfy the projected nodal Dirac algebra: 
\be 
\{ \Gamma_\theta^\mu , \Gamma_\theta^\nu \} = 2 P_-(\theta) 
\ee
where $P_-(\theta)
 \equiv \half \( \Ione + \hat n \cdot \vec \gamma  \Upsilon \) $ is the rank-four projector onto the low-energy subspace.

The relation \eqref{eq:project-gamma-w} continues to hold for $c=3$.
We will also need
\be\label{eq:trgngn} \tr \Gamma N(k) \Gamma N(p) 
=4 (   k_\outplane^\mu  \vec p_{\outplane \mu}
- k_\inplane p_\inplane ) \sin^2 \( { \theta_k - \theta_p \over 2 } \) \ee
(where $\vec k_\outplane  = (k_0, k_z, k_w)$)
and
\be \Gamma N(\vec k_\outplane, k_\inplane, \theta) \Gamma
= - N(\vec k_\outplane, -k_\inplane, \theta + \pi )
= N(- \vec k_\outplane, k_\inplane, \theta + \pi ). \ee
The fermion bubble with four boson insertions will be proportional to 
\bea
&&\tr \Gamma N(k_1) \Gamma N(k_2) \Gamma N(k_3) \Gamma N(k_4) 
\\ \nonumber &=&  4 \sin{ \theta_1- \theta_2 \over 2} 
\sin{ \theta_2- \theta_3 \over 2}
\sin{ \theta_3- \theta_4 \over 2}
\sin{ \theta_4- \theta_1 \over 2} \cdot
\nonumber
\\ && \nonumber ~~~ 
\(  
 ( \tilde k_1 \cdot k_2) ( \tilde k_3 \cdot k_4 ) 
+ ( \tilde k_4 \cdot k_1 ) ( \tilde k_2 \cdot k_3 ) 
- (k_1 \cdot k_3 ) ( k_2 \cdot k_4)
\) 
\eea
where $\tilde k \equiv (\vec k_\outplane, - k_\inplane)$.  
These are the same relations we found for $c=2$, times an overall factor of two from the spinor traces.

Extrapolating to other values of $c$, we will use 
\be\label{eq:trgngnc} \tr \Gamma N(k) \Gamma N(p) 
={s\over 2} (   k_\outplane^\mu  \vec p_{\outplane \mu}
- k_\inplane p_\inplane ) \sin^2 \( { \theta_k - \theta_p \over 2 } \) \ee
\bea\label{eq:trgngngnc}
&&\tr \Gamma N(k_1) \Gamma N(k_2) \Gamma N(k_3) \Gamma N(k_4) 
\\ \nonumber &=&  {s\over 2} \sin{ \theta_1- \theta_2 \over 2} 
\sin{ \theta_2- \theta_3 \over 2}
\sin{ \theta_3- \theta_4 \over 2}
\sin{ \theta_4- \theta_1 \over 2} \cdot
\nonumber
\\ && \nonumber ~~~ 
\(  
 ( \tilde k_1 \cdot k_2) ( \tilde k_3 \cdot k_4 ) 
+ ( \tilde k_4 \cdot k_1 ) ( \tilde k_2 \cdot k_3 ) 
- (k_1 \cdot k_3 ) ( k_2 \cdot k_4)
\) 
\eea
where $s$ is the number of spin components.

\section{RG flow with unconstrained kinematics}
\label{appendix:g-is-dangerously-irrelevant}

The following calculation can be usefully compared to the one-loop renormalization of the Gross-Neveu-Yukawa (GNY) model \cite{Zinn-Justin:2002ecy}.  The set of (six) diagrams contributing at one loop is the same, however not all of them have the same outcome as in the relativistic theory.  
We analyze them in turn.

In studying the renormalization of the boson self-interaction strength $u$, we will restrict attention to the case $d-c=1$ where the Fermi surface and Bose surface are both one-dimensional.  In this case $u = u(\theta)$ is a function of a single variable, and we will find its beta functional, and identify a stable fixed point.
For $d-c >1$, even with rotation invariance, $u$ depends on multiple angles; we leave this generalization for future work.

\begin{figure}
$$ \parfig{.45}{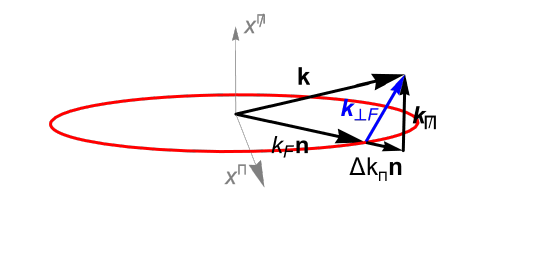} $$
\caption{\label{fig:nodal-line-geometry} 
In the case $d-c=1$, the Fermi surface is one-dimensional.  There are two kinds of directions normal to the FS, the directions in the plane of the FS ($\inplane$), and the directions perpendicular to the plane of the FS ($\outplane$).  
%\JM{FIX LABELS IN FIGURE.}
}
\end{figure}

The boson momentum can be parametrized as  
$ \vec q = \hat m (q_0 + q_{B\inplane} ) + \vec q _{B\outplane} $, 
where $q_0 \hat m$ is a point on the Bose surface, and 
the $(c-1)$-component vector $\vec q _{B\outplane}$ is perpendicular to the subspace containing the Bose and Fermi surfaces (see Fig.~\ref{fig:nodal-line-geometry}).
 We parametrize the boson kinetic term as 
\be\label{eq:introduce-v} 2S_\rho= 
\int_q \rho_q \rho_{-q} ( \omega^2 + q_{B\outplane}^2 + v_B^2 q_{B\inplane}^2 )
\equiv \int_q |\rho_q|^2 Q_B^2
. \ee
A similar statement applies to the fermion kinetic term: 
\be\label{eq:introduce-vF} S_\Psi= 
\int_k \bar \Psi_k \Psi_k \ii ( \omega \Gamma^0  + v_F \Delta k_{B\inplane} \hat n \cdot \vec \Gamma + 
 \vec k_{F\outplane} \cdot \vec \Gamma )
\equiv \int_q \bar \Psi_k \Psi_k  \ii \slashed{K}_F. \ee
In this section we can set $v_B = v_F = 1$, and need not be too careful about using the local nodal line propagator (in contrast to  the next section). In this appendix we use Yukawa coupling $ \bar  \Psi M \Psi \rho$ where $M = 1$; the specific form of $M$
is unimportant here (in contrast, in the next appendix the specific choice of $M$ will be important).

{\bf Boson self-energy.}
The contribution to the boson self energy for $\vec q$ near the Bose surface and low energy $\eps$ only involves the fermion propagators:
\bea \Pi(q) &= &
\parbox{.3\textwidth}{\begin{tikzpicture}[line width=1.0 pt, scale=.7]
%	\draw[fermion] (0,0) circle (1);
	  \draw [fermion]      (0,0)  arc [radius=1, start angle=0, end angle= 180] ;%node[midway,above]{$\omega-\epsilon, q-k$};
    \draw[ line width = .5pt, ->]  (-.8,1.3) -- (-1.2,1.3) node[midway,above]{$k+q$};
	  \draw [fermion]      (-2,0)  arc [radius=1, start angle=180, end angle= 360] ; %node[midway,below]{$\omega, q$};
    \draw[ line width = .5pt, ->]  (-1.2,-1.3) -- (-.8,-1.3) node[midway,below]{$k$};
  
	\draw[dashed] (-2,0)--(-2.5,0) ; %node[midway, below]{$(\epsilon, k)$};
	  \draw [line width = .5pt, ->] (-2.6, -.3) -- node[below] {$q$} ++ (0.4, 0); 
	
	\draw[dashed] (0,0)--(0.5,0) ; %node[midway, below]{$(\epsilon, k)$};
		  \draw [line width = .5pt, ->] (0.3, -.3) -- node[below] {$q$} ++ (0.4, 0); 

%	\draw[fermionbar] (30:1)--(0,0);
%	\draw[vector] (140:1)--(0,0);
%	\draw[vector] (-140:1)--(0,0);
	\end{tikzpicture}
 }
 \\ & 
= & {g^2\over 2} \int \dbar^D k_1\,\dbar^D k_2 ~\tr G(k_1) G(k_2)  \cdot 
\\ && \nonumber \delta^D(k_1 + q - k_2) 
\\ & = & + {s g^2\over 2} \int \dbar^D k  ~  {K_{F} \cdot ( K + Q)_{F} \over 
K_{F}^2 (K+Q)_{F}^2 }~.
\eea

Now we must be careful about the kinematics.  
The momentum of the external boson has components
\be q^\mu = (\varepsilon, q_0 \hat m + q_{B\inplane} \hat m, \vec q_{B\outplane})^\mu , \ee
where by definition $\vec q_{B\outplane}$ is out of the plane containing the Bose and Fermi surfaces.  
So the vector deviation from the Bose surface is $\vec q_B = q_{B\inplane} \hat m + \vec q_{B\outplane} $, and 
$Q_B^2 \equiv  \varepsilon^2  +  v_B^2 q_{B\inplane}^2 + |\vec q_{B\outplane}|^2 $.  
(Note that we include the coupling $v_B^2$ in the definition of $Q_B^2$.)
We make analogous decompositions for the fermion wavevectors: 
$ k^\mu = (\omega, k_F \hat n  + k_{F\inplane} \hat n, \vec k_{F\outplane})^\mu$, and 
$K_F^2 \equiv \omega^2 + k_{F\inplane}^2 + v_F^2 |\vec k_{F\outplane}|^2 $, and similarly for $k' = k+q$. 
%Accordingly, what is $(K+Q)_F$?  
%The tricky point is that the component $k_{F_\inplane} \hat n$ that was normal to the FS at $k_F \hat n$ is not normal to the FS at $k_F \hat n + q_0 \hat m = k_F \hat n'$. 
Since, in the present scheme (in contrast to the next section), the Yukawa vertex does not constrain the momenta beyond overall momentum conservation, $k_\inplane$ and $k'_\inplane$ can be used as independent integration variables.  
Including the frequency with the $\outplane$ directions, the result is
\bea
\Pi(q) 
&\propto& \int \dbar^{c} k_\outplane ~\dbar k_\inplane ~\dbar k'_\inplane {1\over k_\outplane^2 + k_\inplane^2 }
%{1\over k_\outplane^2 + k_\inplane^2 }
{1\over (k+q)_\outplane^2 + k_\inplane^2 }
\nonumber
\\ &\buildrel{c\to 3}\over{\sim }& \int {d^5 k \over k^4 } ~
\eea
which has no logarithmic divergence.

{\bf Fermion self-energy.}  Next we consider the contribution to the self-energy of the fermion
from the bosonic mode with a sphere of minima in its dispersion relation.
Note that a mass for the fermion is not generated; 
if $D$ is even we can say that this is guaranteed by the chiral symmetry, 
$\Psi \to e^{ \ii \alpha \Gamma } \Psi, \Gamma\equiv \ii^{- {D-2\over 2}}\prod_{\mu=0}^{D-1} \Gamma^\mu $. If $D$ is odd we can attribute it to time-reversal symmetry.  
\bea
\Sigma(k) &= &
\parbox{.3\textwidth}{
 \begin{tikzpicture}[line width=1.0 pt, scale=.7]
%	\draw[fermion] (0,0) circle (1);
	  \draw [fermion]  (-1, 0) -- (1,0);
	  \draw [fermion]  (-1.5, 0) -- (-1,0);
	  \draw [line width = .5pt, ->] (-2, -.3) -- node[below] {$k$} ++ (0.4, 0); 
	  \draw [line width = .5pt, ->] (-.5, -.3) -- node[below]{$p$} ++(1,0); 
	  \draw [line width = .5pt, ->] (1.5, -.3) -- node[below] {$k$} ++ (0.4, 0); 
	  \draw [fermion]  (1, 0) -- (1.5,0);
	  \draw [dashed]  %{<[scale=1.5,          length=5,          width=3]}-,line width=1.5pt, opacity=.4] 
    (1,0)  arc [radius=1, start angle=0, end angle= 180];
    \draw[ line width = .5pt, <-]  (-.2,1.3) -- (.2,1.3) node[midway,above]{$p-k$};
	\end{tikzpicture}
}
\\ & =&  \nonumber
g^2 \int \dbar^D p 
%~ \dbar^D q ~\delta^D(k + q - p ) 
G(p)D(p-k) ~.
%g^2 \int_{\star} \dbar^D p D(q) G(p)  
\label{eq:fermion-self-energy1}
\eea
%\\ & = & 
%g^2 \ii \slashed{\tilde K}_{F} \int_\star {d^D q \over ( Q_{B}^2+r) (K+Q)_{F}^2 }
%\eea
Again we can choose $p_{F\inplane}$ and $(p-k)_{B\inplane}$ as independent integration variables.  
Again there is no log divergence.
%\JM{MORE HERE.}

{\bf Boson correction to boson self-interaction.}
In the $s$-channel diagram for forward scattering, the internal lines are independent of the external momenta:
\bea \label{eq:boson-bubble}
\delta u_B(q_1,q_2) &=& \nonumber
\parbox{.045\textwidth}{\begin{tikzpicture}[scale=.6]
\draw[dashed, thick] (-.5,1) node[below, left]{$q_1$}-- (0,0) -- (-.5,-1) node[above, left]{$-q_1$};
\draw[dashed, thick] (.5,0) circle (.5);
	  \draw [line width = .5pt, ->] (.7, .6) -- node[above] {$p$} ++ (-0.4, 0); 
	  \draw [line width = .5pt, <-] (.7, -.6) -- node[above] {$p$} ++ (-0.4, 0); 
\draw[dashed, thick] (1.5,1) node[below, right]{$q_2$} -- (1,0) -- (1.5,-1 ) node[above,right]{$-q_2$};
\draw [line width = .5pt, ->] (-.4, .5) -- (-.2,.1); 
\draw [line width = .5pt, ->] (-.4, -.5) -- (-.2,.-.1); 
\draw [line width = .5pt, <-] (1.4, .5) -- (1.2,.1); 
\draw [line width = .5pt, <-] (1.4, -.5) -- (1.2,.-.1); 
\end{tikzpicture}
}
\\
&=&   4 \int \dbar^D p  u(q_1, p) u(p,q_2)\( { 1 \over P _{B}^2+r  } \)^2  \nonumber
\\ & = & 
%\underbrace{ 
4 {q_0^{d-c} \Omega_{d-c} \over  |v_B| (2\pi)^{d+1} }
%}_{\equiv N_d \gamma} 
\int \dbar \theta' u(\theta')u(\theta-\theta')  \cdot \nonumber 
\\ && \nonumber
~~\int d^{c+1}p_\perp \( { 1\over p_\perp^2 +r } \)^2 
\\ & \buildrel{c= 3 - \eps}\over{=} & 
%\gamma g^2 \int_{\Lambda^2 \lambda^2 }^{\Lambda^2 } d \omega \int _{\Lambda \lambda }^\Lambda dp_\perp p_{\perp}^{c_\rho -1 } \( { 1\over p_\perp^2 + |\omega | + r } \)^2 
%\\ & = & g^2 c 
%\int_{\Lambda^2 \lambda^2 }^{\Lambda^2 } d \omega \( u + |\omega| + r \)|_{u = \Lambda^2}^{u = ( \lambda \Lambda)^2 }  + \CO(\eps g^2 )  \label{eq:setcrhoto2}
%\\ & = & g^2 c \log \( { \omega + ( \lambda \Lambda)^2 + r \over \omega + \Lambda^2 + r } \)|_{\omega = (\lambda \Lambda)^2}^{\omega = \Lambda^2 }  + \CO(\eps g^2 ) 
%\\ & = & g^2 c \log { 2 \lambda^2 \Lambda^2 + r \over 2 \Lambda^2 + r } 
%\\ &\buildrel{\Lambda^2 \gg r } \over{ \simeq}& g^2 c \log \lambda^2 + \CO(\eps g^2 ) 
 4 {N_d\over \eps} { \gamma \over |v_B|}  \int \dbar \theta' u(\theta')u(\theta-\theta')
 . \eea
%Here we draw the diagram with all external momenta ingoing.  
$\gamma \equiv \Omega_{d-c} q_0^{d-c}$ is the volume of the Bose surface, and $N_d \equiv {\Omega_3 \over (2\pi)^D} =  { 1\over 8 \pi^2} {1\over (2\pi)^{ D-4}  } = {1\over 16\pi^3}$.
$\theta$ is the angle between $\vec q_{1\inplane}$ and $\vec q_{2\inplane}$.  The factor of $|v_B|^{-1}$ comes from the change of variables 
$ p_\perp^\mu \equiv  (\omega, \vec p_{B\outplane}, |v_B| p_{B\inplane})^\mu$.
(The factor of $4$ comes from the $- \half$ in the cumulant expansion, times $2 \cdot 2 \cdot 2$ ways to do the contractions in the $s$-channel.)

For generic $\vec q_1, \vec q_2$, we must also consider a possible contribution to the running of $u(\theta)$ from the $t$- and $u$-channel diagrams (which are related to each other by Bose statistics):
\bea
&& \parbox{.045\textwidth}{\begin{tikzpicture}[scale=.6]
\draw[dashed, thick] (-.5,1) node[below, left]{$q_1$}-- (0,0) -- (-.5,-1) node[above, left]{$q_2$};
\draw [line width = .5pt, ->] (-.4, .5) -- (-.2,.1); 
\draw [line width = .5pt, ->] (-.4, -.5) -- (-.2,.-.1); 
\draw[dashed, thick] (.5,0) circle (.5);
	  \draw [line width = .5pt, <-] (.7, .6) -- node[above] {$p_1$} ++ (-0.4, 0); 
	  \draw [line width = .5pt, <-] (.7, -.6) -- node[above] {$p_2$} ++ (-0.4, 0); 
\draw[dashed, thick] (1.5,1) node[below, right]{$q_1$} -- (1,0) -- (1.5,-1 ) node[above,right]{$q_2$};
\draw [line width = .5pt, <-] (1.4, .5) -- (1.2,.1); 
\draw [line width = .5pt, <-] (1.4, -.5) -- (1.2,.-.1); 
\end{tikzpicture}
} 
% \\ & = & \nonumber
%  {N_d \over \eps} {u(\theta)^2\over |v_B|}
%  G(\theta)
%  %F(\theta)} \Omega_{d-c-1} \( q_0 \sin{ \theta \over 2 }\)^{d-c-1} \delta p~
 \label{eq:t-channel-bubble}
\eea
However, with the restriction to forward scattering, for generic $q_{1,2}$, the external momenta completely determine the loop momenta and there is no log divergence.  
% where $\theta$ is the angle between $\vec q_1$ and $\vec q_2$, and 
% \be G(\theta) \equiv {\Omega_{d-c-1} \(q_0 \sin {\theta \over 2} \)^{d-c-1}  \delta p \over  F(\theta)}\ee
% and
% \be\label{eq:def-of-F} F(\theta) \equiv \begin{cases} 
% \sin^2 {\theta \over 2}, ~& \theta \in [\pi/2, 3\pi/2]
% \\ \cos^2 {\theta \over 2}, ~& \text{else}
% \end{cases}
% \ee
% is a continuous, nowhere-zero $2\pi$-periodic function, satisfying $F(\theta) =F(-\theta) = F(\theta + \pi)$.  
% Here is how we arrive at this expression.
% Momentum conservation constrains $p_2 = - p_1 + q_1 + q_2$.
% Writing $ \vec q_i = \hat m_i q_0 + \hat m_i q_{i\inplane} + \vec q_{i\outplane}$, 
% $ \vec p_i = \hat n_i q_0 + \hat n_i p_{i\inplane} + \vec p_{i\outplane}$, 
% this requires $ \hat m_1 + \hat m_2 = \hat n_1 + \hat n_2$, which in turn requires $ \hat n_1 = \hat m_1, 
% \hat n_2 = \hat m_2$ (or vice versa).  
% Thus, in this contribution, there is no convolution.
% Then the $\inplane$ component of $ - \vec p_1 + \vec q_1 + \vec q_2$ at $\hat m_2$
% is $ - p_\inplane \hat m_1 \cdot \hat m_2 = - p_\inplane \cos\theta$. 
% The geometric factor in the numerator is the volume of the sphere on the Bose surface of vectors whose lengths sum to $(\hat m_1 + \hat m_2) q_0$.  

{\bf Fermion correction to boson self-interaction.}
 Up to Bose statistics (which interchanges $q \to -q$ or $q' \to - q'$), there are only two different diagrams where a fermion loop contributes to the boson self-interaction, shown in \eqref{eq:fermion-bubble-with-4-bosons1} and \eqref{eq:light-by-light-diagram1}.
  Because of the difficulty of keeping all the internal lines near the Fermi surface (and the related kinematic constraint on the interaction vertices), these diagrams
only contribute logarithmic singularities for certain values of $\vec q\cdot \vec q' = \cos\theta$.  
Diagrams of both types contribute for the special case where $\vec q = \pm \vec q'$ (in \eqref{eq:light-by-light-diagram1} we show the diagram that contributes for $ \vec q = - \vec q'$).
We refer to this case ($\theta =0, \pi$) as the Brazovskii interaction after \cite{brazovskii1975phase}. 
\bea \label{eq:fermion-bubble-with-4-bosons1}
&&  % \delta u_F^{(1)}(q,q') =
  \parbox{.15\textwidth}{\begin{tikzpicture}[line width=1.0 pt, scale=.7]
%	\draw[fermion] (0,0) circle (1);
	  \draw [fermion]      (0,0)  arc [radius=1, start angle=0, end angle= 90]  node[midway,right]{$k+q'$};
%    \draw[ line width = .5pt, ->]  (-.8,1.3) -- (-1.2,1.3) node[midway,above]{$\omega-\epsilon, q-k$};
	  \draw [fermion]      (-2,0)  arc [radius=1, start angle=180, end angle= 270]  node[midway,below]{$k+q$} ; %node[midway,below]{$\omega, q$};
%    \draw[ line width = .5pt, ->]  (-1.2,-1.3) -- (-.8,-1.3) node[midway,below]{$\omega, q$};
	  \draw [fermionbar]      (0,0)  arc [radius=1, start angle=0, end angle= -90]   node[midway,below right]{$k+q+q'$} ;%node[midway,above]{$\omega-\epsilon, q-k$};
%    \draw[ line width = .5pt, ->]  (-.8,1.3) -- (-1.2,1.3) node[midway,above]{$\omega-\epsilon, q-k$};
	  \draw [fermionbar]      (-2,0)  arc [radius=1, start angle=180, end angle= 90]   node[midway,left]{$k$}; ; %node[midway,below]{$\omega, q$};
%    \draw[ line width = .5pt, ->]  (-1.2,-1.3) -- (-.8,-1.3) node[midway,below]{$\omega, q$};

	\draw[dashed] (-1,-1)--(-1,-1.5) ; %node[midway, below]{$(\epsilon, k)$};
	  \draw [line width = .5pt, ->] (-0.9, -1.5) -- node[right] {$q'$} ++ (0, 0.4); 

	\draw[dashed] (-1,1)--(-1,1.5) ; %node[midway, below]{$(\epsilon, k)$};
	  \draw [line width = .5pt, <-] (-0.9, 1.5) -- node[right] {$q'$} ++ (0, -0.4);

	\draw[dashed] (-2,0)--(-2.5,0) ; %node[midway, below]{$(\epsilon, k)$};
	  \draw [line width = .5pt, ->] (-2.6, -.3) -- node[below] {$q$} ++ (0.4, 0); 
	
	\draw[dashed] (0,0)--(0.5,0) ; %node[midway, below]{$(\epsilon, k)$};
		  \draw [line width = .5pt, ->] (0.3, -.3) -- node[below] {$q$} ++ (0.4, 0); 

%	\draw[fermionbar] (30:1)--(0,0);
%	\draw[vector] (140:1)--(0,0);
%	\draw[vector] (-140:1)--(0,0);
	\end{tikzpicture}
 } 
\\ &&=  
  g^4 \int  \dbar^D k~\tr G(k) G(k+q) G(k+q'+q) G(k+q') . 
%\right. 
\nonumber 
 \eea
\bea\label{eq:light-by-light-diagram1}
%\delta u_F^{(2)}(q,q') &=&
&&
\parbox{.15\textwidth}{\begin{tikzpicture}[line width=1.0 pt, scale=.7]
%	\draw[fermion] (0,0) circle (1);
	  \draw [fermion]      (0,0)  arc [radius=1, start angle=0, end angle= 90]  node[midway,right]{$k+q'$};
%    \draw[ line width = .5pt, ->]  (-.8,1.3) -- (-1.2,1.3) node[midway,above]{$\omega-\epsilon, q-k$};
	  \draw [fermion]      (-2,0)  arc [radius=1, start angle=180, end angle= 270]  node[midway,below]{$k+q$} ; %node[midway,below]{$\omega, q$};
%    \draw[ line width = .5pt, ->]  (-1.2,-1.3) -- (-.8,-1.3) node[midway,below]{$\omega, q$};
	  \draw [fermionbar]      (0,0)  arc [radius=1, start angle=0, end angle= -90]   node[midway,below right]{$k$} ;%node[midway,above]{$\omega-\epsilon, q-k$};
%    \draw[ line width = .5pt, ->]  (-.8,1.3) -- (-1.2,1.3) node[midway,above]{$\omega-\epsilon, q-k$};
	  \draw [fermionbar]      (-2,0)  arc [radius=1, start angle=180, end angle= 90]   node[midway,left]{$k$}; ; %node[midway,below]{$\omega, q$};
%    \draw[ line width = .5pt, ->]  (-1.2,-1.3) -- (-.8,-1.3) node[midway,below]{$\omega, q$};

	\draw[dashed] (-1,-1)--(-1,-1.5) ; %node[midway, below]{$(\epsilon, k)$};
	  \draw [line width = .5pt, <-] (-0.9, -1.5) -- node[right] {$q$} ++ (0, 0.4); 

	\draw[dashed] (-1,1)--(-1,1.5) ; %node[midway, below]{$(\epsilon, k)$};
	  \draw [line width = .5pt, <-] (-0.9, 1.5) -- node[right] {$q'$} ++ (0, -0.4);

	\draw[dashed] (-2,0)--(-2.5,0) ; %node[midway, below]{$(\epsilon, k)$};
	  \draw [line width = .5pt, ->] (-2.6, -.3) -- node[below] {$q$} ++ (0.4, 0); 
	
	\draw[dashed] (0,0)--(0.5,0) ; %node[midway, below]{$(\epsilon, k)$};
		  \draw [line width = .5pt, <-] (0.3, -.3) -- node[below] {$q'$} ++ (0.4, 0); 

%	\draw[fermionbar] (30:1)--(0,0);
%	\draw[vector] (140:1)--(0,0);
%	\draw[vector] (-140:1)--(0,0);
	\end{tikzpicture}
 }
 \\&=&    
g^4  \int  \dbar^D k~\tr  G(k) G(k+q) G(k) G(k+q')    
  \nonumber
  \eea
But again, in the present scheme, there is no logarithm in either of these diagrams.

{\bf Vertex correction.}
\bea
&& \parbox{.3\textwidth}{\begin{tikzpicture}[line width=1.0 pt, scale=.7]
%   \draw[fermion] (0,0) circle (1);
      \draw [fermion]      (2,2) -- node[midway,above]{$k$} (1,1);
      \draw[fermion] (1,1)  -- node[midway,above]{$k_1$}  (0,0) ;%node[midway,above]{$\omega-\epsilon, q-k$};
      \draw [fermion]    (0,0) -- node[midway,below left]{$k_1+q$} (1,-1);
      \draw[fermion] (1,-1) --node[midway,below]{$k' = k+q$} (2,-2) ;%node[midway,above]{$\omega-\epsilon, q-k$};

    \draw[dashed] (-2,0)--node[midway,below]{$q$}(0,0) ; %node[midway, below]{$(\epsilon, k)$};
     \draw [line width = .5pt, ->] (-1.5, -.1) --  (-.5, -.1);
      %-- node[below] {$\epsilon, k$} 
%     ++ (0.4, 0); 
    
    \draw[dashed] (1,-1)--node[midway,right]{$k_1-k$}(1,1) ; %node[midway, below]{$(\epsilon, k)$};
         \draw [line width = .5pt, ->] (1.1, -.5) --  (1.1, .5);

%         \draw [line width = .5pt, ->] (0.3, -.3) 
          %-- node[below] {$\epsilon, k$} 
%         ++ (0.4, 0); 
%   \draw[fermionbar] (30:1)--(0,0);
%   \draw[vector] (140:1)--(0,0);
%   \draw[vector] (-140:1)--(0,0);
    \end{tikzpicture}
     }
    \\ 
    &= & 
   g^3  \int \dbar^D k_1 G(k_1) G(k_1+q) D(k_1- k) ~.
\eea
Now, in order to determine which momenta can give singular contributions to the integral, we encounter a geometry puzzle.  Given a set of vectors $\vec k, \vec k', \vec q$ such that $ \vec k' = \vec k+\vec q$ with
$|\vec k| = |\vec k'| = \vec k_F, |\vec q| = q_0$,
is it possible to choose a vector $\vec k_1$ such that 
all three of the following are true?
\begin{enumerate}
\item $|\vec k_1| = k_F$
\item $|\vec k_1 + \vec q| = k_F $
\item $ |\vec k_1 - \vec k |  = q_0 $
\end{enumerate}
For generic values of $k_F/q_0$ (and in particular for the commensurate value appropriate to the square lattice), the answer is no.
Thus, the vertex correction does not give any contribution to the beta function for the Yukawa coupling\footnote{This is to be contrasted with the result in \cite{Metlitski:2010vm}, which does find a vertex correction. The difference is that \cite{Metlitski:2010vm} studies a critical boson mode at wavenumbers $\vec q= (\pm \pi, \pm \pi)$ with the property that $ 2\vec q =0$ modulo the reciprocal lattice, so that there {\it is} a solution to the relevant geometry problem.  We thank Darius Shi for raising this question.}.  
It is interesting to compare this statement with the Migdal theorem, which forbids vertex corrections to the phonon-FS coupling.
That is also a statement about the kinematical suppression of corrections to a Yukawa interaction between a Fermi surface and a bosonic mode (in that case, a phonon), but only holds in the limit of small ratio of electronic to ionic masses.

{\bf Mass renormalization.}  Finally, the running of $r$ is nearly identical to the GNY theory: 
\bea && \parbox{.3\textwidth}{\begin{tikzpicture}[line width=1.0 pt, scale=.7]
    \draw[dashed] (-1,0) -- (1,0);
    \draw[dashed] (0,.5) circle (.5); 
\end{tikzpicture}
}
\\
&=&  \int { d^D q_1 u(q,q_1)\over Q_{1B}^2 + r}
=  {\gamma\over |v_B|} {N_d \over \eps} \int \dbar \theta u(\theta)  r~.
\eea

{\bf Beta functions.} Our computation of the beta functions has the same general structure as the analysis of 
\cite{Zinn-Justin:2002ecy} (\S11.7) for the GNY theory.
In terms of the renormalized action
\bea
S_r
&=& 
Z_\Psi \int_k \bar \Psi \ii \( \omega \Gamma^0 + \vec k_{F\outplane} \cdot \vec \Gamma
+  k_{F\inplane} \hat n \cdot \vec \Gamma \)  \Psi \nonumber
\\ &+&  Z_\Psi Z_\rho^{1/2} g_0 
\int_x \bar \Psi M \Psi \rho  + Z_\rho^2 u_0 \int_x \rho^4, 
\\ &+& 
{Z_\rho\over 2} \int_q \rho_q \rho_{-q} 
 \left(\omega^2 + q_{B\outplane}^2
 + q_{B\inplane}^2 \right) 
 \nonumber
\eea
we have found 
that the following quantities should be equal to finite terms plus contributions from higher loops:
\bea
\mu^{-2} Z_\rho r_0 + {4 \gamma } \int \dbar \theta u(\theta) r {N_d \over \eps} 
&&\nonumber %...
\\ 
\mu^{-\eps}u_0(\theta) Z_\rho^2 - 
%\left[ 
%\right.  &&  \nonumber \\ 
%\left. \nonumber
%- g^4 ( \alpha(\theta) + \beta(\theta) )
%\right. \nonumber && \nonumber \\ \left. 
 {\gamma} \int \dbar\theta' u(\theta') u(\theta-\theta') 
%+ { u(\theta)^2 \over |v_B| } G(\theta) 
%\right]
{N_d \over \eps}
\label{eq:u0}
 &&  \nonumber
\eea
Since there is no wavefunction renormalization in the present scheme, we can set $Z_\Psi = Z_\rho = 1$.

% Then, demanding 
% $0 = \mu \partial_\mu \lambda^i_0$
% %\propto \eps f_i  - \sum_j \beta_j { %\partial f_i \over \partial %\lambda_j} \ee
% with $ \{ \lambda^i \} = \{ u, g, v_B^2, v_F \}$,
% we can extract the beta functions
% $\beta_i \equiv - \mu\partial_\mu \lambda_i$

Now we compute the beta functional for $u(\theta)$.
\be 
0 = \mu \partial_\mu u_0(\theta) 
\propto  
\eps f_{u(\theta)} - \int_{\theta'} \beta_{u(\theta')} { \delta f_{u(\theta)}  \over \delta u(\theta')}
 - \beta_g { \partial f_{u(\theta) } \over \partial g } + \cdots 
\ee
where $\int_\theta \equiv \int \dbar \theta $, and $\cdots $ is terms of higher order.  
(Note that we use the opposite sign convention for $\beta$ compared to \cite{Zinn-Justin:2002ecy}.)
The operator
\bea
&& { \delta f_{u(\theta)} 
\over \delta u(\theta')} = \delta(\theta- \theta') 
%\\ &+& 
+{ N_d \over \eps} \(
%g^2 c \delta(\theta- \theta') +
{ 2 u(\theta - \theta') }
%\over |v_B|}
\) + \cdots 
\nonumber
\eea
has inverse
\bea
&& \(  { \delta f_{u(\theta)} 
\over \delta u(\theta')} \)^{-1} 
= \delta(\theta- \theta') 
%\\ &-&
-{ N_d \over \eps} \(
%g^2 c \delta(\theta- \theta') +
{ 2 u(\theta - \theta') }
%\over |v_B|}  
\) + \cdots 
\nonumber
\eea
and therefore the beta-functional (for $\theta \in [0, \pi/2]$) is
\bea 
&&\beta_{u(\theta)} 
=  \eps u(\theta)  -N_d 
%\left[
%	 c g^2 u(\theta) 
% 	 \right.
% \\ && \left. + 
{4\gamma} \int_{\theta' } u(\theta') u(\theta- \theta')  
%	 +  {u(\theta)^2 \over |v_B|  }G(\theta)	
%	\right]
% \\ && \left.	
% 	- g^4 \(\alpha \delta_{\theta,0} + \beta \delta_{\theta, \theta_1} \)
% 	\right] + \cdots ~.  \nonumber
\eea
For other values of $\theta$ it is determined by the Bose symmetry relations $\beta_{u(\theta)}=\beta_{u(\theta+\pi)}
= \beta_{u(-\theta)}$ satisfied by $u(\theta)$.

Fourier transforming 
\be u(\theta) = \sum_{\ell \in 2\IZ} e^{ \ii \ell \theta} u_\ell \ee
(the momenta must be even so that $u(\theta) = u(\theta+ \pi)$)
we find that the modes of definite angular momentum decouple: 
\bea \beta_{u_\ell} 
= \eps u_\ell - { 4N_d \gamma  } u_\ell^2 
\eea
and there is a fixed point at 
$ u_\ell = { \eps  \over 4 N_d \gamma } $,
independent of $\ell$.  Therefore 
the fixed point configuration of $u(\theta) $ is a sum of delta functions at $\theta=0$ and $\theta=\pi$.

\section{A scheme to isolate the critical Yukawa interaction}
\label{appendix:sqrt-delta-scheme}

In this appendix, we consider a modification of the Yukawa coupling, analogous to the restriction of the quartic interaction to forward scattering in our theory (and in Fermi liquid theory).  
The idea is to isolate the part of the vertex that keeps the modes on their respective critical surfaces, which seems to be more relevant than for generic kinematics.
%The somewhat disappointing conclusion will be that we do not find a fixed point because of the running of various velocities.  
For a particular choice of Yukawa interaction, we will find a very interesting fixed point of the renormalization group.

The interaction term we study, marginal when $c=3$, is as follows:
\bea 
\nonumber
S_g &=& g \int \dbar^Dk_1 ~\dbar^D k_2 ~\dbar^D q 
\delta^D ( -k_1 + k_2 + q) \cdot 
\\  && 
\bar \Psi_{k_1} M \Psi_{k_2} \rho_{q}
%\Theta\(| -\hat n_1  + \hat n_2 + \hat m| < \delta p \)
~\sqrt{\delta(\theta_1- \theta_2, \theta_0)}~,
\eea
where we decompose each fermion's spatial momentum as $\vec k_i = k_F \hat n_i + \vec k_{i F\perp} $ with $k_F \hat n$ the closest point on the FS to $\vec k_i$, 
the boson momentum as $ \vec q = q_0 \hat m + \vec q_{B\perp}$ with $q_0 \hat m$ the closest point to $ \vec q_i$ on the Bose surface, and $ \hat n_i = (\cos\theta_i, \sin \theta_i, \vec 0)$.
We define the object 
\bea  \sqrt{\delta(\theta, \theta_0)} &\equiv &
\sqrt{\delta(\theta - \theta_0)} + \sqrt{ \delta(\theta + \theta_0)} 
\eea
to impose that the angle between $\vec k_{1\inplane} $ and $\vec k_{2\inplane}$ is $\theta_0$, but we do not specify which is larger.
We will eventually need to discretize the space of angles $ \theta_n = n d\theta, n = 1... N, d\theta= {N \over 2\pi}$.  
The unfamiliar-looking object $\sqrt{\delta(\theta)} $ should be understood as $ {\delta_{n,0}\over \sqrt{d\theta}}$.

Thus, with this scheme, both $u(\theta)$ and $g$ are marginal when $c=3$, and marginally relevant for $c = 3 - \eps$, as we described in App.~\ref{appendix:scaling}. 

The interaction matrix we choose is $ M = \ii \Gamma$, where $\Gamma$ is the chirality operator 
(in \eqref{eq:chirality2} or \eqref{eq:chirality3} for $c=2$ and $c=3$ respectively).  
With this choice of interaction, the associated Hamiltonian term is 
\be H_g = \int d^d x \, \psi^\dagger \ii \gamma^0 \Gamma \psi \, \rho \ee
which is hermitian, \ie, $\ii \gamma^0 \Gamma$ is a hermitian matrix (both for $c=2$ and for $c=3$). Crucially, the operator $\psi^\dagger \ii \gamma^0 \Gamma \psi$ is odd under the PH symmetry (Eq.~\ref{Eq:PHsym}), thereby forbidding terms cubic in $\rho$.

One small subtlety is the following.
The boson momentum can be parameterized as (see Fig.~\ref{fig:nodal-line-geometry})  
$ \vec q = \hat m (q_0 + \Delta q_{B\inplane} ) + \vec q _{B\outplane} $, 
where $q_0 \hat m$ is a point on the Bose surface, and 
the $(c-1)$-component vector $\vec q _{B\outplane}$ is perpendicular to the subspace containing the Bose and Fermi surfaces.
We will find that 
the two parts of the boson kinetic term, $ \rho_q \rho_{-q} ( \omega^2 + q_{B\outplane}^2) $ and $ \rho_q \rho_{-q}\Delta q_{B\inplane}^2 $, run differently. 
Thus, we must keep track of an additional coupling, which we will call $v_B^2$ (we take $v$ to be a positive number).  We parametrize the boson kinetic term as 
\be\label{eq:introduce-v2} 2S_\rho= 
\int_q \rho_q \rho_{-q} ( \omega^2 + q_{B\outplane}^2 + v_B^2 \Delta q_{B\inplane}^2 )
\equiv \int_q |\rho_q|^2 Q_B^2
. \ee
A similar statement applies to the fermion kinetic term: 
\be\label{eq:introduce-vF2} S_\Psi= 
\int_k \bar \Psi_k  \ii ( \omega \Gamma_0^\theta  + v_F \Delta k_{B\inplane} \hat n \cdot \vec \Gamma^\theta + 
 \vec k_{F\outplane} \cdot \vec \Gamma^\theta )\Psi_k
\equiv \int_q \bar \Psi_k  \ii \slashed{K}_F  \Psi_k. \ee
Thus, all velocities below are measured in units of the velocity in the $\outplane$ directions, which does not run, to the order we study.
Note that here we use the local fermion action developed in App.~\ref{sec:kinematics-of-nodal-lines}.

{\bf Boson self-energy.}
The contribution to the boson self energy for $\vec q$ near the Bose surface and low energy $\eps$ only involves the fermion propagators, 
and is essentially Landau damping.
\bea \Pi(q) &= &
\parbox{.3\textwidth}{\begin{tikzpicture}[line width=1.0 pt, scale=.7]
%	\draw[fermion] (0,0) circle (1);
	  \draw [fermion]      (0,0)  arc [radius=1, start angle=0, end angle= 180] ;%node[midway,above]{$\omega-\epsilon, q-k$};
    \draw[ line width = .5pt, ->]  (-.8,1.3) -- (-1.2,1.3) node[midway,above]{$k+q$};
	  \draw [fermion]      (-2,0)  arc [radius=1, start angle=180, end angle= 360] ; %node[midway,below]{$\omega, q$};
    \draw[ line width = .5pt, ->]  (-1.2,-1.3) -- (-.8,-1.3) node[midway,below]{$k$};
  
	\draw[dashed] (-2,0)--(-2.5,0) ; %node[midway, below]{$(\epsilon, k)$};
	  \draw [line width = .5pt, ->] (-2.6, -.3) -- node[below] {$q$} ++ (0.4, 0); 
	
	\draw[dashed] (0,0)--(0.5,0) ; %node[midway, below]{$(\epsilon, k)$};
		  \draw [line width = .5pt, ->] (0.3, -.3) -- node[below] {$q$} ++ (0.4, 0); 

%	\draw[fermionbar] (30:1)--(0,0);
%	\draw[vector] (140:1)--(0,0);
%	\draw[vector] (-140:1)--(0,0);
	\end{tikzpicture}
 }
 \\ & 
= & {g^2\over 2} \int \dbar^D k_1 \dbar^D k_2 ~ \tr M G(k_1)MG(k_2)  \cdot 
\\ && \nonumber \delta^D(k_1 + q - k_2) \delta(\theta_1 - \theta_2,  \theta_0) 
\\ & = &   {g^2\over 2} \int_\star \dbar^D k  ~  %{K_{F} \cdot ( K + Q)_{F} 
{\ii^2  \tr \Gamma N(k) \Gamma N(k+q) 
\over 
K_{F}^2 (K+Q)_{F}^2 }~. \nonumber
\eea
The $\star$ on the integral is there to remind us about the constraint on the angle between $\vec k_\inplane$ and $(\vec k + \vec q)_\inplane$.

A reminder about the kinematics.  
The momentum of the external boson has components
\be q^\mu = (\varepsilon, q_0 \hat m + \Delta q_{B\inplane} \hat m, \vec q_{B\outplane})^\mu , \ee
where by definition $\vec q_{B\outplane}$ is out of the plane containing the Bose and Fermi surfaces.  
So the vector deviation from the Bose surface is $\vec q_B = \Delta q_{B\inplane} \hat m + \vec q_{B\outplane} $, and 
$Q_B^2 \equiv  \varepsilon^2  +  v_B^2 \Delta q_{B\inplane}^2 + |\vec q_{B\outplane}|^2 $.  
(Note that we include the coupling $v_B^2$ in the definition of $Q_B^2$.)
We make analogous decompositions for the fermion wavevectors: 
$ k^\mu = (\omega, k_F \hat n  + \Delta k_{F\inplane} \hat n, \vec k_{F\outplane})^\mu$, and 
$K_F^2 \equiv \omega^2 + \Delta k_{F\inplane}^2 + v_F^2 |\vec k_{F\outplane}|^2 $. 
Accordingly, what is $(K+Q)_F$?  
The component $k_{F_\inplane} \hat n$ that was normal to the FS at $k_F \hat n$ is not normal to the FS at $k_F \hat n + q_0 \hat m = k_F \hat n'$.

\begin{figure}
$$ \includegraphics[width=.4\textwidth]{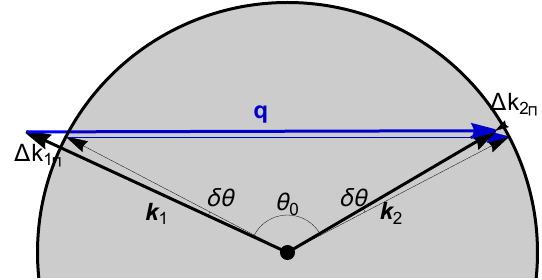}$$
\caption{ \label{fig:delta-theta}The dispersion of the fermions depends on the distance of $\vec k$ to the Fermi surface.  
We parametrize a general momentum $\vec k = \hat n k_F + \hat n k_{F\inplane} + \vec k_{F\outplane}$, 
where $\vec k_{F\outplane}$ lies in the $c-1$ directions perpendicular to the plane (or more general subspace) containing the Fermi surface (not shown here).  
When a fermion with momentum $\vec k_1$ absorbs a boson with momentum $\vec q$ via the Yukawa coupling, we need to decompose its resulting momentum $\vec k_2 = \vec k_1 + \vec q$ according to the same scheme.  
The marginally-relevant vertex only couples modes that satisfy $\hat n_1 \cdot \hat n_2 = \cos\theta_0$.  In the boson self-energy, the  external boson momentum $\vec q = \hat m (q_0 + q_\perp)$ is fixed (in blue).  Momentum conservation then determines the remaining angle $\delta \theta$ in terms of the deviations from the critical surfaces, $ \Delta k_{1\inplane}, \Delta k_{2\inplane}, \dqinplane$ according to \eqref{eq:argument-of-boson-delta}.
%\JM{FIX LABELS IN FIG}
}
\end{figure}

Consider the momentum conservation condition in the plane of the FS (see Fig.~\ref{fig:delta-theta}: 
\be \delta^2 \( \vec k_{1\inplane} + \vec q_\inplane - \vec k_{2\inplane} \) ~.\ee
We focus on one of the two solutions for $\theta_2 = \theta_1 - \theta_0$; the other will give the same contribution.
WLOG, we take the boson momentum to point in the $\hat x$ direction, 
$\vec q  = \hat m (q_0 + \dqinplane)$, $\hat m = (1,0)$.    
If for a moment we ignore the deviations from the critical surfaces, the momentum conservation condition says 
$ \hat m q_0 + \hat n_1 k_F - \hat n_2 k_F = 0 $ 
(with $\hat n_1 = (\cos \theta_1, \sin \theta_1), \hat n_2 = (\cos \theta_2, \sin \theta_2) = 
(\cos (\theta_1 - \theta_0) , \sin (\theta_1 - \theta_0) )) $
is solved when $ \theta_1 = { \pi + \theta_0 \over 2 }$ 
(so that $\theta_2= { \pi - \theta_0 \over 2}$).
%(The second solution for $\theta_1$ will multiply the final result of the integral by $2$).
Allowing for small deviations from this solution in the approximation $\dkinplane/k_F \ll 1$, we write 
\be \theta_1 = { \pi + \theta_0 \over 2 } + \delta \theta \ee
and expand in $\delta \theta$, $\Delta k_{1\inplane}, \Delta k_{2\inplane}, \dqinplane$ (regarding them all as the same order) to find that the argument of the delta function is (to linear order)
\be 
\label{eq:argument-of-boson-delta}
\(\dqinplane - \sin {\theta_0\over 2} (\Delta k_{1\inplane} + \Delta k_{2\inplane} ) , 
\cos {\theta_0\over 2 } (\Delta k_{1\inplane} - \Delta k_{2\inplane} ) -  { q_0 } \delta \theta \) \ee
The component along $\hat m$ then determines $\Delta k_{2\inplane}$ in terms of the remaining integration variable $\Delta k_{1\inplane}$
and the external momentum $\dqinplane$.
This part of the delta function is only a function of scaling variables, as in the scaling analysis above.  
Thus we can eliminate
\be 
\Delta k_{2\inplane} = - \Delta k_{1\inplane} + {\Delta q_\inplane \over \sin {\theta_0\over 2}} 
\label{eq:projection-into-FS}.
\ee
We will drop the subscript $1$ on $k_1$ from now on.
The second component of the delta function eliminates the $\delta \theta$ integral; note that $\delta \theta$ does not appear in the integrand anywhere else.  
% The actual point on the FS closest to $\vec k + \vec q$ will not be exactly $k_F \hat n'$.
% But since $ |k_{F\inplane}| \ll k_F$, we can approximate the vector between $ \vec k + \vec q$ and the FS as 
% (see Fig.~\ref{fig:kplusq-perp})\JM{Note that this is still an approximation, because $\hat n$ should change.}
% \be \vec k_{F\outplane} + \hat n' ( \cos \theta_0 k_{F\inplane} + \sin { \theta_0 \over 2 } q_{B\inplane}).
% \label{eq:projection-into-FS}
% \ee 
Using \eqref{eq:projection-into-FS}, we have 
\bea 
K_F^2 &=& \omega^2 + |\vec k_{F\outplane}|^2 + v_F^2 k_{\inplane}^2, 
\\
(K+Q)_F^2 &=& ( \omega +\varepsilon)^2 + | \vec k_{F\outplane} + \vec q_{B\outplane}|^2 \nonumber
\\ && ~~\nonumber+ 
v_F^2 \(- k_{\inplane} + { q_\inplane \over \sin {\theta_0\over 2}} \)^2 
\\ 
K_F \cdot (K+Q)_F 
 & =&  \omega ( \omega +\varepsilon) +
 \vec k_{F\outplane} \cdot (\vec k_{F\outplane} + \vec q_{B\outplane}). ~\nonumber
 \\ &+& \nonumber
 v_F^2 |\cos \theta_0| k_{\perp} \( k_{1\inplane} - { q_\inplane \over \sin {\theta_0\over 2}} \)
\eea\bea
  &=&  \half \( \nonumber
%K_F^2 
   \omega^2 + |\vec k_{F\outplane}|^2 + |\cos\theta_0| v_F^2 k_{\inplane}^2   ~\nonumber
\right. 
 \\  & + &\left. \nonumber
  %(K+Q)_F^2  
    (\omega +\eps)^2 + (k_{\outplane} + q_\outplane)^2 + v_F^2 |\cos\theta_0| 
    \(k_\inplane - { q_\inplane \over \sin {\theta_0\over 2}}\)^2 
 \right. 
\\ &-& ~~~~\left. 
  \( \epsilon^2 + |\vec q_{B\outplane}|^2 +   {v_F^2 |\cos\theta_0| \over \sin^2{\theta_0\over 2}} q_{\inplane}^2\) 
   ~~\) 
  ~. 
 \label{eq:three-terms-of-Pi}
 %( \omega + \varepsilon)^2 + |\vec k_{F\outplane} + \vec q_{B\outplane}|^2  +  \cos^2 \theta_0 k_{F\inplane}^2 
\eea

Writing
\be {1\over K_{F}^2 (K+Q)_{F}^2 } = \int_0^1 { dx \over \CD^2 }, \ee
the quantity being squared in the denominator is 
\be \CD \equiv (  \omega + x \varepsilon)^2 + | \vec k_{F\outplane} + x \vec q_{B\outplane}|^2 
+ v_F^2 \(\dkinplane - { x \over \sin {\theta_0 \over 2} } \dqinplane \)^2 - \Delta~
\ee
where $\Delta$ is independent of the integration variables and vanishes when the external momentum is on the Bose surface.
Notice that we can include the frequency components with the $\outplane$ components and we do so from now on.
We will make the change of variables
$ \tilde \omega = \omega + x \varepsilon, 
\vec {\tilde  {k}}_{F\outplane} \equiv  \vec k_{F\outplane}+ x \vec q_{B\outplane}, 
%\tilde k_{F\inplane} \equiv \sqrt{ \Upsilon} k_{F\inplane} $, 
\tilde k_\inplane \equiv |v_F| \( \dkinplane - { x \over \sin {\theta_0 \over 2} } \dqinplane\) $, so that 
\be \CD = \tilde k_\outplane^2 + \tilde k_\inplane^2 - \Delta. \ee

At this point we commit to using dimensional regularization to identify the logarithmic dependence on the ultraviolet cutoff.  
Using \eqref{eq:trgngnc}, the calculation of the boson self-energy now gives 
\bea && \Pi(q) \\ \nonumber
&=& \ii^2 {g^2\over 2} \int_\star \dbar^D k 
\frac{\tr \Gamma N(k) \Gamma N(k+q)}
{(K^2)_F (K+Q)^2_F}
\\\nonumber
&=& -{\cor 2} g^2 k_F { s \sin^2 { \theta_0 \over 2}  \over 4 }
\int dx {1\over |v_F|} \int {\dbar^{c+1}\tilde k \over (\tilde k^2 + \Delta)^2 }
\cdot  
\\ &\cdot & \nonumber 
\(   \( \tilde k_\outplane - x q_\outplane\) \( \tilde k_\outplane + (1-x) q_\outplane\) 
\right.
\\  
&-& \left. \nonumber v_F^2 \( {\tilde \dkinplane\over |v_F| } + { x \over \sin {\theta_0 \over 2} } \dqinplane \) 
\( - { \tilde \dkinplane \over |v_F| } + (1-x) { q_\inplane \over \sin {\theta_0 \over 2 } } \)  \) 
\\ &=& \nonumber
-{sg^2 k_F\over 2 |v_F| }   { \sin^2 { \theta_0 \over 2 }   }
\int_0^1 dx x (1-x)  {N_d \over \eps} \(  - q_\outplane^2 -v_F^2 { \dqinplane^2 \over \sin^2 { \theta_0 \over 2}} \)
\\ &=& 
{sg^2 k_F\over |v_F| }   { \sin^2 { \theta_0 \over 2}  \over 12 }
  {N_d \over \eps} \(  q_\outplane^2 + { v_F^2  \over \sin^2 { \theta_0 \over 2}} \dqinplane^2 \) .
\eea
Here $\eps \equiv 3- c$, and
$N_d = { 1\over 8 \pi^2} {1\over (2\pi)^{ D-4}  }$ as above.
A factor of {\cor 2} comes from the fact that there are two solutions for $\theta_2$.

\begin{figure}
    $$ \includegraphics[width=.45\textwidth]{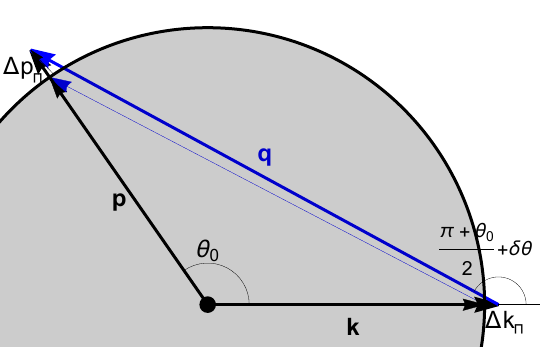} $$
    \caption{\label{fig-fermion-self-energy-kinematics}If we fix the external fermion momentum $\vec k$, momentum conservation and the constraint on the vertex leave a single degree of freedom $dp_\perp$ from the $\inplane$ momenta of the fermion self-energy diagram.}
\end{figure}

{\bf Fermion self-energy.}  Next we consider the contribution to the self-energy of the fermion
from the bosonic mode with a sphere of minima in its dispersion relation.
Note that a mass for the fermion is not generated; 
if $D$ is even we can say that this is guaranteed by the chiral symmetry, 
$\Psi \to e^{ \ii \alpha \Gamma } \Psi, \Gamma\equiv \ii^{- {D-2\over 2}}\prod_{\mu=0}^{D-1} \Gamma^\mu $. If $D$ is odd we can attribute it to time-reversal symmetry.  
\bea
\Sigma(k) &= &
\parbox{.3\textwidth}{
 \begin{tikzpicture}[line width=1.0 pt, scale=.7]
%	\draw[fermion] (0,0) circle (1);
	  \draw [fermion]  (-1, 0) -- (1,0);
	  \draw [fermion]  (-1.5, 0) -- (-1,0);
	  \draw [line width = .5pt, ->] (-2, -.3) -- node[below] {$k$} ++ (0.4, 0); 
	  \draw [line width = .5pt, ->] (-.5, -.3) -- node[below]{$p$} ++(1,0); 
	  \draw [line width = .5pt, ->] (1.8, -.3) -- node[below] {$k$} ++ (0.4, 0); 
	  \draw [fermion]  (1, 0) -- (1.5,0);
	  \draw [dashed]  %{<[scale=1.5,          length=5,          width=3]}-,line width=1.5pt, opacity=.4] 
    (1,0)  arc [radius=1, start angle=0, end angle= 180];
    \draw[ line width = .5pt, <-]  (-.2,1.3) -- (.2,1.3) node[midway,above]{$q$};
	\end{tikzpicture}
}
\\ & =&  \nonumber
g^2 \int \dbar^D p~ \dbar^D q ~\delta^D(k + q - p ) 
\cdot 
\\ & & \nonumber
M G(p)M D(q)\delta(\theta_k-\theta_p,\theta_0) ~.
%g^2 \int_{\star} \dbar^D p D(q) G(p)  
\label{eq:fermion-self-energy2}
\eea
%\\ & = & 
%g^2 \ii \slashed{\tilde K}_{F} \int_\star {d^D q \over ( Q_{B}^2+r) (K+Q)_{F}^2 }
%\eea
In terms of the deviations of the momenta from the Fermi surface, the measure for the $\inplane $ components is 
\be d^2 p_\inplane d^2 q_\inplane = k_F q_0 d\dpinplane d\dqinplane d\theta_p d\theta_q . \ee
The kinematic constraint from the vertex allows two values of $\theta_p$ over which we must sum; let's study the solution where $\theta_p = \theta_k + \theta_0$ first.  
If $\vec k_\inplane,\vec p_\inplane,\vec q_\inplane$ all lay on their respective critical surfaces, the $\inplane$ component of the delta function would be solved by 
$\theta_q = { \pi + \theta_0 \over 2}$.  
So we regard $ \dpinplane, \dkinplane, \dqinplane \ll q_0, k_F$ and expand $\theta_q = { \pi + \theta_0 \over 2} + \delta\theta$ (see Fig.~\ref{fig-fermion-self-energy-kinematics}), in terms of which the argument of the delta function in the $\inplane$ directions can be written as 
\be (q_0 \delta \theta +  \cos {\theta_0 \over 2} (\dpinplane - \dkinplane) , 
\dqinplane - \sin {\theta_0\over 2} (\dkinplane + \dpinplane ) ). \ee
Again, one of the arguments constrains the scaling variables.
Using this to do the integrals over $\dqinplane$ and $\theta_q$, the contribution from 
 $\theta_p = \theta_k + \theta_0$ is 
\bea 
 \Sigma_+ = {k_F g^2\over (2\pi)^{D}} \int d\dpinplane d^c p_\outplane
{ M N(p) M 
%\ii  \slashed{P}_{F} 
\over P_F^2 (P-K)_B^2 }~.
\eea
Note that we include the frequency with the $\outplane$ directions.  
The propagators are
\bea 
{1\over P_F^2 (P-K)_B^2}
= \int_0^1 dx {1\over \CD^2 } \eea
with
\bea 
\CD &\equiv& (1-x)P_F^2 + x (P-K)_B^2 
\\ &=&  (1-x) p_\outplane^2 + x (p_\outplane -  k_\outplane)^2 
\\ && \nonumber + (1-x)v_F^2\dpinplane^2 + 
x v_B^2 \sin^2 {\theta_0 \over 2} (\dpinplane + \dkinplane )^2 
\nonumber
\\ &=& ( p_\outplane - x k_\outplane)^2 
+ \sun \( \dpinplane + {x v_B^2 \sin^2 {\theta_0\over 2} \over \Upsilon } \dkinplane \)^2 
- \Delta 
\nonumber 
\eea
where 
\be 
\sun \equiv (1-x)v_F^2 + x v_B^2 \sin^2{\theta_0 \over 2}
\ee
and $\Delta$ is independent of the integration variables and vanishes when the external momentum lies on the Fermi surface.

Changing integration variables to 
\be 
\tilde p_\outplane \equiv p_\outplane - x k_\outplane, 
\tilde p_\inplane \equiv \sqrt{\sun}\( \dpinplane + {x v_B^2 \sin^2{\theta_0\over 2} \over \sun } \dkinplane  \), \ee 
the momentum integral becomes Lorentz invariant 
\bea 
\Sigma_+ &=& {k_F g^2} \int_0^1 {dx \over \sqrt{\sun}} 
 \underbrace{\int {d^{c+1} \tilde p 
 \over (2\pi)^D(\tilde  p^2 - \Delta)^2 } }_{= {N_d \over \eps}}   \cdot 
\\ 
&& \nonumber 
\( 
  p_\outplane^\mu \( \gamma_\mu + \hat n_p \cdot \vec D_\mu \) 
-  v_F \dpinplane \( - \hat n_p \cdot \gamma  - \Upsilon \)  \) 
% \( x k_\outplane\cdot \Gamma 
% - {x v_B^2 \sin^2{\theta_0 \over 2} \over \Upsilon } v_F \hat n_{\theta_p} \cdot \Gamma 
%  k_\inplane \)~.
\eea
The sum of the contributions $\Sigma_\pm $ from 
$\theta_p = \theta_k \pm \theta_0 $ cancel out the components not along $\hat n $, leaving 
\bea 
\Sigma &=& \Sigma_+ + \Sigma_- 
% = {2 k_F g^2} {N_d \over \eps} 
% \int_0^1 {dx \over \sqrt{\Upsilon}} \cdot 
% \\ && \nonumber 
% \( x k_\outplane\cdot \Gamma 
% - \cos \theta_0 {x v_B^2 \sin^2 {\theta_0\over 2}  \over \Upsilon } v_F k_\inplane \cdot \Gamma 
%   \)~.
%  \eea
\\ &=&  {\cor 2} g^2 \int_0^1 dx {N_d \over \eps}
\( 
% p_0 \( - \gamma_0 - \ii \vec n_p \cdot \vec D_0 \) 
  k_\outplane^\mu \( \gamma_\mu + {\cor \cos \theta_0} \hat n_p \cdot \vec D_\mu \) 
\nonumber \right.
\\  &&   \left.
+  v_F k_\inplane 
\( x v_B^2 \sin^2 {\theta_0\over 2} \over \sun \)
\( - {\cor \cos \theta_0} \hat n_p \cdot \gamma  - \Upsilon \)  \) 
 \nonumber
% \\ &  = & {\cor 2} g^2 k_F {N_d \over \eps} \int dx {1\over \sqrt{\Upsilon} } 
%  \left[ x  \( k_0 \( - \gamma_0 + {\cor  \cos\theta_0 } \ii \vec n_k \cdot \vec D_0 \) 
% + \ii k_\outplane^i \( \gamma_i - {\cor  \cos\theta_0 }  \ii \hat n_k \cdot \vec D_i \)  \) 
% \right.  \nonumber
% \\ && \left. ~~~~~~~
% +  \ii v_F \( {\cobl +} { x v_B^2 \sin^2 {\theta_0 \over 2 } \over \Upsilon} k_\perp \) 
% \( {\cor  \cos\theta_0 }  \hat n_k \cdot \vec \gamma  - \Upsilon \)  \right] 
\\ & = & 
{\cor 2} g^2 k_F {N_d \over \eps} 
\left[ 
\int_0^1 { x dx \over \sqrt{ \sun}}
k_\outplane^\mu \Gamma^{\theta+\pi}_\mu({\cor \cos\theta_0}) 
\nonumber
\right. 
\\ && \left. ~~~~~~\nonumber
 + \int_0^1 {x  dx \over  \sun^{3/2}} 
v_F k_\inplane \( { v_B^2 \sin^2  {\theta_0 \over 2 } \over \sun}\) \Gamma^{\theta+\pi}_\inplane({\cor \cos\theta_0}) 
\right]
\\ & = &  \nonumber
2 g^2 k_F {N_d \over \eps} 
\left[ 
\int_0^1 { x dx \over \sqrt{ \sun}}
%\(
k_\outplane^\mu \Gamma^\theta_\mu(-\cos\theta_0) 
%+ \ii k_z \Gamma^\theta_z(-\cos\theta_0) \) 
\nonumber
\right. 
\\ && \left. ~~~~~~ \nonumber
 + \int_0^1 {x  dx \over  \sun^{3/2}} 
v_F k_\inplane \( { v_B^2 \sin^2  {\theta_0 \over 2 } \over \sun}\) \Gamma^\theta_\inplane(-\cos\theta_0) 
\right]
 \eea
 The red factor of $ {\cor 2 \cos\theta_0 }$ comes from summing over the contributions
 of the two solutions of the angular constraint (which also plays the important role of cancelling the components of the in-plane momentum in directions transverse to $ \hat n$).  
 In the penultimate step we used the definitions 
 \eqref{eq:deformed}.
 In the last step we used the fact that $\Gamma^{\theta+\pi}_\mu(w) = \Gamma^\theta_\mu(-w)$.

Now we project the self-energy into the subspace spanned by the propagator. 
Using \eqref{eq:project-gamma-w}, the result is 
\bea P_-(\theta) \Sigma(k) P_-(\theta) 
&=&
2 g^2 k_F 
{ 1 - \cos\theta_0\over 2 }
{N_d \over \eps} 
\left[ 
\int_0^1 { x dx \over \sqrt{ \sun}}
%\( k_0 \Gamma_0^\theta + \ii k_z \Gamma_z^\theta \) 
k_\outplane^\mu \Gamma_\mu^\theta
\nonumber
\right. 
\\ &+& \left. 
  \int_0^1 {x  dx \over  \sun^{3/2}} 
v_F k_\inplane v_B^2 \sin^2  {\theta_0 \over 2 } \Gamma_\inplane^\theta
\right]
 \eea

 The $x$ integrals are
(using $\sin\theta_0>0$)
\bea \int_0^1 { x dx \over \sqrt{ \sun}}
&=& { 2 \over 3 } { 2 |v_F| + |v_B| \sin {\theta_0\over 2} 
\over (|v_F| + |v_B| \sin {\theta_0\over 2})^2 }
\\ \int_0^1 {x  dx \over  \sun^{3/2}}
 &=& { 2 \csc{\theta_0\over 2} \over |v_B| (|v_F| + |v_B| \sin {\theta_0\over 2} )^2 }.
\eea

{\bf Boson correction to boson self-interaction.}
This does not involve the Yukawa vertex and is therefore the same as in the analysis of App.~\ref{appendix:g-is-dangerously-irrelevant}.

\begin{figure}
\parfig{.45}{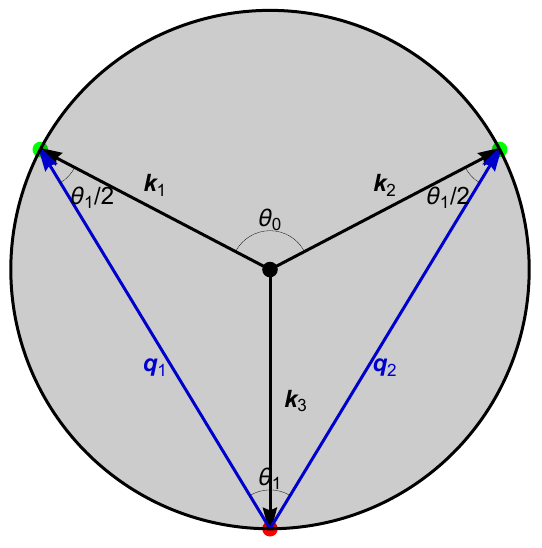}
\caption{\label{fig-theta1}$\theta_1 = \theta_0/2$ is defined to be the angle between two vectors $\vec q_1, \vec q_2$ that connect a point $\vec k_3$ on the FS to two other points on the FS, $\vec k_1, \vec k_2$.  When the Fermi surface is one dimensional, this determines a unique angle; when the Fermi surface is higher-dimensional, the story is more complicated.}
%cated, see Fig.~\ref{fig:3d-geometries}.}
\end{figure}

{\bf Fermion correction to boson self-interaction.}
 Up to Bose statistics (which interchanges $q \to -q$ or $q' \to - q'$), there are only two different diagrams where a fermion loop contributes to the boson self-interaction, shown in \eqref{eq:light-by-light-diagram2} and \eqref{eq:fermion-bubble-with-4-bosons}.
  Because of the difficulty of keeping all the internal lines near the Fermi surface (and the related kinematic constraint on the interaction vertices), these diagrams
can only possibly contribute logarithmic singularities for certain values of $\vec q\cdot \vec q' = \cos\theta$.  
Diagrams of both types contribute for the special case where $\vec q = \pm \vec q'$ (in \eqref{eq:fermion-bubble-with-4-bosons} we show the diagram that contributes for $ \vec q = - \vec q'$).
We refer to this case ($\theta =0, \pi$) as the Brazovskii interaction after \cite{brazovskii1975phase}. 
\bea\label{eq:light-by-light-diagram2}
%\delta u_F^{(2)}(q,q') &=&
&&
\parbox{.15\textwidth}{\begin{tikzpicture}[line width=1.0 pt, scale=.7]
%	\draw[fermion] (0,0) circle (1);
	  \draw [fermion]      (0,0)  arc [radius=1, start angle=0, end angle= 90]  node[midway,right]{$k+q'$};
%    \draw[ line width = .5pt, ->]  (-.8,1.3) -- (-1.2,1.3) node[midway,above]{$\omega-\epsilon, q-k$};
	  \draw [fermion]      (-2,0)  arc [radius=1, start angle=180, end angle= 270]  node[midway,below]{$k+q$} ; %node[midway,below]{$\omega, q$};
%    \draw[ line width = .5pt, ->]  (-1.2,-1.3) -- (-.8,-1.3) node[midway,below]{$\omega, q$};
	  \draw [fermionbar]      (0,0)  arc [radius=1, start angle=0, end angle= -90]   node[midway,below right]{$k$} ;%node[midway,above]{$\omega-\epsilon, q-k$};
%    \draw[ line width = .5pt, ->]  (-.8,1.3) -- (-1.2,1.3) node[midway,above]{$\omega-\epsilon, q-k$};
	  \draw [fermionbar]      (-2,0)  arc [radius=1, start angle=180, end angle= 90]   node[midway,left]{$k$}; ; %node[midway,below]{$\omega, q$};
%    \draw[ line width = .5pt, ->]  (-1.2,-1.3) -- (-.8,-1.3) node[midway,below]{$\omega, q$};

	\draw[dashed] (-1,-1)--(-1,-1.5) ; %node[midway, below]{$(\epsilon, k)$};
	  \draw [line width = .5pt, <-] (-0.9, -1.5) -- node[right] {$q$} ++ (0, 0.4); 

	\draw[dashed] (-1,1)--(-1,1.5) ; %node[midway, below]{$(\epsilon, k)$};
	  \draw [line width = .5pt, <-] (-0.9, 1.5) -- node[right] {$q'$} ++ (0, -0.4);

	\draw[dashed] (-2,0)--(-2.5,0) ; %node[midway, below]{$(\epsilon, k)$};
	  \draw [line width = .5pt, ->] (-2.6, -.3) -- node[below] {$q$} ++ (0.4, 0); 
	
	\draw[dashed] (0,0)--(0.5,0) ; %node[midway, below]{$(\epsilon, k)$};
		  \draw [line width = .5pt, <-] (0.3, -.3) -- node[below] {$q'$} ++ (0.4, 0); 

%	\draw[fermionbar] (30:1)--(0,0);
%	\draw[vector] (140:1)--(0,0);
%	\draw[vector] (-140:1)--(0,0);
	\end{tikzpicture}
 }
 \\&=&    
g^4  \int_\star  \dbar^D k~\tr M G(k)M G(k+q)M G(k)M G(k+q')    
  \nonumber
  \eea
\bea \label{eq:fermion-bubble-with-4-bosons}
&&  % \delta u_F^{(1)}(q,q') =
  \parbox{.15\textwidth}{\begin{tikzpicture}[line width=1.0 pt, scale=.7]
%	\draw[fermion] (0,0) circle (1);
	  \draw [fermion]      (0,0)  arc [radius=1, start angle=0, end angle= 90]  node[midway,right]{$k+q$};
%    \draw[ line width = .5pt, ->]  (-.8,1.3) -- (-1.2,1.3) node[midway,above]{$\omega-\epsilon, q-k$};
	  \draw [fermion]      (-2,0)  arc [radius=1, start angle=180, end angle= 270]  node[midway,below]{$k+q$} ; %node[midway,below]{$\omega, q$};
%    \draw[ line width = .5pt, ->]  (-1.2,-1.3) -- (-.8,-1.3) node[midway,below]{$\omega, q$};
	  \draw [fermionbar]      (0,0)  arc [radius=1, start angle=0, end angle= -90]   node[midway,below right]{$k+q+q'$} ;%node[midway,above]{$\omega-\epsilon, q-k$};
%    \draw[ line width = .5pt, ->]  (-.8,1.3) -- (-1.2,1.3) node[midway,above]{$\omega-\epsilon, q-k$};
	  \draw [fermionbar]      (-2,0)  arc [radius=1, start angle=180, end angle= 90]   node[midway,left]{$k$}; ; %node[midway,below]{$\omega, q$};
%    \draw[ line width = .5pt, ->]  (-1.2,-1.3) -- (-.8,-1.3) node[midway,below]{$\omega, q$};

	\draw[dashed] (-1,-1)--(-1,-1.5) ; %node[midway, below]{$(\epsilon, k)$};
	  \draw [line width = .5pt, ->] (-0.9, -1.5) -- node[right] {$q'$} ++ (0, 0.4); 

	\draw[dashed] (-1,1)--(-1,1.5) ; %node[midway, below]{$(\epsilon, k)$};
	  \draw [line width = .5pt, <-] (-0.9, 1.5) -- node[right] {$q$} ++ (0, -0.4);

	\draw[dashed] (-2,0)--(-2.5,0) ; %node[midway, below]{$(\epsilon, k)$};
	  \draw [line width = .5pt, ->] (-2.6, -.3) -- node[below] {$q$} ++ (0.4, 0); 
	
	\draw[dashed] (0,0)--(0.5,0) ; %node[midway, below]{$(\epsilon, k)$};
		  \draw [line width = .5pt, ->] (0.3, -.3) -- node[below] {$q'$} ++ (0.4, 0); 

%	\draw[fermionbar] (30:1)--(0,0);
%	\draw[vector] (140:1)--(0,0);
%	\draw[vector] (-140:1)--(0,0);
	\end{tikzpicture}
 } 
\\ &&=  
  g^4 \int_\star  \dbar^D k~\tr MG(k) MG(k+q) MG(k+q'+q)M G(k+q) . 
%\right. 
\nonumber 
 \eea  
The  diagram indicated in \eqref{eq:light-by-light-diagram2} 
contributes also when $q$ and $q'$ are such that there exists $k$ with 
$k$,  $k+q$ and $k+ q'$ all on the FS.
When the FS is one-dimensional, this happens only when $ \vec q \cdot \vec q'= \cos \theta_1$, with $\theta_1 = \theta_0/2$, 
the angle between two points on the Bose surface that connect a given point on the Fermi surface to two other points on the Fermi surface (see Fig.~\ref{fig-theta1}).  
When this condition on the external momenta is satisfied, there is an additional contribution to the running of $u(\theta)$.
The  diagram indicated in \eqref{eq:fermion-bubble-with-4-bosons} 
contributes when $q$ and $q'$ are such that there exists $k$ with 
$k$,  $k+q$ and $k+ q+ q'$ all on the FS.  This is the same condition on $\vec q\cdot \vec q'  \cos \theta_1$.  (See Fig.~\ref{fig-theta1}.)
We note that the analogous condition for $d-c >1$ is quite complicated.

\begin{figure}
$$\parfig{.4}{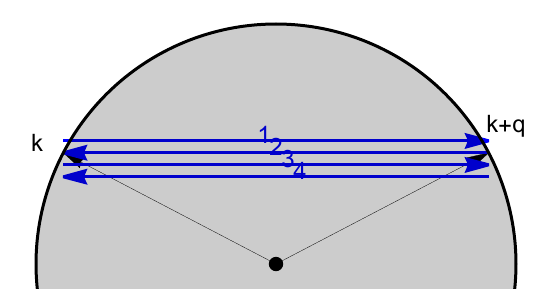}$$
$$
\parfig{.25}{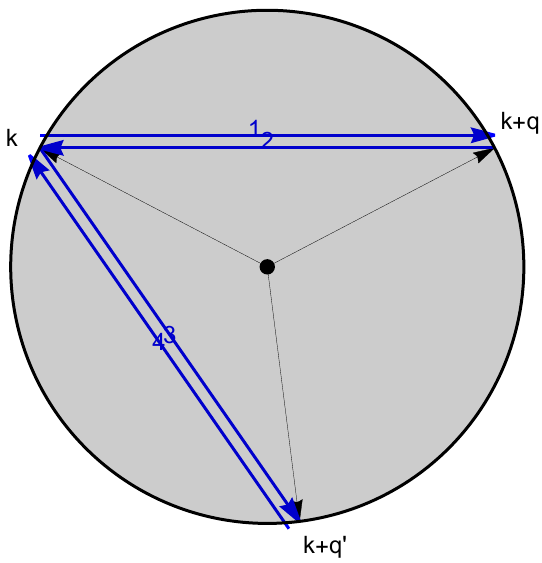}\parfig{.25}{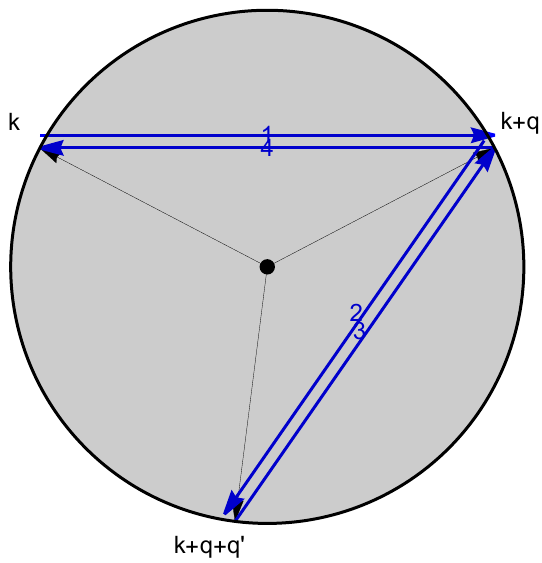}$$
\caption{\label{fig-four-boson-processes}The diagram \eqref{eq:light-by-light-diagram2} contributes when $q=q'$, in which case it describes the process indicated in the top diagram.  
It also contributes when the angle between $q$ and $q'$ is $\theta_1$,  in which case it describes the lower left process.  
The diagram in \eqref{eq:fermion-bubble-with-4-bosons} describes the lower right process; this only contributes when the angle between $q$ and $q'$ is $\theta_1$.  The last two processes are related by a relabelling of momenta.}    
\end{figure}

In both \eqref{eq:fermion-bubble-with-4-bosons} and \eqref{eq:light-by-light-diagram2} we can put the external boson momentum on the Bose surface.
Using \eqref{eq:projection-into-FS},
the vector deviations of the fermion momenta from the Fermi surface are 
\be \vec k_1  = ( \Delta k_\inplane, \vec k_\outplane), 
 \vec k_2  = ( - \Delta k_\inplane, \vec k_\outplane), 
 \vec k_3 = \vec k_1, \vec k_4 = \vec k_2. \ee
 Thus, using \eqref{eq:trgngngnc}, the numerator is 
 \bea 
\CN &\equiv &\tr \Gamma N(k_1)\Gamma N(k_2)\Gamma N(k_3)\Gamma N(k_4)
\\&=& {s\over 2} \sin^4 { \theta_0 \over 2}
%(2-1)
 \( \Delta k_\inplane^2 + k_\outplane^2 \) 
 \equiv {s\over 2} \sin^4 { \theta_0 \over 2} k^2.
\nonumber
 \eea
The same analysis applies to the second diagram, so the two diagrams only differ by a symmetry factor. 
For both, the integral gives
\bea  && g^4 \int_\star \dbar^Dk
{\CN \over (k^2)^4 }
%\\&=& \nonumber
= {s\over 2} \sin^4 { \theta_0 \over 2} {g^4 k_F\over |v_F|}{N_d \over \eps}
\eea

The result is of the form 
\be \delta u(\theta) = g^4 {N_d \over \eps} \left( \alpha(\theta)
%\( \delta(\theta) + \delta(\theta- \pi) \)
+ \beta(\theta) \right)~,
   % + \beta \(\delta_{\theta, \theta_1} 
   % + \delta_{\theta, \theta_1+ \pi}
   % + \delta_{\theta, -\theta_1}
   % + \delta_{\theta, -\theta_1+ \pi}
   % \)  \)  
\ee
% where 
% \be \delta_{\theta,0} \equiv \begin{cases} 0, &  \theta  \neq 0 \\ 
% 							  1 , & \theta = 0 
% 							  \end{cases}~,
% \ee
% is a {\it Kronecker} delta function, with zero measure.
%We must specify $\alpha, \beta$.
where 
(for the case of a one-dimensional Fermi surface ($d-c=1$))
\be \alpha(\theta)  = \sin^4 { \theta_0 \over 2} {k_F s \over 8 |v_F|} \(\delta(\theta) + \delta(\theta- \pi) \) \ee
and
%, where $V$ is again the ($d-c-1$-dimensional) volume of the intersection of the Fermi surface with the Bose surface centered at a point on the Fermi surface.
% The factor of $6$ is because of the $6$ possible orderings of external legs of the 
% diagram \eqref{eq:light-by-light-diagram2}.
\bea
    \beta(\theta) &=& \sin^4 { \theta_0 \over 2} {k_F s\over 8 |v_F|}  
    \( \delta(\theta- \theta_1)
    + \delta(\theta - \theta_1+ \pi)
\right.  \nonumber \\  &&    \left.
    + \delta(\theta + \theta_1)
    + \delta(\theta +\theta_1+ \pi)
   \)
\eea
(To understand the numerical factor: there are 36 possible contractions, 1/3 of which contribute to one of the terms in $\alpha$ and $2/3$ of which contribute to a term in $\beta$, but $\beta$ has twice as many terms.  The fourth order in the cumulant expansion contributes a factor of $-{1\over 24}$.)
 %\Omega_{d-c-1} \( k_F \sin{\theta_1\over 2}\)^{d-c-1}$ 
%It is equal to $\theta_1=\theta_0/2 \in [0, \pi/2]$. 
Interestingly, for $d-c>1$, $\beta(\theta)$ has support in an interval about $\theta=0$; we don't pursue this case further here.

We also observe in passing that if we did not include the factor of $\ii \Gamma$ in the interaction vertex, the result for both $g^4$ diagrams would be zero.
For both, the integral would give
\bea  && g^4 \int_\star \dbar^Dk
{(k^2)^2 - 2 (\tilde k \cdot k)^2 \over (k^2)^4 }
\\ &=& 
 {g^4 k_F\over |v_F|} {N_d \over \eps}
- 2 {g^4 k_F \over (2\pi)^{D}}
\int k_\outplane^{c-1} dk_\outplane dk_\inplane 
{\( k_\outplane^2 - v_F^2 k_\inplane^2 \)^2
\over 
\( k_\outplane^2 +  v_F^2 k_\inplane^2 \)^4
}\nonumber
\\ & = & 
\nonumber 
{k_F g^4 \over |v_F|} {N_d \over \eps} \(1 -  { \Omega_2 \over \Omega_3} {\pi \over 2} \)~
 = 0 ,
\eea
where we used
\be \int_{-\infty}^\infty dx 
{ \(x^2 - a^2 \)^2 
\over 
\(x^2 + a^2 \)^4
} = { \pi \over 4 a^3 } \ee
to do the $k_\inplane$ integral.
We would like to know the physical meaning of this unexpected cancellation.

{\bf Vertex correction.}  The analysis of the vertex correction is the same: there is no value of the loop momenta where all propagators are on their respective critical surfaces.

{\bf Beta functions.} 
% The computation of the beta functions for this model has the same general structure as the analysis of 
% \cite{Zinn-Justin:2002ecy} (\S11.7) for the Gross-Neveu-Yukawa theory.
In terms of the renormalized action
\bea
S_r
&=& 
Z_\Psi \int_k \bar \Psi \ii \(  k_{F\outplane}^\mu   \Gamma_\mu^\theta
+  v_{F0} k_{F\inplane} \Gamma_\inplane^\theta \)  \Psi \nonumber
\\ &+&  Z_\Psi Z_\rho^{1/2} g_0 
\int_x \bar \Psi M\Psi \rho  + Z_\rho^2 u_0 \int_x \rho^4, 
\\ &+& 
{Z_\rho\over 2} \int_q \rho_q \rho_{-q} 
 \left(\omega^2 + q_{B\outplane}^2
 + v_{B0} ^2 q_{B\inplane}^2 \right) 
 \nonumber
\eea
we have found 
that the following quantities should be equal to finite terms plus contributions from higher loops:
\bea
  k_{F\outplane}^\mu \cdot \Gamma_\mu^\theta
\( Z_\Psi + {{4 \over 3} g^2 k_F  
\(2|v_F| + |v_B|\sin {\theta_0 \over 2 }\)
\over (|v_F| + |v_B|\sin {\theta_0 \over 2 })^2   } {N_d \over \eps} \)
&& \nonumber
%&=&  ...
%\text{finite} + \text{2 loop}.
\\ 
{ k_{F\inplane} \Gamma_\inplane^\theta
} 
\( Z_\Psi  v_{F0} + 
 v_F |v_B|  {4 g^2 k_F \sin{\theta_0\over 2} \over \(|v_F| + |v_B|\sin {\theta_0 \over 2 }\)^2  } {N_d \over \eps} \)
&&  \nonumber
%...
\\ 
 \half (\omega^2 + q_{B\outplane}^2) \( Z_\rho +
 \cos^2 {\theta_0 \over 2 }    {  g^2 s k_F  \over
 6 |v_F| } 
% |v_F| \sin^2 {\theta_0 \over 2 }}  
 {N_d \over \eps} \)
 &&  \nonumber
%...%\text{finite} + \text{2 loop} 
\\ 
 \half q_{B\inplane}^2 \( Z_\rho v_{B0}^2 + 
 v_B^2 {\cos^2{\theta_0 \over 2}
 \over    \sin^2 {\theta_0 \over 2 }}
 {   g^2 s  k_F  |v_F|  \over 6 v_B^2 } 
{N_d \over \eps} \)
 &&  \nonumber
%... %\text{finite} + \text{2 loop} 
\\ 
%\mu^{-2} Z_\rho r_0 + \gamma u r {N_d \over \eps} &=& ...
\mu^{-2} Z_\rho r_0 + { \gamma } \int \dbar \theta u(\theta) r {N_d \over \eps} 
&&\nonumber %...
\\ 
\mu^{- \eps/2 } g_0 Z_\Psi Z_\rho^{1/2} - 0 g^3 
 &&  \nonumber
%...%\text{finite} + \text{2 loop} 
\label{eq:g0}
\\ 
\mu^{-\eps}u_0(\theta) Z_\rho^2 - \left[ 
%\right.  &&  \nonumber \\ 
%\left. \nonumber
- g^4 ( \alpha(\theta) + \beta(\theta) )
\right. \nonumber && \nonumber
\\ 
\left. 
+ {4 \gamma\over |v_B|} \int \dbar\theta'u(\theta') u(\theta-\theta') 
%+ { u(\theta)^2 \over |v_B| } G(\theta) 
\right] {N_d \over \eps}
 &&  \nonumber
 \eea

This determines
\bea Z_\Psi &=& 1 - 
a_\Psi g^2 {N_d \over \eps}, 
a_\Psi \equiv 
{{4 \over 3} k_F } {
2|v_F| + |v_B| \sin {\theta_0 \over 2 } \over \(|v_F| + |v_B| \sin {\theta_0 \over 2 } \)^2  }
\nonumber
\\ Z_\rho &=& 1 -  a_\rho g^2 {N_d \over \eps}, 
a_\rho \equiv 
{  s k_F \cos^2 {\theta_0 \over 2} \over  6 |v_F| }   \eea
and therefore the bare couplings are 
\bea \nonumber 
 v_{B0} &=& v_B\( 1  + g^2 e_B {N_d \over \eps} \)
 +... \equiv f_{v_B}(g,u,v_B,v_F)
 \\ 
 v_{F0} &=& v_F\( 1 + g^2 e_F {N_d \over \eps}\)
 +... \equiv f_{v_F}
 \\ 
 r_0 &=& \mu^2 r \( 1 + {N_d \over \eps} \( 
{ g^2 a_\rho } + {\gamma \over |v_B|}\int \dbar \theta u(\theta)   \)\) 
\nonumber
\\
 g_0 &=& \mu^{\eps/2}( g + b g^3 {N_d \over \eps})  + ... \equiv \mu^{\eps/2} f_g,
\\ 
%u_0(\theta)  &=& \mu^\eps ( u(\theta)  +{N_d \over \eps} ) + ... 
u_0(\theta) &=& \mu^{\eps} \( 
	u(\theta) + { N_d \over \eps} 
		\left[ g^2 c u(\theta) 
\right.\right. 
\\ 
&-&\left.\left.
 g^4 \( \alpha(\theta) 
 %\( \delta_{\theta,0} + \delta_{\theta, \pi}\)
   + \beta(\theta)
   % \(\delta_{\theta, \theta_1} 
   % + \delta_{\theta, \theta_1+ \pi}
   % + \delta_{\theta, -\theta_1}
   % + \delta_{\theta, -\theta_1+ \pi}
   % \)  
   \)  
   \right. \right. \nonumber
\\ &&  			\left. \left.
			+ {4 \gamma\over |v_B|} \int \dbar \theta'' u(\theta'') u(\theta- \theta'') 
   %+ { u(\theta)^2 \over |v_B|} G(\theta)
   \right]  \) \nonumber
\\ & \equiv&  \mu^\eps f_{u(\theta)}  \nonumber
 \eea
with 
\bea 
%a &=& {\gamma\over |v_B|} , 
c &=& { s k_F \over  |v_F| } , 
%~c = {2s V\delta p \over |v_F| \sin^2{\theta_0 \over 2} } ,
%\delta_{qq'}
\nonumber
\\ 
 b &=& a_\Psi + a_\rho/2
 %
 %k_F   \( 
%{1\over |v_B|+ |v_F| \sin {\theta_0 \over 2 }}+ {s \over 2 v_F \sin^2 {\theta_0 \over 2} } \),~
\\ e_F&=& 
a_\Psi - 
|v_B|  {4 k_F \sin {\theta_0\over 2} \over \(|v_B| + |v_F|\sin {\theta_0 \over 2 }\)^2  }
% {s k_F \over |v_F|} \(
% {1\over  \sin^2 {\theta_0 \over 2}}
% - {1\over v_B^2 }
% %{s V\delta p\over v_F} \( {1\over 2} + 1 %\cot^2{\theta_0 \over 2} 
%  \), 
\nonumber
\\ 
 e_B &=& 
 {a_\rho\over 2} - 
 {\cos^2{\theta_0 \over 2}
 \over    \sin^2 {\theta_0 \over 2 }}
 {  s  k_F  |v_F|  \over 12 v_B^2} 
 %  k_F  \( 
% { 2 - \cos \theta_0 /|v_F| 
% \over |v_B| + |v_F| \sin { \theta_0 \over 2 }}
% \)
%{1\over 2 v_F \sin^2 {\theta_0 \over 2 } } -  { 2   \cos \theta_0   \over v_B + v_F\sin {\theta_0 \over 2 }  }\)
 ~.
\eea
Then, demanding 
$0 = \mu \partial_\mu \lambda^i_0$
%\propto \eps f_i  - \sum_j \beta_j { %\partial f_i \over \partial %\lambda_j} \ee
with $ \{ \lambda^i \} = \{ u, g, v_B^2, v_F \}$,
we can extract the beta functions
$\beta_i \equiv - \mu\partial_\mu \lambda_i$
(note that we use the opposite sign convention for $\beta$ compared to \cite{Zinn-Justin:2002ecy}).  
The resulting equations 
determining the beta functions for the ordinary couplings $g, v_B, v_F$ are 
\bea
0 &=& \mu \partial_\mu g_0  \nonumber
\\ &=& \mu^{\eps/2}\( { \eps \over 2 } f_g 
- \beta_g { \partial f_g \over \partial g} - \beta_{v_B^2} { \partial f_g \over \partial v_B^2 }
- \beta_{v_F} { \partial f_g \over \partial v_F} ~
  \)
  \label{eq:determine-betag}
%\\ & \propto & 
%\\ 
%0 &=& \mu \partial_\mu u_0 
%= \mu^{\eps}\( { \eps } f_u
%- \beta_u { \partial f_u \over \partial u} - \beta_{g} { \partial f_u \over \partial g } 
%- \beta_{v^2} { \partial f_u \over \partial v^2} \nonumber
%\)
\\
0 &=& \mu \partial_\mu v^2_{B0} 
=  0 f_{v_B^2} 
- \beta_{v_B^2} { \partial f_{v_B^2} \over \partial v_B^2} - \beta_{g} { \partial f_{v_B^2} \over \partial g }
- \beta_{v_F} { \partial f_{v_B^2} \over \partial v_F } 
\nonumber \\
0 &=& \mu \partial_\mu v_{F0} 
=  0 f_{v_F} 
- \beta_{v_F} { \partial f_{v_F} \over \partial v_F} - \beta_{g} { \partial f_{v_F} \over \partial g }
- \beta_{v_B^2} { \partial f_{v_F} \over \partial v_B^2}~~~~\nonumber
~.
\eea

{\bf Running of velocities.}
Now we analyze the seemingly-innocuous flow of the velocities.  
We can use the last two equations of \eqref{eq:determine-betag} to eliminate 
\bea \beta_{v_B} &=& 
- \beta_g {\partial f_{v_B}\over \partial g} 
= - 2 g  {N_d \over \eps} \beta_g e_B
= - g^2 N_d e_B 
\label{eq:betav}
\\ 
\beta_{v_F} &=& 
- \beta_g {\partial f_{v_F}\over \partial g} 
= - 2 g  {N_d \over \eps} \beta_g e_F
= - g^2 N_d e_F 
\eea
up to higher-order terms.
 Thus, the beta functions for $v_B$ and $v_F$ vanish 
when 
\bea 
 0 & = & e_B = 
{  s k_F  \cos^2 \theta_0  \over 12 |v_F| } \( 1 - { v_F^2 \over v_B^2  \sin^2 {\theta_0 \over 2} }\)
\nonumber 
\\ 0 & = & e_F = { 4 k_F \over (|v_F|+|v_B|\sin{\theta_0\over 2 } )^2 } \cdot 
\\\nonumber & & \( 
{2\over 3} |v_F| - |v_B| \sin { \theta_0 \over 2} \( 1 - {1\over3 } \) \)  ,
\nonumber
\eea
These two a priori independent conditions impose the {\it same} condition on $|v_F|/|v_B|$.
The first demands that $ |v_F| = \pm |v_B| \sin {\theta_0 \over 2} $,
in which case the second says that 
\be 0 \buildrel{!}\over{=} |v_F| \( {2\over 3} + {1\over 3} - 1 \)  = 0 .\ee
We comment at this point that if we used a non-local action for the fermion field, such as 
the seemingly-appealing $\int \dbar^D k \bar\Psi \Gamma_\mu K_F^\mu \Psi$, we would not have found a simultaneous fixed point for $v_B$ and $v_F$.

Evaluated at the fixed point $|v_F| = |v_B| \sin {\theta_0\over 2} $, 
the constants in the wavefunction renormalization simplify to
$ a_\Psi = { k_F \over |v_F|}, 
a_\rho = {sk_F \cos^2{\theta_0\over 2} \over 6 |v_F|} $, so that 
\be b = {k_F \over 12 |v_F|}\( 12 + s{ \cos^2 {\theta_0 \over 2}  } \) \equiv b_0 {k_F \over |v_F|}. \ee

{\bf Finding fixed points.}
Plugging \eqref{eq:betav} into \eqref{eq:determine-betag}, 
the terms coming from $\beta_{v_B}, \beta_{v_F}$ only contribute at two-loop order (\ie~in the same way as the neglected two-loop boson self-energy correction), and we find
\bea \nonumber
\beta_g &=& \half \eps g 
%- N_d g^2 F(v)
-   
% \( {1\over v_B+v_F\cos {\theta_0 \over 2 }}+ {s \over 4 } \) 
b N_d g^3 
\eea
Note that the $g^3$ term in the beta function for the Yukawa coupling $g$ comes entirely from the wavefunction renormalization.  
The sign of the contribution of the wavefunction renormalization to the running of $g$ is fixed by the fact that $Z_\Psi < 1 $ is guaranteed by unitarity.  
The beta function for $g$ has a fixed point at
\be
g_\star^2 = { \eps \over 2 b N_d }~.
%{ - F(v) N_d + \sqrt{ (F(v)N_d)^2 + 2 b N_d \eps} \over 2 b N_d  } \buildrel{\eps \ll 1}\over{\approx} { \eps \over 2 F(v) N_d }~. 
\ee

Following a similar analysis to App.~\ref{appendix:g-is-dangerously-irrelevant}, the beta-functional for $u$ (for $\theta \in [0, \pi/2]$) is
\bea  \label{eq:integro-diff}
&&\beta_{u(\theta)} 
=  \eps u(\theta)  -N_d \left[
	 c g^2 u(\theta) 
	 \right.
\\ && \left. + {4\gamma \over |v_B|} \int_{\theta' } u(\theta') u(\theta- \theta')  
%	 +  {u(\theta)^2 \over |v_B|  }G(\theta)	
	\right. 	
\\ && \left.	
	- g^4 \(\alpha \delta_{\theta,0} + \beta \delta_{\theta, \theta_1} \)
	\right] + \cdots ~.  \nonumber
\eea
For other values of $\theta$ it is determined by the Bose symmetry relations $\beta_{u(\theta)}=\beta_{u(\theta+\pi)}
= \beta_{u(-\theta)}$ satisfied by $u(\theta)$.

Now we must analyze this integro-differential equation for the boson self-coupling function $u(\theta)$.
The linear term  in $u(\theta)$, at the fixed point for $g$, is 
\be u (\eps  - N_d c g_\star^2 )  
= u \eps \( 1 - { c \over 2 b } \)  \equiv u R
\label{eq:def-of-R}\ee
where $c, b$ are defined above.  
%As a function of $v_B, v_F$ (setting $s=1$) this looks like Fig.~\ref{fig-critical-v}. 
At the fixed point for the velocities, $|v_F| = |v_B| \sin {\theta_0\over 2}$, 
\be R=\eps \(  1 - {c \over 2 b }\) 
 = \eps \( 1 - { 6 \over 12 + \cos^2 { \theta_0\over 2 } } \) . \ee

We will analyze this equation \eqref{eq:integro-diff} by discretizing the range of $\theta$ into $N$ bins of width $ d\theta = { 2\pi/N}$.  
We rewrite the equation to make all the dependence on $d\theta$ explicit:
\bea 
\dot u(\theta) &=& R u(\theta) 
%- d\theta B(\theta) u(\theta)^2 
- C \sum_{\theta'} d\theta u(\theta') u(\theta-\theta')  \nonumber
\\ &&   + {1\over  d\theta}\( D \delta_{\theta,0} + E  \delta_{\theta, \theta_1}\) . 
\label{eq:solve-me}\eea
where $\dot u(\theta) \equiv \beta_{u(\theta)}$.
%\JM{Specify $B,C,D,E$ in terms of ...}
In terms of the QFT data, the coefficients are
as follows.  $R$ was defined in \eqref{eq:def-of-R}, and 
(at the fixed point for the velocities)
\bea
%B(\theta)&=& { 2 q_0 N_d \over |v_B| F(\theta)}, 
~ && C= { 8 \pi q_0 N_d \over |v_B|} = {8 \pi q_0 N_d \sin^2 {\theta_0 \over 2} \over |v_F| }, 
\\ \nonumber
D&=& { \eps^2 |v_F| \over N_d^2 k_F } \sin^4 {\theta_0 \over 2}{\alpha_0 \over 4 b_0^2 }
= { \eps^2 |v_F| \over N_d^2 q_0 } \sin^3 {\theta_0 \over 2}{\alpha_0 \over 2 b_0^2 }
, 
~E= D {\beta_0 \over \alpha_0}.
\eea
where we wrote 
$\alpha \equiv {k_F \over |v_F|} \alpha_0, 
\beta \equiv {k_F \over |v_F|} \beta_0, 
b= {k_F \over |v_F|} b_0$, 
$\alpha_0 = \beta_0 = {1\over 8}$.
Multiplying the BHS of \eqref{eq:solve-me} by $q_0$, we see that only the dimensionless combination $ uq_0$ appears.  Henceforth we set $q_0=1$ by redefining $u$, 
and choose units with $v_F = 1$.
$C,D,E$ are then pure numbers.  

We make the ansatz that 
\def\sing{\text{sing}}
\be 
u(\theta) = { u^{\sing}(\theta) \over d\theta} 
+ \tilde{u}(\theta) \label{eq:ansatz}
\ee
where 
\be u^{\sing}(\theta) = u^{\sing}_0 \delta_{\theta,0}
+ u^\sing_1 \delta_{\theta,\theta_1}
\ee
(for $\theta \in [0, \pi/2]$)
is nonzero only at the special angles.
If we allow support of $u_\sing$ at any other angle, the leading order equation will have no solution.

With this setup, we can expand the equation in powers of $d\theta$.  
The leading term, at
order $d\theta^{-1}$, says:
\bea\nonumber
\dot u^\sing(\theta)
&=& R u^\sing(\theta) 
- {2C } u^{\sing}_0 u^\sing(\theta) 
%- B(\theta) \left(u^\sing(\theta)\right)^2 
\\ &+& D \delta_{\theta,0}+E \delta_{\theta,\theta_1}~.
\label{eq:singular-terms}
\eea
At $\theta = 0$ this says
\be \dot u^{\sing}_0 = R u^{\sing}_0 - 2 C \left(u^{\sing}_0\right)^2 
%- B(0) \left(u^{\sing}_0\right)^2 
+ D .\ee
This has fixed points at 
\be
u^{\sing}_{0\star} = { R \pm \sqrt{ R ^2 + 8 D C} \over 4C }
\ee
of which the upper sign is positive, and corresponds to a stable fixed point, as discussed below.
The other root gives an unstable fixed point.

Similarly, at $\theta = \theta_1$, Eq. \eqref{eq:singular-terms} implies
\be 
\dot u^{\sing}_1 = R u^{\sing}_1 - 2 C u^{\sing}_0 u^{\sing}_1 
%- B(\theta_1) \left(u^{\sing}_1\right)^2 
+E. \ee
Plugging in the positive fixed-point value of $u^{\sing}_0$, the fixed-point value of $u^{\sing}_1$ (denoted as $u^{\sing}_{1\star} = E/(2 Cu_{0\star}^\sing - R)$) is also real and positive.

Before we discuss the flow of the smooth part of the coupling $\tilde{u}(\theta)$, let us check the stability of the fixed point values of $u^{\sing}_0$ and $u^{\sing}_0$. That is, we write $u^{\sing}_0 = u^{\sing}_{0\star} + \delta u^{\sing}_0$, and $u^{\sing}_1 = u^{\sing}_{1\star} + \delta u^{\sing}_1$, and ask whether the deviations $\delta u^{\sing}_0, \delta u^{\sing}_1$ shrink or grow at the linear order. One finds,
\be 
\dot{\delta u^{\sing}_0} = 
\(R - 
%\(2 B(0) +2C \)  
4C
u^{\sing}_{0\star} \) \delta u^{\sing}_0
\ee 
% \be 
% \dot{\delta u^{\sing}_0} = \left(R - 2 B(0) - 2C\right) u^{\sing}_{0\star} \delta u^{\sing}_0
% \ee 
One may verify that the expression under the brackets is always negative, irrespective of the sign of $R$. Therefore, the fixed point value of $u^{\sing}_0$ found above is stable. Similarly, one finds 
\be 
\dot{\delta u^{\sing}_1} = \left(R - 2C  u^{\sing}_{0\star} 
%-2 B(\theta_1) u^{\sing}_{1\star}  
\) \delta u^{\sing}_1
\ee 
% \be 
% \dot{\delta u^{\sing}_1} = \left(R - 2 B(\theta_1) - 2C u^{\sing}_{0\star} \right) \delta u^{\sing}_1
% \ee 
The expression under the brackets is again negative and therefore, the fixed point value of $u^{\sing}_1$ is also stable.

Next we consider our ansatz, Eq.~\ref{eq:ansatz}, at order $d\theta^0$. We find for $\theta \in [0,\pi/2]$
\bea \dot{\tilde{u}}(\theta)  \nonumber
&=& R \tilde{u}(\theta) 
- 4 C  u^{\sing}_{0\star} \tilde{u}(\theta)  
- 8 C  u^{\sing}_{1\star} \tilde{u}(\theta-\theta_1)
\\ &-&  C d\theta \sum_{\theta'}  \tilde{u}(\theta')\tilde{u}(\theta-\theta')  ~.\label{eq:smooth-part}
%\\ &-& B(0) u^{\sing}_{0\star} \delta_{\theta,0}\tilde{u}_(0)
%-B(\theta_1) u^{\sing}_{1\star} \delta_{\theta,\theta_1} \tilde{u}(\theta_1) \nonumber ~.
% + E\delta_{\theta,\theta_1}
\eea
There are a few points worth noting here. First, we have substituted the fixed point values of $u^{\sing}_{0}$ and $u^{\sing}_{1}$ in this equation. This is because as just discussed, these values are stable against small deviations. Second, we regard the sum over $\theta'$ as order $d\theta^0$ because although it has a prefactor of $d\theta$, it has $N = {2\pi \over d\theta}$ terms. We observe that the term linear in $\tilde{u}(\theta)$ is driven toward zero by a positive value of $u^{\sing}_{0\star}$ -- the Brazovskii interaction competes with $\tilde{u}(\theta)$. 
% Lastly, if one writes the above equation at a generic angle $\theta \neq 0, \theta_1$, then the last two terms involving $B(0), B(\theta_1)$ do not contribute. This suggests that one may neglect these terms. We will see below that this is indeed the case, as these terms are actually suppressed by a factor of $d\theta$ relative to other terms in this equation.

Discrete Fourier transforming the BHS via $\tilde{u}_\ell = \frac{1}{N} \sum_{\theta} \tilde{u}(\theta) e^{- \ii \ell \theta }$, and $\tilde{u}(\theta) = \sum_\ell \tilde{u}_\ell e^{\ii \ell \theta}$, we have:
\bea 
\dot{\tilde{u}}_\ell 
&=& \( R - 4 C u^{\sing}_{0\star}  - 8 C u^{\sing}_{1\star}  \cos\( \ell \theta_1\) \)\tilde{u}_\ell - 2\pi C \tilde{u}_\ell^2  \nonumber
% \\ &-& \frac{2}{N} B(0) u^{\sing}_{0\star} \tilde{u}(0) 
% %+ ( E
% - \frac{4}{N} B(\theta_1) u^{\sing}_{1\star} \tilde{u}(\theta_1)
% \cos \theta_1
% \nonumber 
\eea
for $\ell$ even; the condition $u(\theta )=u(\theta+\pi) $ implies that $u_\ell \neq 0$ only for even $\ell$. 

Now there are two possibilities for a given angular momentum mode $\ell$:

\begin{enumerate}
    \item $R - 4 C u^{\sing}_{0\star}  - 8 C u^{\sing}_{1\star}  \cos(\ell \theta_1) > 0$: 
    In this case, if one starts with $\tilde{u}_\ell \approx 0$ and positive, $\tilde{u}_\ell$ grows until it reaches the fixed point value of 
    \be\tilde{u}_{\ell\star} = \left(R - 4 C u^{\sing}_{0\star} - 8 C u^{\sing}_{1\star}  \cos(\ell \theta_1)\right)/ (2\pi C) .\ee 
    This is a stable fixed point as one may readily verify by writing $\tilde{u}_{\ell} = \tilde{u}_{\ell\star} + \delta \tilde{u}_{\ell}$ and noticing that the perturbation $\delta \tilde{u}_{\ell}$ always shrinks towards zero. On the other hand, if $\tilde{u}_\ell$ starts out negative, then it runs away to negative infinity (of course, our equations are valid only for small $\tilde{u}_\ell$, but it suffices to say that in this case, we don't find a perturbatively accessible stable fixed point).
    \item $R - 4 C u^{\sing}_{0\star}  - 8 C u^{\sing}_{1\star}  \cos(\ell \theta_1) < 0$: In this case, if one starts with $\tilde{u}_\ell \approx 0$, irrespective of its sign, then $\tilde{u}_\ell$ flows back to zero. On the other hand, if one starts out with $\tilde u_\ell$ negative and large in magnitude, then one again experiences a run-away flow.
\end{enumerate}

\begin{figure}
    \centering
 \includegraphics[width=0.5\textwidth]{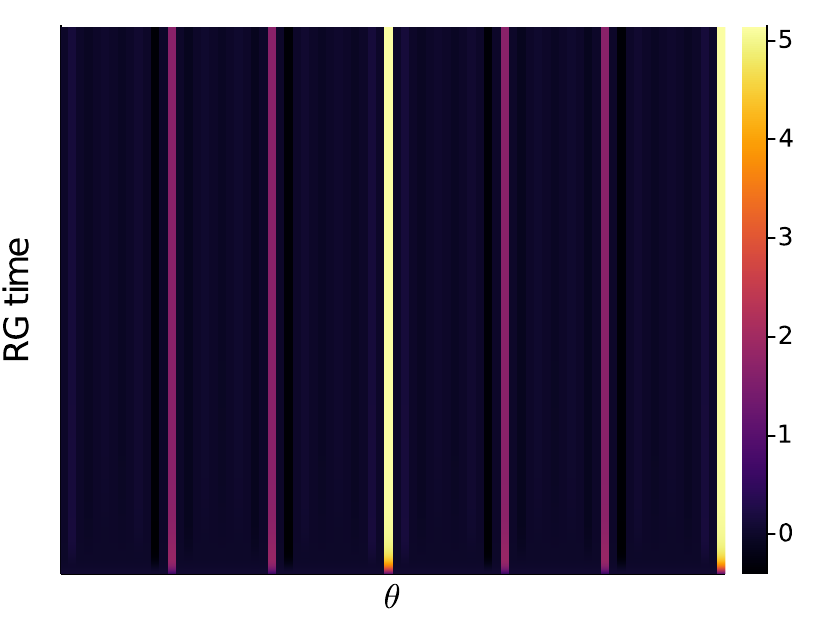}
 \parfig{.5}{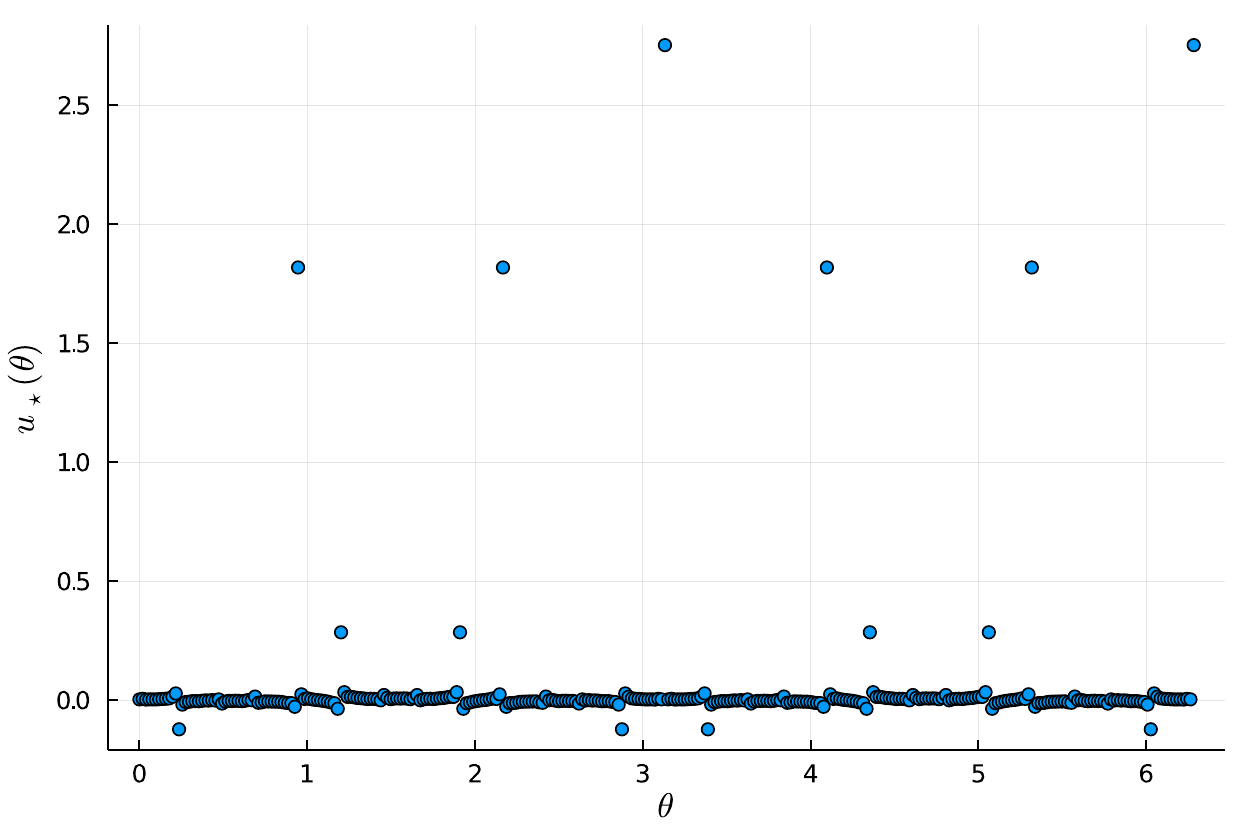}
    \caption{Top: The results of a numerical solution to the RG equation for $u(\theta)$ starting from the analytic fixed-point configuration. 
    The bright yellow lines are $\theta =0, \pi$, and the other lines are at $\theta_1$ and its images under the Bose symmetry relations.  
    Bottom: The fixed-point configuration $u_\star(\theta)$ with $N=320$.  Both figures use the value of $\theta_0$ for the commensurate filling of the square lattice.}
    \label{fig:brazovskii-wins}
\end{figure}

The above calculation implies that the stable fixed point corresponds to
\be 
u_\star(\theta) = { u_\star^\sing(\theta) \over d\theta} 
+ \tilde{u}_\star(\theta)
\ee 
where $u_\star^\sing(\theta) = u^\sing_{0\star} \delta_{\theta,0}
+ u^\sing_{1\star} \delta_{\theta,\theta_1}$, and 
$\tilde{u}_\star(\theta) = \sum'_\ell \tilde{u}_{\ell\star} e^{\ii \ell \theta}$ where the sum over $\ell$ runs over those values of $\ell$ that satisfy $R - 4 C u^{\sing}_{0\star}  - 8 C u^{\sing}_{1\star}  \cos(\ell \theta_1) > 0$. Here we have again restricted the angle $\theta \in [0, \pi/2]$, and the values for other angles follows from symmetry. The linear stability of this fixed point follows from the aforementioned considerations, but let us repeat the argument for the sake of completeness. We have already shown explicitly that the fixed point value of the leading term in $d\theta$, namely, $u_\star^\sing(\theta)$, is stable against small perturbations. The linear stability of the term $\tilde{u}_\star(\theta)$ follows from the fact that an arbitrary perturbation may be decomposed into its angular momentum modes, and since we are only considering linear stability, one may consider the stability of each angular momentum mode separately. We have shown above that for any $\ell$, the fixed point corresponding to $\tilde{u}_{\ell} = \tilde{u}_{\ell\star}$ mentioned above is stable to linear order.
The fixed-point solution for $u(\theta)$ and the RG behavior in its neighborhood are shown in Fig.~\ref{fig:brazovskii-wins}.

%%%%%%%%%%%%%%%%%%%%%%%%%%%%%%%%%%%%

\subsection{Critical exponents}

The anomalous dimension for coupling or field $a$ can be found by 
\be\eta_a(\lambda) = 
- \lambda_i \partial_{\lambda_i} \alpha^{(1)}_a(\lambda),\ee 
where $\alpha^{(1)}_a(\lambda)$ is the coefficient of $1/\eps$ in the corresponding renormalization coefficient.  Using this, we have 
\bea\nonumber
\eta_\rho &=& a_\rho g_\star^2 N_d 
%= 0.0345 \eps
={ s \eps \cos^2 {\theta_0 \over 2}\over 12 + s \cos^2 { \theta_0 \over 2} }=\begin{cases} 0.0176 \eps,&  s=1\\
0.0345 \eps, & s=2
\end{cases}
\\ \nonumber
\eta_\Psi &=& a_\Psi g_\star^2 N_d
={ 6 \eps \over 12 + s \cos^2 { \theta_0 \over 2} }
%= 0.483 \eps 
=\begin{cases} 0.491 \eps,&  s=1\\
0.483 \eps, & s=2\end{cases}~.
\eea
Recall that $N_d = {1\over 16 \pi^3}$.
The numerical values are taken at the commensurate value of $q_0/k_F$ for the square lattice.  

\begin{figure}[h]
    \centering
 \includegraphics[width=0.5\textwidth]{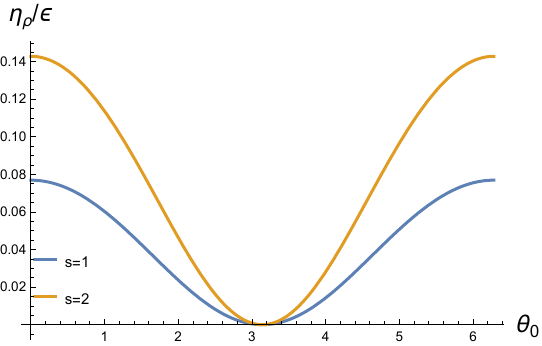}
 \includegraphics[width=0.5\textwidth]{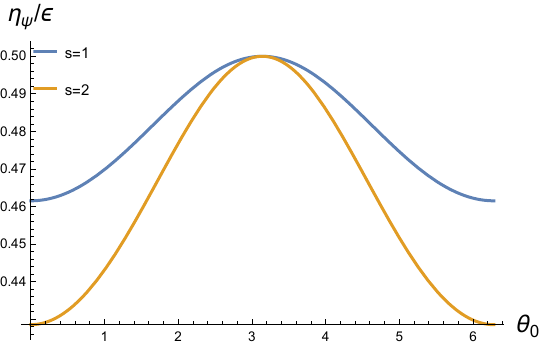}
 \includegraphics[width=0.5\textwidth]{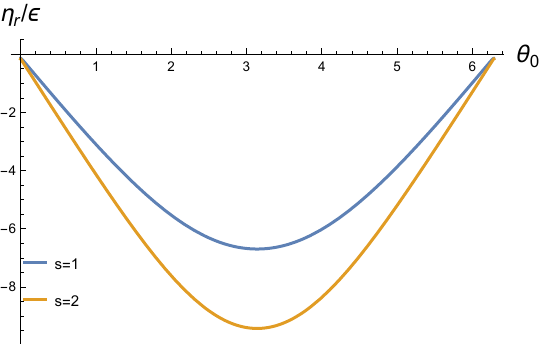}
    \caption{Critical exponents as functions of $\theta_0$ for $s=1,2$.  
Top: $\eta_\rho$ Middle: $\eta_\psi$    Bottom: The correlation length critical exponent $\eta_r$ in \eqref{eq:etar}.}
    \label{fig:etar-vs-theta0}
\end{figure}

From the running of $r$, we find that the correlation length critical exponent is
\bea
\eta_r &=& - N_d\( g_\star^2 a_\rho
+ {\gamma\over |v_B|} \int \dbar \theta u_\star(\theta) \)~.
\eea
At the fixed point, 
\bea \gamma \int \dbar \theta u_\star(\theta)
&=& 2 u_{0\star}^\sing + 4 u_{1\star}^\sing + 2\pi \tilde u_{\ell=0,\star} 
\\ &=&  { 4 D + 8 E \over  \sqrt{8 CD + R^2} - R }.
%= 265.2 \eps. 
\nonumber
\eea
The resulting function of $\theta_0$ and $s$ is rather unwieldy:
\begin{widetext}
\be \label{eq:etar}
\eta_r = - \eps { 
  16 \pi s (1+\cos \theta_0 )  
+ 3  \( 24 - 11 s + s \cos\theta_0  + \sqrt{ ( 24 - 11 s + s \cos\theta_0)^2 - 36864 \pi^4 s ( \cos \theta_0 - 1 ) } \)
\over 16 \pi (24 + s(1+ \cos \theta_0)}
\ee
\end{widetext}
This function of $\theta_0$ is depicted in Fig.~\ref{fig:etar-vs-theta0}.
For the commensurate value of $\theta_0$ for the square lattice, we find
\be \eta_r = 
\begin{cases} 2.31 \eps,&  s=1\\
3.18 \eps, & s=2\end{cases}.\ee

Although we believe that all critical exponents should be determined by the dimensions of the relevant operators, because this critical point comes with not just one but two dimensionful quantities ($k_F$ and $q_0$), we do not expect the usual hyperscaling relations to hold.  
We leave for the future a direct calculation of other exponents, such as the order parameter exponent $\beta$.

\section{An unsuccessful scheme for implementing $\mathsf{LU}(1)$ symmetry}
\label{appendix:LU1}

\begin{figure}
    \centering
\parfig{.23}{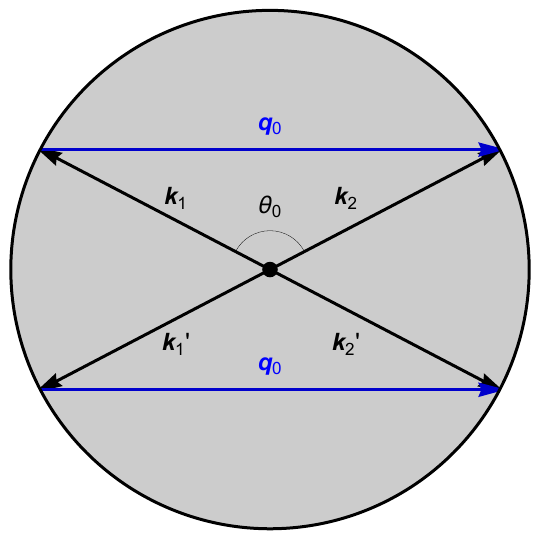}  
\parfig{.23}{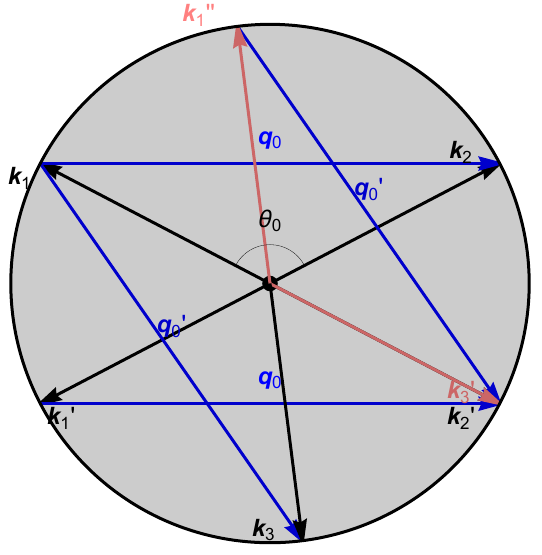}  
\caption{Left: The labelling of boson momenta by pairs of points on the Fermi surface $\vec q = \vec k_1 - \vec k_2$ is not unique. 
Right: Each point $\vec k_1$ participates in two such decompositions of a boson momentum $\vec q$.  Combining these two facts, the $\mathsf{LU}(1)$ transformation parameter at each point on the Fermi surface must be the same as every other. }
    \label{fig:LU1}

\end{figure}

Recall that an ersatz Fermi liquid is 
defined to be a system (in two spatial dimensions) with a Fermi surface that, like a Fermi liquid, has a $\mathsf{LU}(1)$ symmetry, associated with independent fermion number conservation at each point on the Fermi surface.
It seems that the only way such a symmetry can be preserved by the Yukawa coupling to $\rho$ is if $\rho(x)$ also transforms under the symmetry.  
In particular, a mode of $\rho$ labelled by momentum
$ \vec q = \vec k_1 - \vec k_2$ connecting two points $\vec k_{1,2}$ on the Fermi surface must transform as a bifundamental 
under the associated $\gU(1)$ factors:
\be \rho_{\vec k_1 - \vec k_2 } \to e^{ \ii \( \varphi(\vec k_1)- \varphi(\vec k_2)\) } \rho_{\vec k_1 - \vec k_2}. \ee
Interestingly, such a labelling of boson momenta by the points on the Fermi surface that they connect is also an ingredient in the double-line notation introduced by Sung-Sik Lee to account for the $1/N$ expansion in various non Fermi liquids \cite{Lee:2009epi}. 
However, this labelling is not unique: 
each vector $\vec q$ connects {\it two} pairs of points on the Fermi surface:
there is always a second pair of points:
$ \vec q_0  = \vec k_2' - \vec k_1'$, as in Fig~\ref{fig:LU1}.
But this implies that only the transformations where the $\gU(1)_{k_1}$ parameter is the same as the $\gU(1)_{k_1'}$ parameter can be symmetries.  
In examples studied in \cite{2023AnPhy.45069221B, 2024PhRvB.110o5142K}, similarly, only a subgroup of the full $\mathsf{LU}(1)$ is a symmetry.

However, in our model, the point $k_1$, in turn, also appears in another difference, namely:
$\vec q_0' = \vec k_3 - \vec k_1 $, as in Fig~\ref{fig:LU1}.
And this other vector $\vec q_0'$ can also be decomposed as 
$ \vec q_0' = \vec k_3' - \vec k_1''$, as in Fig~\ref{fig:LU1}.
(In fact, the points $k_3'$ and $k_2'$ are the same point.)
In this way we learn that the only allowed $\mathsf{LU}(1)$ transformation has parameter
$ \alpha_{k_1} = \alpha_{k_1'} = \alpha_{k_1''} $.  
But now $k_1'$ and $k_1''$ both also appear in another difference, and in this way it seems we can relate all of the transformations at each point on the Fermi surface, so that the symmetry is in fact just $\gU(1)$.

\bibliographystyle{ucsd}
\bibliography{references}

\clearpage

\end{document}